\input amstex
\documentstyle{amsppt}
\magnification=1200

\loadbold
\define \hl{\;\hat\ll\;}
\define \pl{\ll^p}
\define \h{$\;\widehat{}\;$-}
\define \I{\Cal I\Cal P}
\define \lo{\Bbb L}
\define \r{\Bbb R}
\define \s{\Bbb S}
\define \<{\langle}
\define \>{\rangle}
\define \bm{\bar M_\phi^+}
\define \bdm{\partial_\phi^+(M)}

\topmatter
\title
Topology of the Future Chronological Boundary: \\
Universality for Spacelike Boundaries 
\endtitle
\rightheadtext{Topology of the Future Chronological
Boundary}
\author
Steven G. Harris
\endauthor
\address
Department of Mathematics, Saint Louis University,
St. Louis, MO 63103, USA\endaddress
\email harrissg\@slu.edu\endemail
\abstract
A method is presented for imputing a topology for
any chronological set, i.e., a set with a
chronology relation, such as a spacetime or a
spacetime with some sort of boundary.  This
topology is shown to have several good properties,
such as replicating the manifold topology for a
spacetime and replicating the expected topology for
some simple examples of spacetime-with-boundary; it
also allows for a complete categorical
characterization, in topological categories, of the
Future Causal Boundary construction of Geroch,
Kronheimer, and Penrose, showing that construction
to have a universal property for future-completing
chronological sets with spacelike boundaries. 
Rigidity results are given for any reasonable
future completion of a spacetime, in terms of the
GKP boundary:  In the imputed topology, any such
boundary must be homeomorphic to the GKP boundary
(if all points have indecomposable pasts) or to a
topological quotient of a closely related boundary
(if boundaries are spacelike).  A large class of
warped-product-type spacetimes with spacelike
boundaries is examined, calculating the GKP and
other possible boundaries, and showing that the
imputed topology gives expected results; 
included among these are the Schwarzschild
singularity and those Robertson-Walker singularities
which are spacelike.
\endabstract
\endtopmatter

\head 1. The Future Chronological
Boundary
\endhead

\subhead 1.1 Introduction 
\endsubhead

In 1973, Geroch, Kronheimer,
and Penrose  introduced in \cite{GKP} the notion of
the Causal Boundary for a strongly causal
spacetime.  This is a seemingly very natural
method of appending a future endpoint to each
future-endless timelike curve in a spacetime (and
a past endpoint to each past-endless timelike
curve), in a manner which is conformally invariant
and which depends, for the future endpoints, only
on the past of each curve (and dually for the past
endpoints).  If
$M$ denotes the spacetime and $\partial^\#(M)$ its
Causal Boundary, then we speak of $M^\# = M \cup 
\partial^\#(M)$ as the ``completion" of $M$ by the
causal boundary.  This comes equipped with  a
chronology relation $\ll^\#$ and a causality
relation $\prec^\#$ which extend those ($\ll$ and
$\prec$) on $M$; however,
$x \ll^\# y$ or $x \prec^\# y$ may possibly 
obtain in $M^\#$ for $x$ and $y$ in $M$, even
though $x \ll y$ or $x \prec y$ do not hold
(though this cannot happen if $M$ is globally
hyperbolic).  There is also defined a
topology on $M^\#$ which induces the original
topology on $M$ as a subspace and makes the
elements of
$\partial^\#(M)$ actual endpoints to the
appropriate curves in $M$.

For those interested in understanding the
large-scale structure of spacetimes in terms of a
natural boundary, this all sounds quite good; but 
there are problems:  The construction of the Causal
Boundary consists of first defining what may be
called the Future Causal Boundary (the future
endpoints of timelike curves) and, separately, the
Past Causal Boundary.  Then these two are melded
together in an elaborate procedure (though this is
quite simple if there are only spacelike components
of the boundary and no naked singularities).  A very
complicated topology is defined on this combined
boundary, which is further adumbrated with quotients
so as to make it Hausdorff.  The total complications
are quite formidable, and very little has been done
in the way of computing the Causal Boundary for
specific spacetimes (an example of an explicit
calculation for a flat, multiply-connected
two-dimensional spacetime is given in
\cite{HD}; this also gives an example of how the
Causal Boundary may be utilized to help
understand the behavior of a spacetime).   Besides
this, the topology as defined by GKP is not,
actually, at all what one might expect; for
instance, for Minkowski space, the Causal Boundary
is a pair of  null cones (one each for the future
and the past), just as in the standard conformal
embedding of Minkowski space into the Einstein
static universe (see
\cite{HE})---except that each cone element (null
line) is an open set!

But, still, the notion  of the Causal Boundary
seems, somehow, very natural.  Can that be
established in a rigorous sense, and, if so, can
information about other boundaries of spacetimes
be derived from knowledge of the Causal Boundary? 
It is the purpose of an intended series of
articles, of which this is the second, to give
affirmative answers to these questions.

The first article, \cite{H}, established a
rigorous sort of ``naturalness" for a portion of
the Causal Boundary: the Future Causal Boundary
(without melding with the Past Causal Boundary),
with its chronology relation only (no causality
relation and no topology).  To emphasize that this
is only a portion of the full GKP Causal Boundary,
which has both chronology and causality
relations, it was called the Future Chronological
Boundary, and that nomenclature is preserved
here.  The form of ``naturalness" explicated in
that article (and this) is categorical: showing
that the constructions defined are functorial,
natural, and universal, in the sense of category
theory.  The benefit derived from this is that one
then is assured that the constructions are
categorically unique for having the requisite
properties---any other functorial construction
with the same naturalness and universality must be
naturally equivalent to the one at hand, in the
strict categorical sense.

Here is what all that means in simplified
outline:  First there was defined a category of
objects with morphisms between those objects, a
category large enough to contain both spacetimes
and spacetimes-with-boundaries (and
chronology-preser\-ving continuous functions between
spacetimes); these objects are called
chronological sets, and the morphisms,
future-continuous functions.  Then it was shown
that adding the Future Chronological Boundary
$\partial^+(X)$ to a chronological set
$X$ produces a future-complete chronological  set
$X^+$; and that this process is a functor from the
category of chronological sets to the subcategory
of future-complete chronological sets (so for any
future-continuous $f: X
\to Y$ between chronological sets, we obtain the
extension $f^+: X^+ \to Y^+$ of $f$).   The
inclusion  $\iota_X^+ : X \to X^+$ was shown to be
a natural transformation (i.e, it commutes with $f$
and $f^+$: $f^+ \circ \iota_X^+ = \iota_Y^+ \circ
f$).  Finally, the appropriate universality
property was established:  For any
future-continuous
$f: X \to Y$ for which $Y$ is already 
future-complete, $f^+$ is the unique
future-continuous extension of $f$ to $X^+$, i.e.,
so that $f^+ \circ \iota_X^+ = f$.  The upshot is
that the future completion process is thus  left
adjoint to the forgetful functor from
future-complete chronological sets to
chronological sets; and left adjoints, by category
theory, are unique up to natural equivalence
(natural transformations consisting of
isomorphisms)---see \cite{M}.  It is the
universality principle---$f^+ \circ \iota_X^+ =
f$---that is the most useful piece of information,
allowing comparison of any other way of
future-completing
$X$ (i.e., of mapping $X$ into any future-complete
object $Y$) with the GKP future completion (i.e.,
$X^+$).

It must be emphasized that despite the usage of
the phrase ``future-continuous function", there
was no real topology in the constructions above: 
The term ``future-continuous" refers to
preservation of ``future limits", a generalization
of future endpoints for timelike curves; there was
no information in those constructions relating to
more general notions of convergence.  It is the
purpose of this paper to fill that lacuna: to
establish functoriality, naturalness, and
universality for the Future Chronological Boundary
construction, including appropriate topology---at
least in the case of spacelike boundaries.

One of the reasons why categorical properties 
such as universality are important for a boundary
construction process, is that they allow one to
deduce much about all possible boundaries---at
least, those meeting the same basic conditions, such
as being completing objects in the same category. 
As a consequence, there are developed here several
important rigidity (or quasi-rigidity) results for
future-completing boundaries on spacetimes; these
take the form of saying that any
``reasonable" future-completing boundary on a
spacetime must be topologically identical to (or,
depending on the hypotheses, a topological quotient
of) the GKP Future Causal Boundary---and the GKP
boundary is fully explicated here for a number of
classical spacetimes, such as interior Schwarzschild
and some Robertson-Walker spaces.  However, these
results depend crucially on accepting a specific
topology for the chronological sets (spacetime {\it
cum\/} boundary) that one is looking at; and a large
burden of this paper is to make the case for the
acceptance of this topology.

\medpagebreak

The construction of $X^+$ in \cite{H} was divided
into two procedures, each independently functorial,
natural, and universal:  First, the construction of the
Future Causal Boundary $\hat\partial(X)$ of a
chronological set $X$, with a   ``first
approximation" of the needed chronology relation
on $\hat X = X \cup \hat\partial(X)$ (i.e., a
relation, $\hat\ll$, that extends that of $\ll$ on
$X$ without relating any elements of $X$ that
$\ll$ does not); this is called the
future completion functor.  Second, a way of
extending a chronology relation to relate
additional elements in a chronological set which
is necessary for certain purposes; this is called
the past-determination functor.  It is, in
essence, the composition of these two functors
that results in $X^+$, the addition to $X$ of what
amounts to the GKP Future Causal Boundary.

The second part of Section 1 gives an overview of
the results presented here in general form, with
some discussion of significance.  The third part
gives a detailed summary of the constructions and
conventions previously established in \cite{H},
upon which the rest of this paper builds.  A
rigidity theorem for ``nice" boundaries (in the
chronological category) is added in the fourth
part.  

The basic ideas for the Future Chronological
Boundary are contained in the future completion
functor, and it is the inculcation of topology
with that functor that will be discussed first, in
Section 2.  However, it must be noted that the
future completion functor operates only on
chronological sets that are already
past-determined.  Section 3 explores how topology
works with the past-determination functor.  Both
these  Sections assume ``regularity": that every
point in the chronological sets under discussion has
an indecomposable past (true for spacetimes and for
some spacetimes-with-boundary).  Section 3 ends with
important rigidity theorems for regular chronological
sets

But there are spacetimes for which the Causal
Boundary construction yields non-regular
points: elements of the boundary whose pasts are not
indecomposable past sets (typically for boundaries
which are timelike and two-sided in the spacetime). 
This presents a considerable complication for the
introduction of topology by the means outlined in
Section 2, but it is important to cover such
boundaries, as they are sometimes the most natural
ones.  This forms the subject of Section 4, which
concludes with quasi-rigidity theorems covering
chronological sets that are not necessarily regular,
but have spacelike boundaries.

Finally, Section 5 looks at examples, including
some classical spacetimes.  Included is a rigidity
result for boundaries formed by embedding into
larger manifolds.

\subhead
1.2 Discussion and overview
\endsubhead
 
The import of the results in \cite{H} was
two-fold:  First, to show how the GKP construction
of the (Future) Causal Boundary is a formally
natural and universal construction, in an
appropriately (albeit limited) categorical sense;
and second, to show how causal and set-theoretic
information can be derived about any purported
(and ``reasonable") future boundary of a
spacetime, via a map from the GKP boundary.  The
idea followed here is to recast these results in a
topological frame, by this means:  

There will be defined a way to impute a topology
for any chronological set $X$; then the results
from \cite{H} will be shown to hold in appropriate
topological categories:  For instance, morphisms
in the new categories will be required to be
continuous with respect to the topologies inferred
from the chronology relations, and then it will be
shown that the same functors as from \cite{H},
applied to continuous morphisms, result in
continuous morphisms.  That will allow the
categorical results from \cite{H} to stand in the
new topological categories, as well.

But an important part of the problem is to ensure
that one has the right topology.  What is required
for this?  At a minimum, we certainly need that if
the chronological set $(X, \ll)$ under discussion
is actually a spacetime, then the topology
inferred from $\ll$ had better be the original
manifold topology; the topology defined here
(called the \h topology) satisfies this basic
requirement.

Beyond that, one may wish to require that if $(X,
\ll)$ is a spacetime with some ``natural"
topological boundary, coming equipped
with an extension of the spacetime chronology
relation, then the inferred
\h topology ought to match this natural topology. 
It will be shown that this is the case for a class
of warped-product spacetimes with ``obvious"
spacelike boundary, and also with some
modifications of that boundary; and that this
generalizes to similar modifications of general
spacetimes with spacelike boundary.

The spacetimes considered for examples are
generalizations of interior Schwarz\-schild, with
its spacelike singularity as the boundary: a
product of a portion of the Lorentzian line with a
product of complete Riemannian factors $(K_i,h_i)$,
$M = (a,b) \times \prod_i K_i$, with a warped product
metric $-(dt)^2 \,+\, \sum_i f_i(t)h_i $; if the warping
functions $f_i$ satisfy an integral condition, then the
Future Chronological Boundary of $M$ will be
spacelike.  (Examples of classical spacetimes with
this conformal structure are interior
Schwarzschild, the Kasner spacetimes, and many
standard static spacetimes.) One anticipates that
the boundary of $M$ ``ought" to be $K = \prod_i
K_i$, attached to $M$ so as to form $\bar M \cong (a,b]
\times K$; and this is precisely what $M^+ =
\hat M$ is in the \h topology. This is some
confirmation that the inferred topology is doing
what it ought to do.
 
Our anticipation of that form for the
boundary of $M$ comes from a canonical embedding
of $M$ into $\r \times K$.  But one can
imagine many other embeddings $\phi : M \to N$, yielding
some construct $\bar M_\phi$ as some sort of
topological completion of $M$.  There is also a
natural chronology relation on $\bar M_\phi$, so
we can ask if the \h topology from that chronology
relation replicates the natural topology on $\bar
M_\phi$ coming from the embedding $\phi$.  So long
as $\phi$ extends continuously to $\hat
\partial(M)$, and that extension is proper onto
its image, the answer is yes.  Furthermore, this
is not restricted to the warped product spacetimes
of the previous paragraph, but to any strongly
causal spacetime with $\hat\partial (M)$ spacelike
(one may speak of $\phi$ extending either to
$M^+$ or to $\hat M$, as these are naturally
homeomorphic in the \h topologies; past-determination
makes only minor changes in the chronology relation,
not affecting the \h topology).

The insistence on a spacelike boundary for these
examples is a crucial element in the exposition
featured here; in fact, it is only spacelike
boundaries that can be admitted in the category for
which the chronological categorical results can be
extended to topological ones.  Timelike boundaries
have intrinsic problems for the categorical
extension:  There are simple examples of a
spacetime $M$ with timelike boundary (such as the
$x > 0$ portion of Minkowski 2-space) and
continuous,  future-continuous map $f : M \to N$,
such that the extension of $f$ to the boundary of
$M$ (\i.e., $\hat f$, the application of the
future completion functor to $f$) is
discontinuous; indeed, there is no continuous
extension of $f$ to the boundary.  But
future-continuity of $f$ prevents this happening
for spacelike boundaries (what happens for null
boundaries is an open question at this time). 
Thus, the topological category featured will be
chronological sets $X$ with $\hat\partial(X)$
spacelike.

An important reason for believing in the correctness
of the topology construction featured here---the \h
topology for any chronological set---is the fact
that it does lead to strongly categorical results. 
The fullest statement of these results is Theorem
3.4, applicable in the category of
past-distinguishing, regular chronological sets with
spacelike boundaries.  The truly interesting
results---the ones with the most clear physical
impact---are the rigidity theorems which, utilizing
the \h topology, sharply curtail the allowable
boundaries on spacetimes.  But these results are
interesting only in so far as one believes that the
\h topology is the way to understand the topological
behavior of the boundaries in question when defined
only by their chronological characteristics.  Thus,
in a somewhat curious fashion, the categorical
results both help establish the {\it bona fides\/}
of the \h topology construction, and also yield some
of the more interesting topological consequences in
terms of that \h topology.

We then have these imports for the topological
results:  First, the GKP construction, with
topology, is seen to be the categorically unique
way to future-complete a spacetime in a manner
which has a universal relation to any other
future completion (at least, with respect to
spacetimes with spacelike boundaries).  Second, that
universality property gives us a way of relating---as
a topological quotient---any proposed future boundary to
the GKP boundary (in case that boundary is spacelike);
this is a sort of quasi-rigidity for future boundaries.
Third, that universality (and associated
quasi-rigidity) is applicable in cases beyond the
strictly categorical ones, including situations
involving ``non-regular" spaces, where the past of a
boundary point may not be indecomposable (these must
be accounted as reasonable boundary constructions). 
Fourth, for regular boundaries, there is a strong
rigidity result applicable even without the assumption
of spacelike boundaries.

The most important results of this paper probably
are the rigidity results of Sections 3, 4, and 5:  

Theorem 3.6 (foreshadowed by Theorem 1.1) has
this implication:  For any spacetime $M$, any
regular, past-distinguishing future completion of
$M$ must be homeomorphic (in the \h topology)  to
$\hat M$, and the future-completing boundary must
be similarly homeomorphic to the Future
Chronological Boundary.  (Note that this result is
not restricted to spacelike boundaries.)  As an
example:  Proposition 5.2, applied to interior
Schwarzschild, shows that the Schwarzschild
singularity, in its GKP formulation, has the topology
of $\r^1 \times \Bbb S^2$, and Theorem 3.6 asserts
that this is the only possible topology for a
regular, past-distinguishing, future-completing
boundary on that spacetime (assuming the \h topology
is used for such a boundary).  
 
Theorem 4.8 addresses
the same question with respect to allowing for
non-regular, ``generalized" past-distinguishing,
``generalized" future completions (where
``generalized" accommodates for non-regular
points):  Restricting to the situation of spacelike
boundaries, this theorem implies that the completing
object must (in the \h topology) be a topological
quotient of $\hat M$; from Corollary 4.9, the
completing boundary must be a topological quotient
of what I call the Generalized Future Chronological
Boundary (a subset of the Future Chronological
Boundary, removing IPs corresponding to non-regular
points).  Applied to the Schwarzschild
singularity, this means that anything even
approaching a reasonable future completing
object can be only a quotient of $\r^1 \times \Bbb
S^2$---again, using the \h topology.  

Theorem 5.3,
showing that embeddings (of a proper sort) yield
future completions with natural topology the same as
the \h topology, then yields a quasi-rigidity for
this most common means of completing a spacetime. 
For Schwarzschild, this means that if we attempt to
derive a topology for the singularity, not by
imposing the \h topology on a boundary with some
chronology relation, but by the purely topological
means of embedding the spacetime into a larger
manifold, then the same result still holds (if the
embedding is proper onto its image):  The only
possibilities for the singularity are quotients of
$\r^1 \times \Bbb S^2$.

\subhead
1.3 Summary of previous nomenclature and
constructions
\endsubhead
 
The following is all from \cite{H}:

A {\it chronological set\/} is a set $X$ together
with a relation $\ll$ which is transitive and
non-reflexive, such that every point in $X$ is
related to at least one other point, and such that
there is a countable subset
$S$ of $X$ so that for all $x \ll y$ in $X$,
there is some $s \in S$ with $x \ll s \ll y$.  For
any $x \in X$, the $past$ of $x$ is $I^-(x) = \{y
\;|\; y \ll x\}$; for $A \subset X$, the past of
$A$ is $I^-[A] = \bigcup_{a \in A}I^-(a)$.  (The
use of square brackets for a subset is to be
noted, as subsets of a chronological set $X$ will
often be used as elements of an extension of $X$,
and it is needful to distinguish whether one is
speaking of the past of a subset {\it qua\/}
subset of $X$ or {\it qua\/} element of the
extension; in general, parentheses denote the
application of a function to an element of its
proper domain, while brackets denote the usage of
that function as an operator on subsets of the
function's domain.)  Futures are defined dually,
using $I^+$. A chronological set $X$ is called
{\it past-distinguishing\/} if $I^-(x) = I^-(y)$
implies $x = y$.  Any strongly causal spacetime is
a past-distinguishing chronological set (Theorem 4
in \cite{H}). 

A {\it future chain\/} in a chronological set $X$
is a sequence $\{x_n\}$ of elements of $X$ obeying
$ x_n \ll x_{n+1}$ for all $n$; in a chronological
set, future chains serve in the role of timelike
curves in a spacetime.  A point $x$ is a {\it
future limit\/} of a future chain $c$ if $I^-(x) =
I^-[c]$; if $X$ is past-distinguishing, then a
future chain can have at most one future limit.  A
function $f: X \to Y$ between chronological sets
is {\it chronological \/} if $x \ll y$ implies
$f(x) \ll f(y)$; note that this implies a future
chain gets mapped to a future chain.  A
chronological function is {\it
future-continuous\/} if it preserves future limits
of future chains.  For strongly causal spacetimes,
a future limit of a future chain is precisely the
same as a topological limit of the sequence;
functions which are both past- and
future-continuous are the same as continuous
functions which preserve future-directed timelike
curves (Theorem 4 in \cite{H}).

A non-empty subset $P \subset X$ is a {\it past
set\/} if $I^-[P] = P$; a past set $P$ is {\it
indecomposable\/} if it cannot be written as the
union of proper subsets which are past sets. For
each indecomposable past set (or IP) $P$ there is a
(non-unique) future chain $c$ such that $P =
I^-[c]$ ($c$ is said to {\it generate\/} $P$), and
any set of the form $I^-[c]$, for $c$ a future
chain, is an IP (Theorem 3 in \cite{H}).  This is
in strict analogy with IPs in strongly causal
spacetimes being precisely those subsets which are
the pasts of timelike curves.  A past set $P$ is
an IP if and only if for every pair of points in
$P$, there is a third point in $P$ to the future
of each of the first two (Theorem 2 in \cite{H}).

A chronological set is called {\it
future-complete\/} if every future chain has a
future limit.  No strongly causal spacetime is
future-complete.  It is the job of the
future completion functor to provide a
future-complete chronological set $\hat X$ for
each chronological set $X$.  This is accomplished
by adding to the point-set $X$ the {\it Future
Chronological Boundary\/} of $X$, $\hat\partial(X)
= \{P \;|\; P$ is an IP in $X$ such
that $P$ is not $I^-(x)$ for any
point $x \in X\}$; we let $\hat X = X
\cup \hat\partial(X)$, the {\it future
(chronological) completion\/} of $X$.  By Theorem
5 in \cite{H}, this is a chronological set under
the following relation, where
$x$ and $y$ are any points in $X$, $P$ and $Q$ any
elements of $\hat\partial(X)$:

\roster
\item $x \hl y$ iff $x \ll y$
\item $x \hl Q$ iff $x \in Q$
\item $P \hl y$ iff for some $w \in I^-(y)$,
$P \subset I^-(w)$
\item $P \hl Q$ iff for some $w \in Q$,
$P \subset I^-(w)$.
\endroster
To avoid confusion, $\hat I^-$ will sometimes be
used to denote the past in
$\hat X$, to distinguish from the past in $X$.

The future completion of a chronological set is, as
the name suggests, future-complete:  For any
future chain $c$ in $X$, $I^-[c]$ is an IP which
either is $I^-(x)$ for some $x$ which is a future
limit for $c$ (in $X$), or is itself a future
limit for $c$ in $\hat X$; for a chain $c$
including a subsequence of elements
of $\hat\partial(X)$, there are interpolated
elements of $X$ (as per (3) and (4) above), and
these generate the future limit for $c$.  The
inclusion map $\hat\iota_X : X \to \hat X$, the
{\it standard future injection\/} for
$X$, is future-continuous.  If $X$ is
past-distinguishing, then so is its future
completion.  If $X$ is itself future-complete,
then $\hat X = X$.

Let $f: X \to Y$ be a chronological function
between chronological sets with $Y$
past-distinguishing; we need to define an
extension of $f$ to the future completions, $\hat
f: \hat X \to \hat Y$.  For $x \in X$, we define
$\hat f(x) = f(x)$.  For $P \in \hat\partial(X)$,
generated by a future chain $c$, consider $Q =
I^-[f[c]\,]$, an IP in $Y$:  Either $Q = I^-(y)$
for a unique $y \in Y$, in which case we define
$\hat f(P) = y$; or there is no such point in
$Y$, in which case $Q \in \hat\partial(Y)$, and we
define $\hat f(P) = Q$. 

In order that $\hat f$ also be chronological, we
need an additional assumption on $Y$:  We define a
chronological set to be {\it past-determined\/} if
whenever $I^-(y) \subset I^-(w)$ and $w \ll x$,
we also have $y \ll x$ (mimicking the definition of
$\hl$ for future completions).  This is true for
globally hyperbolic spacetimes, but false for many
spacetimes with ``holes" (such as Minkowski
2-space with a spacelike half-line removed).  If a
chronological set is past-determined, then so is
its future completion. 

Now let $f: X \to Y$ be a chronological map with
$Y$ past-distinguishing and past-determined; then
$\hat f$ is also chronological.  Moreover, if $f$
is future continuous then $\hat f$ is the unique
future-continuous map satisfying $\hat f \circ
\hat\iota_X = \hat\iota_Y
\circ f$.  An alternative formulation:  For $f: X
\to Y$ future-continuous with $Y$ future-complete
as well as past-determined and past-distinguishing,
$\hat f: \hat X \to Y$ is the unique 
future-continuous function with $\hat f \circ
\hat\iota_X = f$ (Proposition 6 in \cite{H}); this
is the universality property.

This last allows us to define things categorically: 
Let {\bf PdetPdisChron} be the category of
past-determined, past-distinguishing chronological
sets with future-continuous functions as the
morphisms, and let {\bf FcplPdetPdisChron} be the
subcategory with future-complete objects (and the
same morphisms).  We then have that future
completion is a functor
$\;\widehat{}\; : \text{\bf PdetPdisChron} \to
\text{\bf FcplPdetPdisChron}$ (that
$\widehat{g \circ f} = \hat g \circ \hat f$
follows from the uniqueness of extending a function to
the future completions, since $\hat g
\circ \hat f$ has the requisite properties for the
extension of $g \circ f$).  The standard future
injections $\hat\iota_X$ yield a natural 
transformation $\hat{\boldsymbol\iota}$ from the
identity functor on {\bf PdetPdisChron} to
$\;\widehat{}\;$, and $\;\widehat{}\;$ is
left-adjoint to the ``forgetful" (i.e., inclusion)
functor from {\bf FcplPdetPdisChron} to {\bf
PdetPdisChron} (that just means that the
universality property above holds).  This is
important, since left-adjoints are categorically
unique (i.e., unique up  to natural equivalence): 
future completion is the categorically unique way
to create a future-complete chronological set from
a given past-determined, past-distinguishing one.

A substantial awkwardness is that
future completion requires a past-determined
object if it is to act functorially:  If the
target of a future-continuous function $f$ is not
past-determined, $\hat f$ may very well not be
chronological, and many spacetimes are not
past-determined.  The remedy is the
past-determination functor, which (categorically)
extends the chronology relation to additional
pairs of points.  Specifically:  For any
chronological set $X$ with chronology relation
$\ll$, the {\it past-determination\/} of $X$, 
written $X^p$, is $X$ with the chronology relation
$\pl$, defined by $x \pl y$ if $x \ll y$ or if
$I^-(x)$ is non-empty and for some $w \ll y$,
$I^-(x) \subset I^-(y)$; $I^{-p}$ will be used to
denote the past in $X^p$, as necessary.  $X^p$ is
the {\it past-determination\/} of $X$; it is
past-determined, and if $X$ is already
past-determined, then $X^p = X$.  If $X$ is,
respectively, past-distinguishing or
future-complete, then so is $X^p$ (Proposition 10
in \cite{H}).  The function
$\iota^p_X : X \to X^p$ which, on the set level, is the
identity, is future-continuous.

For any future chain $c$ in $X^p$ (i.e., a chain
with respect to the relation $\pl$) there is a
(non-unique) future chain $c'$ in $X$, said to be
{\it associated\/} to $c$, such that a point $x$
is a future limit of $c$ in $X^p$ if and only if
it is a future limit of $c'$ in $X$ (Proposition
11 in \cite{H}); for instance, for
$\cdots x_n \pl x_{n+1} \cdots$, we have
$I^-(x_n) \subset I^-(w_n)$ for some $w_n \ll
x_{n+1}$; then $\{w_n\}$ is an associated chain in
$X$.

To be properly functorial, past-determination
requires just a bit more in the way of
hypothesis:  In \cite{H} a chronological set $X$ 
was called {\it past-connected\/} if every point
is a future limit of some future chain; this
includes strongly causal spacetimes.  As mentioned
in the beginning of Section 2 below, a better
nomenclature is this:  Call a point $x \in X$
{\it regular} if $I^-(x)$ is indecomposable; and
call $X$ regular if all its points are regular
(actually ``past-regular" would be more
appropriate, from the standpoint of
time-duality).  This is equivalent to being
past-connected:  For $x$ to be the future limit of
a future chain is precisely to say that $I^-(x) =
I^-[c]$ for some future chain $c$, and that is
equivalent to $I^-(x)$ being indecomposable.  The
notion of this being an ordinary, every-day sort
of behavior to expect from a point is helpful to
associate with this concept; thus, what was called
past-connected in \cite{H} will be called regular
here, with {\bf Preg} substituting for {\bf Pcon}
in the naming of categories.

If $f: X \to Y$ is a future continuous function and $X$
is regular, then $X^p$ is still regular and
$f^p : X^p \to Y ^p$ is future continuous, where
$f^p$ is the same set-function as $f$.  Thus we
have a functor $\bold p : \bold{PregChron} \to
\bold{PdetPregChron}$, where the infix {\bf -Preg-}
denotes ``(past-)regular".  The maps $\iota^p_X$
form a natural transformation $\boldsymbol
\iota^{\text{\bf p}}$, i.e., for $f : X \to Y$ in
{\bf PregChron}, $f^p
\circ \iota^p_X = \iota^p_Y \circ f$.  Finally,
we have the requisite universality property:  For 
$f : X \to Y$ in {\bf PregChron} and $Y$ already
past-determined, $f^p : X^p \to Y$ is the unique
future-continuous function satisfying $f^p \circ
\iota^p_X = f$ (Corollary 12 in \cite{H}).  This
means the past-determination functor is also a
left-adjoint to a forgetful functor (inclusion of
{\bf PdetPregChron} in {\bf PregChron}); thus,
past-determination is the categorically unique way
to create a past-determined chronological set from
a given regular one. 

These two functors, past-determination and
future completion, compose, as do the respective
natural transformations, yielding another
left-adjoint functor $\;\widehat{}\; \circ \bold
p:$  {\bf Preg\-Pdis\-Chron} $\to$ {\bf
Fcpl\-Pdet\-Preg\-Pdis\-Chron}, the categorically
unique way to create a future-complete and
past-determined chronological set from a given
past-distinguishing, regular one.  However, this
is not quite the GKP Future Causal Boundary:  That
construction is actually $(\hat X)^p$ (which we
will write as $X^+$, as above), rather than
$\widehat{X^p}$.  These two are actually isomorphic
via $j_X : \widehat{X^p} \to (\hat X)^p$, defined
by $j_X(x) = x$ for $x \in X$, and, for $P \in
\hat\partial(X^p)$ generated by a future chain
$c$ in $X^p$, $j_X(P) = I^-[c']$, where $c'$ is any
future chain in $X$ associated to $c$; $j_X$ is
future-continuous and has a future-continuous
inverse ($(j_X)^{-1}$ maps $I^-[c]
\in \hat\partial(X)$ to $I^{-p}[c]$) (Proposition
13 in \cite{H}).  

We need to have a functor associated with the
$X^+$ construction, but it cannot be done by
setting $f^+ = (\hat f)^p$, because that is not, in
general, future-continuous (one needs the target
of $f$ to be past-determined in order for $\hat f$
to be future-continuous).  We finesse this
difficulty by defining, for $f: X \to Y$ in {\bf
PregPdisChron},  $f^+ = j_Y \circ \widehat{f^p}
\circ (j_X)^{-1} : X^+ \to Y^+$; then we have the
functor $\boldkey + :
\bold{PregPdisChron} \to
\bold{Fcpl\-Pdet\-Preg\-Pdis\-Chron}$.  The maps
$\iota^+_X = j_X
\circ \hat\iota_{X^p} \circ
\iota^p_X : X \to X^+$ define a natural
transformation
$\boldsymbol\iota^{\boldkey +}$, and we have the
universality property:  For any $f : X \to Y$ in
{\bf PregPdisChron} with $Y$ future-complete and
past-determined, $f^+ : X^+ \to Y$ is the unique
future-continuous map satisfying $f^+ \circ
\iota^+_X = f$ (Theorem 14 in \cite{H}).  Thus,
$\boldkey +$ is left-adjoint to the same forgetful
functor as is
$\;\widehat{}\; \circ \text{\bf p}$, so the two are
naturally equivalent (the maps
$j_X$ provide the natural equivalence $\bold j :
\;\widehat{}\; \circ \bold p \;\dot\to\; \boldkey 
+$).

\subhead 1.4 A Chronological Rigidity Theorem
\endsubhead 

Here is a result that could have been mentioned in
\cite{H}, as it has to do solely with the 
chronological category, but was not:  In essence,
the Future Chronological Boundary is the {\it
only\/} way to future-complete a regular
chronological set.  

Suppose we begin with a regular chronological set
$X$ and ask how we might define a 
future completion for it.  What we are seeking is
a future-complete, past-distinguishing
chronological set $Y$ together with a map $i : X
\to Y$ such that the restriction $i_0: X
\to i[X]$ is an isomorphism of chronological
sets, and such that the remainder of $Y$---that
which is not in $Y_0 = i[X]$---consists solely of
future limits of future chains in $Y_0$.  Then,
except for past-determination,
$Y$ can only be $X^+$ and $i$ must be
$\iota^+_X$, up to isomorphism (specifically:
$\iota_y^p \circ i = i^+ \circ \iota_X^+$, and
$i^+$ is an isomorphism):

\proclaim{Theorem 1.1} Let $X$ and $Y$ be
chronological sets with $X$ regular and $Y$
future-complete and past-distinguishing, and $i: X \to
Y$ a future-continuous map obeying

\roster
\item with $Y_0 = i[X]$, the chronology relation
on $Y$, restricted to $Y_0$, yields a
chronological set;
\item with $i_0: X \to Y_0$ the restriction of $i$,
$i_0$ is a chronological isomorphism; and
\item with $\partial(Y) = Y - Y_0$, every
element of $\partial(Y)$ is a future limit of a
future chain in $Y_0$.
\endroster

Then $i^+ : X^+ \to Y^+$ is a chronological
isomorphism.
\endproclaim

\demo{Proof} First note that since $Y$ is
future-complete, $\hat Y = Y$, so $Y^+ = (\hat Y)^p =
Y^p$.  Also, since $Y$ is future complete, so is
$Y^p$,  so $\widehat{Y^p} = Y^p$.  Therefore, $j_Y
: \widehat{Y^p} \to (\hat Y)^p$ is just the
identity map on $Y^p$.  Thus, $i^+ = j_Y \circ
\widehat{i^p} \circ (j_X)^{-1} = \widehat{i^p}
\circ (j_X)^{-1} : (\hat X)^p \to Y^p$.

Obviously, $i^+$ is onto $Y_0$; it is also onto
$\partial(Y)$:  For any $y \in \partial(Y)$, there
is some future chain $c$ in $Y_0$ with $y$ the
future limit of $c$.  Since $i_0$ is an
isomorphism, $c' = i^{-1}[c]$ is a future chain in
$X$.  If $c'$ has a future limit $x \in X$, then
$i(x) = y$ by future-continuity, so $y \in Y_0$. 
Therefore, $c'$ has no future limit in $X$, but it
has a future limit $P
\in \hat\partial(X)$; $\widehat{\iota_X^p}(P)$ is
also the future limit of $c'$ in $X^p$ (actually,
in $\hat\partial(X^p)$).  Then
$\widehat{i^p}(\widehat{\iota_X^p}(P)) = y$ (it
must be the future limit of $i[c'] = c$).  Note
that $j_X(\widehat{\iota_X^p}(P)) = I^-[c']$; thus,
$\widehat{i^p}((j_X)^{-1}(I^-[c'])) = y$.

Clearly, $i^+$ is injective on $X$; it is also
injective on $\hat\partial(X)$:  Suppose $i^+(P_1)
= y = i^+(P_2)$ for $P_1$ and $P_2$ in
$\hat\partial(X)$.  Let $P_k$ be generated by a
future chain $c_k$ in $X$; then $(j_X)^{-1}(P_k)$ is
generated by $c_k$ in $X^p$, hence, is its future
limit in $X^p$.  Then, by future-continuity,
$\widehat{i^p}((j_X)^{-1}(P_k)) = y$ is the future
limit of $i[c_k]$.  Thus, $I^-(y) =
I^-[\,i[c_1]\,] = I^-[\,i[c_2]\,]$.  This means
that for each $n$, there is some $m$ with $i(c_1(n))
\ll i(c_2(m))$, and {\it vice versa} for $1$ and 
$2$ exchanged.  Then the same relationship obtains
between $c_1$ and $c_2$, which means $P_1 = P_2$.

We already know that $j_X$ is an isomorphism of the
chronology relation; we must show the same for
$\widehat{i^p} : \widehat{X^p} \to Y^p$:  

Suppose for $x_1$ and $x_2$ in $X$, $i(x_1) \ll^p
i(x_2)$; then there is some $w \ll i(x_2)$ with
$I^-(i(x_1)) \subset I^-(w)$.  If $w \in
\partial(Y)$, we can replace it with $z \in Y_0$
obeying the same relationship:  There is some $u$
with $w \ll u \ll i(x_2)$ and if $u$ is not itself
in $Y_0$, then $u$ is the future limit of a chain
in $Y_0$, so there is some
$z \in Y_0$ with $w \ll z \ll u \ll i(x_2)$. 
Thus, we can  find $x \in X$ with $I^-(i(x_1))
\subset I^-(i(x))$ and $i(x) \ll i(x_2)$.  Then by
the isomorphism of
$i_0$, the same relationship obtains in $X$ among
$x_1$, $x$, and $x_2$, so $x_1 \ll^p x_2$.  

Suppose for $x \in X$ and $y \in \partial(Y)$, 
$i(x) \ll^p y$.  Since $y$ is the future limit of
$i[c]$ in $Y$, it is also the future limit of 
$i[c]$ in $Y^p$, {\it i.e.\/},  $I^{-p}(y) =
I^{-p}[\,i[c]\,]$; that says precisely that
$\widehat{i^p}(P) = y$, so we need to show $x \ll^p
P$. Let $c$ be a chain in
$X$ such that $y$ is the future limit of $i[c]$; for
$n$ sufficiently high, $i(x) \ll^p i(c(n))$.  By the
result above, $x \ll^p c(n)$.  Let $P$ be the IP in
$X$ generated by $c$; then we have $x \ll^p
c(n) \ll P$ in $\hat X$, from which it follows
that $x \ll^p P$. 

Suppose for $x \in X$ and $y \in \partial(Y)$, $y
\ll^p i(x)$.  As above, we have $y =
\widehat{i^p}(P)$ for $P$ generated by $c$ in
$X$, where $y$ is the future limit of $i[c]$.  We
can find some $w \in Y$ with $y \ll^p w \ll^p
i(x)$, and we can take $w = i(z)$ for some $z \in
X$ (if $w \in \partial(Y)$, then $w$ is the future
limit in $Y$ of some chain in
$Y_0$, so it is the future limit in $Y^p$ of such a
chain).  From $i(c(n)) \ll y \ll^p i(z)$, we derive
$i(c(n)) \ll i(z)$, whence $c(n) \ll z$, for all 
$n$.  From knowing $i(z) \ll^p i(x)$, we learn, by
the results above, that $z \ll^p x$, i.e., $I^-(z)
\subset  I^-(u)$ for some $u \ll x$.  Then all
$c(n) \ll u$, so $P \subset I^-(u)$.  This gives
us $P \ll x$, so
$P \ll^p x$.

Suppose for $y_1$ and $y_2$ in $\partial(Y)$, $y_1
\ll^p y_2$; say $y_k = \widehat{i^p}(P_k)$.  As
in the paragraph above, we can find
$z \in X$ with $y_1 \ll^p i(z) \ll^p y_2$.  Then
applying the previous two paragraphs shows us that
$P_1 \ll^p z \ll^p  P_2$, so $P_1 \ll^p P_2$. \qed
\enddemo

In case $Y$ is regular, there is an equivalent
formulation of the hypotheses of this theorem that
is worth mentioning: that $Y$ have a subset $Y_0$
which, under the restriction of the chronology
relation from $Y$, is a chronological set in its
own right; that $i[X] = Y_0$ and the restriction
$i_0 : X \to Y_0$ is a chronological isomorphism;
and that $Y_0$ is ``chronologically dense" in $Y$,
in the sense that for any $y_1 \ll y_2$ in $Y$,
there is some $z \in Y_0$ with $y_1 \ll z \ll
y_2$.  One direction doesn't require regularity: To
find $i(x)$ with $y_1 \ll i(x) \ll y_2$, locate
$y$ with $y_1 \ll y \ll y_2$.  Either $y$ is
already in $Y_0$ or it is the the future limit of
$i[c]$ for some chain $c = \{x_n\}$ in $X$; in the
latter case, for some $n$, $y_1 \ll i(x_n)$, and
$i(x_n) \ll y$. For the other direction:  For $y
\in Y$, if $y$ is regular, then $I^-(y)$ is
generated by a chain $\{y_n\}$, and for each $n$ we
pick $z_n \in Y_0$ with $y_n \ll z_n \ll y_{n+1}$.  

Theorem 1.1 shows that any past-distinguishing,
future-completing boundary on regular $X$ must be
isomorphic to $\hat\partial(X)$---at least in the
past-determinations---both in terms of its own
structure and in how it is related to $X$.  This
amounts to a rigidity theorem for these sorts of
boundaries on regular chronological sets.  But
this does not mean that there is only one way to
put a reasonable future boundary on a spacetime: 
Important flexibility lies in relaxing insistence
on regularity for the completion---in looking,
instead, at what were called in \cite{H}
generalized future-complete objects.  The
generalized notions for future limits,
future-continuous, future-complete,
past-distinguishing, and past-determined, all with
applicability to non-regular chronological sets,
will be recalled and explored in Section 4.

\head 2. Regular Chronological Sets and Future
Completion
\endhead

In any strongly causal spacetime $M$,  for any
point in $x \in M$, $I^-(x)$ is an IP.  This is
also true for the elements of the Future
Chronological Boundary of any chronological set
$X$:  If $P \in \hat\partial(X)$ is generated by
the future chain $c$ (in $X$), then $\hat I^-(P)$
is also generated by $c$ (in $\hat X$).  The
ubiquity of this property is what suggests the
nomenclature of ``regular" for it, i.e., that a
point $x$ is regular if $I^-(x)$ is an IP, and a
chronological set $X$ is regular if all its points
are regular.  Then we have that if $X$ is regular,
so are $\hat X$ and $X^p$ (if $I^-(x)$ is generated
by a future chain $c$, then $c$ also generates
$\hat I^-(x)$ and $I^{-p}(x)$).

Non-regular points are typically encountered when
combining the Past and Future Causal Boundaries to
produce the full Causal Boundary of GKP.  A typical
example would be with $X$ being Minkowski 2-space
$\Bbb L^2$ with the negative time-axis $\{(0,t) 
\;|\; t \le 0\}$ removed (see Figure 1):  $X^+$
produces boundary points on either side of the
missing semi-axis,
$P_s^+ = \{(x,t) \;|\; t < -x+s \text{ and } x >
0\}$ and $P_s^- = \{(x,t)  \;|\; t < x+s
\text{ and } x < 0\}$ for $s \le 0$.  The dual
construction, $X^-$, produces analogous boundary
points $F_s^+ =
\{(x,t) \;|\; t > x+s \text{ and } x > 0\}$ and
$F_s^- =
\{(x,t) \;|\; t > -x+s \text{ and } x < 0\}$, for
$s < 0$, and a single $F_0 =
\{(x,t) \;|\; t > |x|\}$.  The combining of these
in $X^\#$ identifies $P_s^+ \equiv F_s^+$ (call it
$B_s^+$)  and $P_s^- \equiv F_s^-$ (call it
$B_s^-$), both for $s<0$, and also $P_0^+
\equiv F_0 \equiv P_0^-$ (call it $B_0$).  Then 
$I^{-\#}(B_0)$ consists of the
points of $X$ making up $P_0^+ \cup P_0^-$ (plus
boundary points $B_s^+$ and $B_s^-$ and others at
timelike and null past infinity), two separate
components of the past of $B_0$.  Thus,
$B_0$ is a non-regular point in $X^\#$.

The topology to be defined in this Section on a
chronological set $X$ works rather well if $X$ is
regular.  But it can fail to give expected results 
for non-regular points such as $B_0$ in $X^\#$
above:  It fails to have $B_0$ as a limit of the
sequence $\{(0,1/n)\}$.  A more complicated version
of the topology will be detailed in Section 4 to
take care of non-regular points.

\medpagebreak

What is an appropriate topology to place on a
chronological set $X$?  One that might come to 
mind is the Alexandrov topology:  Declare $I^-(x)$
and $I^+(x)$ to be open sets for all $x \in X$, as
a sub-basis for a topology.  For spacetimes, being
strongly causal is equivalent to the Alexandrov
topology being the same as the manifold topology
(see \cite{BE}), which makes this seem a natural
choice for the topology on a chronological set. 
However, this will not serve:  Although $I^-(x)$
is always open in any spacetime, there are
instances of spacetime-with-boundary in which we
do not want that to be so.  

A typical example would be any
of a number of completions of Minkowski 2-space 
with a spacelike half-line deleted, say, $X =\Bbb
L^2 - \{(x,0)| x \le 0\}$ (see Figure 2); as
completion
$\bar X$ take $X$ plus points on the future and past
edges of the slit: $\{p_s^+ \;|\; s<0\}$ on the
future side (essentially, $p_s^+ = (s,0^+)$),
$\{p_s^-
\;|\; s<0\}$ on the past edge ($p_s^- = (s,0^-)$),
and $p_0$ joining the two (functioning as
$(0,0)$).  The topology of the added points is
that of two half-lines conjoined at the ends,
glued to the $\{t>0\}$ and $\{t<0\}$ regions in
the obvious manner.  As chronology relation on
$\bar X$, set $p \ll q$ iff there is a 
future-directed timelike curve from $p$ to $q$. 
Then the $\bar X$-past of $(0,1)$ contains
$\{p_s^+ \;|\; -1 < s < 0\}$ and
$p_0$, but not a single $p_s^-$:  $p_0 \in 
I^-_{\bar X}((0,1))$ but no neighborhood of $p_0$
is in that past set.

So while the Alexandrov topology is sometimes
reasonable (for instance, in some
space\-times-with-boundary with only timelike
boundaries), we must do something else for a 
general chronological set.

The procedure followed here will be to define a
topology on a chronological set $X$ by defining
what the limits of sequences are; then a subset of
$X$ is defined to be closed if and only if it
contains the limits of all its sequences.  This is
not quite as straight-forward as it sounds, if it
turns out that a sequence can have more than one
limit---and, unfortunately, that is a very live
possibility, even for a chronological set which
is, say, the future completion of a reasonable
spacetime (an example will be given of such).  In
such a case, one can conceivably have a point
which is in the closure of a sequence but is not a
``limit" of that sequence as given in the
definition.

Perhaps the best way to be clear about what is
going on is to eschew the word ``limit" and speak
simply of a function $L: \Cal S(X) \to \frak
P(X)$, where $\Cal S(X)$ denotes the set of
sequences in
$X$ (i.e., maps $\sigma : \Bbb Z^+ \to X$) and
$\frak P(X)$ is the power set of
$X$; $L(\sigma)$ is to be thought of as the set of
points which are ``first-order" limits of the
sequence $\sigma$ (we can call $L$ the
``limit-operator" for
$X$, without prejudice to the term ``limit").   So
long as $L$ has the property that for subsequences
$\tau \subset \sigma$, $L(\tau)
\supset L(\sigma)$, then a topology is defined on
$X$ by defining a subset $A$ of $X$ to be closed
if and only if for every sequence $\sigma \subset
A$, $L(\sigma) \subset A$.  Every second-countable
topological space can have its topology 
characterized in this fashion---and it is the
existence of the countable subset $S$, in the
definition of chronological set given in Section
1, that makes it possible to treat chronological
sets as second-countable topological spaces.

A singleton set $\{x\}$ will be closed so
long as $L(\hat x)$ is either $\{x\}$ or 
$\emptyset$, where $\hat x$ is the constant
sequence $\hat x(n) = x$, all $n$.  Let us assume
that for all $x$, $L(\hat x)$ contains $x$. 
Then with $L[A]$, for $A \subset X$, denoting
$\bigcup \{L(\sigma) \;|\; \sigma \in \Cal S(A)\}$,
we have $L[A] \supset A$ for any subset $A$.  So
long as for any sequence $\sigma$, $L(\sigma)$ is
finite, then for any set $A$,
$\slanted{closure}(A) = L[A]$.  More generally, we
must look to iterations of the set-function
$L[\;]$:  $L^1 = L$; for any ordinal $\alpha$ let
$L^{\alpha + 1}[A] = L[\,L^\alpha[A]\,]$; and for
any limit-ordinal
$\alpha$, let $L^\alpha[A] = \bigcup_{\beta <
\alpha}L^\beta[A]$.  Then  $\slanted{closure}(A) =
L^\Omega[A]$, where $\Omega$ is the first
uncountable ordinal (reason: Any point $x \in
L^{\Omega + 1}[A]$ lies in $L(\sigma)$ for some
sequence $\sigma$ lying in $L^\Omega[A]$.  For all
$n$, there is some ordinal $\alpha_n < \Omega$ with
$\sigma(n) \in L^{\alpha_n}[A]$.  But there is
some $\beta < \Omega$ with all $\alpha_n < \beta$. 
Thus, $x \in L^\beta[A] \subset L^\Omega[A]$. 
Therefore, $L^{\Omega+1}[A] = L^\Omega[A]$.).  For a
function $f: X \to Y$, where $X$ and $Y$ are
topological spaces defined using, respectively,
the limit-operators $L_X$ and $L_Y$, $f$ is 
continuous if and only if for every sequence
$\sigma$ in $X$, $f[L_X(\sigma)] \subset
L_Y^\Omega[\,f[\sigma]\,]$.  (This follows thus: $x$
is a limit of a sequence $\sigma$ if and only if
every open set containing $x$ eventually contains
$\sigma$, i.e, every closed set excluding $x$
eventually excludes $\sigma$, i.e., every closed
set containing a subsequence of $\sigma$ contains
$x$; this is equivalent to $x$ being in the
closure of every set containing a subsequence of
$\sigma$, which is equivalent to $x$ being in the
closure of every subsequence of $\sigma$, i.e., $x
\in \bigcap_{\tau \subset
\sigma}L^{\Omega}[\tau]$.  Then $f: X \to Y$ is
continuous if and only if for every sequence
$\sigma$ and point $x$ in $X$, $x \in \bigcap_{\tau \subset
\sigma}L_X^{\Omega}[\tau]$ implies $f(x) \in 
\bigcap_{\tau \subset
f[\sigma]}L_Y^{\Omega}[\tau] = \bigcap_{\tau
\subset \sigma}L_Y^{\Omega}[f\,[\tau]\,]$; and
this is equivalent to having, for every sequence
$\sigma$ in $X$, $f[\,L_X^{\Omega}[\sigma]\,]
\subset L_Y^{\Omega}[f\,[\sigma]\,]$.  A bit of
transfinite induction shows that if for every
sequence $\sigma$, $f[L_X(\sigma)] \subset
L_Y^{\Omega}[\,f[\sigma]\,]$, then for every
sequence $\sigma$,
$f[\,L_X^{\Omega}[\sigma]\,] \subset
L_Y^{\Omega}[\,f[\sigma]\,]$.) In
particular, $f[L_X(\sigma)] \subset L_Y(f[\sigma])$
for all $\sigma$ implies $f$ is continuous. 

\definition{Limit-operator in a regular
chronological set} For $X$ a regular
chronological set, the limit-operator $L$ is
defined thus:
 
For any sequence $\sigma = \{x_n\}$ and any point
$x$, $x
\in L(\sigma)$ if and only if

\roster
\item for all $y \ll x$, eventually $y \ll x_n$
(i.e., for some $n_0$, $y \ll x_n$ for all $n >
n_0$), and
\item for any IP $P$ containing $I^-(x)$, if for
all $y \in P$, there is some subsequence
$\{x_{n_k}\}$ with
$y \ll x_{n_k}$ for all $k$, then $P = I^-(x)$.
\endroster
\enddefinition

Note, first of all, that this obeys $L(\tau) 
\supset L(\sigma)$ for $\tau \subset \sigma$: 
Clause (2) says, essentially, that $I^-(x)$ is a
maximal IP for obeying clause (1)---i.e., that all
of its elements are eventually in the past of the
sequence---but this is generalized to
subsequences.  (Sometimes the contrapositive of
clause (2) is more natural:  Any IP
properly containing $I^-(x)$ contains a point $y$
such that eventually $y \not\ll x_n$.)
It follows that
$L$ defines a topology on $X$; as this topology is
designed especially with $\hat X$ in mind, let us 
call it the \h{\it topology\/} on $X$.  We will
call the elements of $L(\sigma)$ the
\h{\it limits\/} of $\sigma$.  (See Figure 3.)

Next, note that $L(\sigma)$ depends only on the
points which make up the sequence $\sigma$, and
not on the order in which they appear in the
sequence (save for whether a given point appears a
finite number of times or an infinite number of
times in the sequence):  We could reformulate the
definition as $x \in L(\sigma)$ if and only if (1)
for all $y \ll x$, $y$ is in the past of all but a
finite number of points of $\sigma$ (more
precisely: $\{n \;|\; y \not\ll x_n\}$
is finite), and (2) for any IP $P \supset I^-(x)$,
if for all $y \in P$, $y$ is in the past of an
infinite number of points of $\sigma$ (more
precisely: $\{n \;|\; y \ll x_n\}$ is infinite),
then $P = I^-(x)$.

\proclaim{2.1 Proposition} For any regular
past-distinguishing chronological set $X$, for any 
$x \in X$, $L(\hat x) =
\{x\}$\rom; thus, all points are closed in the
$\;\widehat{}\;$-topology.\endproclaim

\demo{Proof} 
First, $x \in L(\hat x)$:  For any $y \ll x$, $y 
\ll \hat x(n)$ for all $n$ (i.e, $y \ll x$).   Let
$P$ be any IP with $I^-(x) \subset P$.  To say
that for any $y \in P$, $y$ is in the past of each
element of a subsequence of $\hat x$ is to say $y
\ll x$ for all $y \in P$, i.e., that $P \subset 
I^-(x)$; thus, this implies $P = I^-(x)$.

Second, if $z \in L(\hat x)$, then $z = x$:   We
have for all $y \ll z$, $y \ll x$, i.e., $I^-(z)
\subset I^-(x)$.  Thus we can apply part (2) of
the definition of $z \in L(\hat x)$ to the IP
$I^-(x)$:  For all $y \in I^-(x)$, $y$ is in the
past of each element of a subsequence of $\hat x$,
so $I^-(x) = I^-(z)$.  Thus, by
past-distinguishment, $z = x$. \qed
\enddemo 

An immediate corollary is that  in a
past-distinguishing regular chronological set, if
a sequence $\sigma$ has more than one point that
appears an infinite number of times, then
$L(\sigma) = \emptyset$ (if both $x$ and $y$
appear infinitely often, then $\hat x$ and $\hat
y$ are both subsequences, so $L(\sigma) \subset
\{x\}$ and $L(\sigma) \subset \{y\}$).

Here is an example showing that a most reasonable
chronological set need not be Hausdorff (see figure
4):  Let
$M$  be the the spacetime considered before, $\Bbb
L^2$ with the negative time-axis removed, and let $X
=
\hat M$, its future completion; this introduces
the boundary points mentioned before, $P_s^+$ and
$P_s^-$, for $s \le 0$, representable respectively
by $(s,0^+)$ and $(s,0^-)$; however, we will not
meld $P_0^+$ and $P_0^-$ together, but keep them
separate, staying precisely with $\hat M$. 
Consider the sequence $\sigma = \{(0, 1/n)\}$:  The
past of $P_0^+$ (in $X$) consists of $\{(x,t) \;|\;
0 < x < -t\} \cup \{P_s^+ \;|\; s < 0\}$; for all
$p \in I_X^-(P_0^+)$, for all $n$, $p \ll
\sigma(n)$.  Furthermore, for any IP $P$ (in $X$)
properly containing $I_X^-(P_0^+)$, eventually
$\sigma(n) \in P$, and we can find $p \in P$ so
that for $n$ sufficiently high, $p \ll
\sigma(n)$ fails.  Therefore $P_0^+ \in L(\sigma)$. 
It follows that any closed set not containing 
$P_0^+$ must also omit all but a finite portion of
$\sigma$:  Any neighborhood of $P_0^+$ must
contain a tail-end of
$\sigma$.  The same, of course, is true for $P_0^-$,
so any two neighborhoods of these points intersect. 
(Actually, any neighborhood of either point must
contain an $\Bbb L^2$-neighborhood of the origin
intersected with
$\{(x,t) \,|\; |x| < t\}$.)

We want to know that the concept of \h limit is
compatible with future limit of a future chain.  
This is the case:

\proclaim{2.2 Proposition} Let $c = \{x_n\}$ be a
future chain in a regular chronological space.  
Then a point
$x$ is a future limit for $c$ if and only if it is 
a \h limit for $c$; furthermore, if for every
point $x$, $L(\hat x) = \{x\}$, then
$L^\Omega[c] = L[c]$.\endproclaim

\demo{Proof} Suppose that $x$ is a future limit 
for $c$, i.e., that $I^-(x) = I^-[c]$; we need to
show that it is also a \h limit.  For any $y \ll
x$, $y \in I^-[c]$, so $y \ll x_n$ for all $n$
large enough.  Consider any IP
$P \supset I^-(x)$:  For any point $y$, if,
for all $k$, $y \ll x_{n_k}$ for some subsequence, 
then $y \in I^-[c] = I^-(x)$; thus, if this is true
for all $y \in P$, it follows that $P = I^-(x)$.

Now suppose that $x \in L(c)$; we must show it a 
future limit of $c$.  For all $y \ll x$, we have
eventually $y \ll x_n$; thus $I^-(x) \subset
I^-[c]$.  Now, for all $y \in I^-[c]$, eventually
$y \ll x_n$, and $I^-[c]$ is an IP containing
$I^-(x)$; therefore, applying clause (2) of the
definition of $L(c)$, we obtain $I^-[c] = I^-(x)$.

To show $L^\Omega[c] = L[c]$, we need only show
that $L^2[c] = L[c]$ (which is $c \cup L(c)$). 
Consider any sequence of points $\sigma = \{x_n\}$
in $L(c)$.  We just need to show that 
for any $x \in L(\sigma)$, $x$ is a future limit of
$c$, for then (from what we've just seen) $x \in 
L(c)$.  Consider such $x \in L(\sigma)$:  For any
$y \ll x$, eventually $y \ll x_n$; since $I^-(x_n)
= I^-[c]$, that shows $y \in I^-[c]$.  Therefore,
$I^-(x) \subset I^-[c]$.  Thus, $I^-[c]$ is an IP
containing $I^-(x)$.  If it properly contains
$I^-(x)$, then it must contain some $y$ which is
eventually not in the past of $x_n$; but that is
impossible, since $x_n$ and $c$ have the same
past, for any $n$.  Therefore, $I^-[c] = I^-(x)$,
i.e., $x$ is a future limit of $c$.
\qed\enddemo

It follows that a chronological function $f : X 
\to Y$ between regular chronological sets with $Y$
past-distinguishing, which is continuous in the
respective \h topologies, is also 
future-continuous:  For any future chain $c$ in
$X$ with future limit $x$, $x$ is also a \h limit
for $c$; therefore, $f(x) \in L[\,f[c]\,]$. 
Now, $f[c]$ is also a future chain in
$Y$, so $L(f[c])$ consists of the future limits of
$f[c]$; with $Y$ past-distinguishing, there can be 
no more than one of these:  There is at most a
single \h limit for $f[c]$.  Thus, $f(x)$ is that
unique \h limit for
$f[c]$, which must be the (unique) future limit of
$f[c]$.  However, the converse does not hold:  A
future-continuous function can fail to
be \h continuous.  A simple example is $f: \Bbb L^1 \to
\Bbb L^1$ with $f(t) = t$ for $t \le 0$ and $f(t) 
= t + 1$ for $t > 1$.
 
There are many ways to define a topology for
chronological sets; why should this \h topology be
considered a good choice, especially since it need
not even be Hausdorff in reasonable instances?  One of
the prime prerequisites of a good topology
construction is that it replicate the manifold
topology in the case of a spacetime; that is obeyed by
the \h topology:

\proclaim{2.3 Theorem} Let $M$ be a strongly
causal spacetime; then the \h topology induced by 
the spacetime chronology relation is the same as
the manifold topology on $M$.\endproclaim

\demo{Proof} It suffices to show that for
any sequence $\sigma = \{x_n\}$ in $M$, a point $x$
is the usual (manifold-)topological limit of
$\sigma$ if and only if it is a \h limit of
$\sigma$ (for that establishes that the map $1_M :
(M, \text{manifold topology}) \to (M, \text{\h
topology})$ is bicontinuous).

Suppose $x$  is the usual topological limit of
$\sigma$.  Clause (1):  For any $y \ll x$, $V =
I^+(y)$ is an open set containing $x$; therefore,
the sequence is eventually inside $V$, i.e.,
eventually $y \ll x_n$.  Clause (2) (see Figure 5): 
Let
$P$ be any IP containing $I^-(x)$.  Since $M$ is
strongly causal, it has the Alexandrov topology, and
we can pick points $z \ll x$ (note that
$z \in P$) and $w \gg x$ so that $U = I^+(z) \cap
I^-(w)$ is geodesically convex:  For any $u \ll v$
both in $U$, there is a future-timelike geodesic in
$U$ from $u$ to $v$.  Suppose that every point  in
$P$ is in the past of every element of a
subsequence of $\sigma$.  For any $p \in P$, 
there is some $q \in P$ with $q \gg p$ and $q \gg
z$.  For some subsequence, for all $k$,
$q \ll x_{n_k}$; also, $x_{n_k} \in U$. 
Therefore, for all $k$, $q$ and $x_{n_k}$
are both in $U$, so there is a future-timelike 
geodesic $\gamma_k$ from $q$ to $x_{n_k}$, lying in
$U$.  Then the curves $\{\gamma_k\}$ have a limit
curve $\gamma$ (lying in $U$) from $q$ to $x$, and
$\gamma$ is a future-causal geodesic.  Then we have $p
\ll q \prec x$, so $p \ll x$: $P = I^-(x)$.   

Now suppose $x$ is a \h limit of $\sigma$ (see
Figure 6).   If
$x$ is not the manifold-topological limit of
$\sigma$, then there are a relatively compact
neighborhood $U$ of $x$ and a subsequence
$\{x_{n_k}\}$ which never enters $U$; by strong
causality we can pick $U$ so that no timelike
curve exits and re-enters $U$.  Pick a future chain
$\{z_n\}$ with $x$ as future limit (which is the 
same as saying the manifold-topological limit). 
For $n$ sufficiently large, $z_n \in U$; then,
since $z_n \ll x$, for $k$ sufficiently
large---say, $k \ge K_n$---$z_n \ll x_{n_k}$, so
there is a future-timelike curve $c^n_k$ from
$z_n$ to $x_{n_k}$.  Since $x_{n_k}$ is not in
$U$, $c^n_k$ exits $U$ at a point $y^n_k \in
\partial U$.  For each $n$, the curves  $\{c^n_k
\;|\; k \ge K_n\}$ have a future-causal limit curve
$c^n$ from $z_n$ to some point $y^n \in \partial
U$.   The curves $\{c^n\}$ have a future-causal
limit curve $c$ from $x$ to some point $y \in
\partial U$.

Let $P = I^-[c] = I^-(y)$ and apply clause (2) of 
the \h limit definition (applicable because any
point in the past of $x$ is in the past of $y$): 
For any $p \in P$, we have $I^+(p)$ is a
neighborhood of $y$; therefore, for $n$
sufficiently large $y^n \in I^+(p)$; therefore,
for $k$ sufficiently large---say, $k \ge
J_n$---$y^n_k \in I^+(p)$.  Since $y^n_k \ll
x_{n_k}$ (via $c^n_k$), we have that for $n$
sufficiently large and $k \ge J_n$, $p \ll
x_{n_k}$:  Clause (2) allows us to conclude $P =
I^-(x)$.  But this is impossible, since $y \succ
x$ and $y \ne x$, since $y \in
\partial U$ with $U$ a neighborhood of
$x$.  Therefore, $x$ must be the manifold-limit of
$\sigma$. \qed \enddemo

Another desirable property of a topology 
construction is that it accord well with boundary
constructions:  If $\bar X$ is $X$ with some sort
of future boundary, then $\bar X$ should have a
topology appropriately related to that of $X$.  
Here is how the
\h topology fares in this regard:

\proclaim{2.4 Theorem} Let $\bar X$ be a regular
chronological set with $X$ a subset of $\bar X$
satisfying the following:
\roster
\item The restriction of $\ll$ to $X$ yields 
another regular chronological set; and
\item for any $p \ll q$ in $\bar X$, there is some $x
\in X$ so that $p \ll x \ll q$.
\endroster
Then the \h topology on $X$ (as a
chronological  set in its own right) is the
same as the subspace topology it inherits from the
\h topology on $\bar X$, and $X$ is dense in $\bar
X$. \endproclaim

\demo{Proof} We must first establish a 
correspondence between the IPs of $X$ and those of
$\bar X$:

For any IP $P$ in $X$, generated by a future chain
$c$ in $X$, let $\bar P = I_{\bar X}^-[c]$, an IP 
in $\bar X$.  Since $\bar P = I_{\bar X}^-[P]$,
this is independent of the choice of the
generating chain $c$.

For any IP $Q$ in $\bar X$, generated by a future
chain $c$ in $\bar X$, use condition (2) of the
hypotheses to construct an interweaving chain 
$c_0$ in $X$ (i.e., if $c = \{p_n\}$, then pick
$c_0 = \{x_n\}$ with $p_n \ll x_n \ll p_{n+1}$);
let $Q_0 = I_X^-[c_0]$, an IP in $X$.  Since $Q_0
= Q \cap X$, this is independent of the choices of
the chains $c$ and $c_0$.

\proclaim{Lemma} The maps $P \mapsto \bar P$ and 
$Q \mapsto Q_0$ establish an isomorphism between
the IPs of $X$ and of $\bar X$, as partially
ordered sets under inclusion. \endproclaim

\demo{Proof of Lemma} Let $P$ be an IP in $X$,
generated by a future chain $c$.  Then $\bar P$ is
generated (in $\bar X$) by the same chain $c$.  
Thus, for a generating chain for $(\bar P)_0$, we
may take yet again the chain $c$ (or, following
strictly the defining construction, a chain $c_0$
in $X$ interweaving $c$---having, therefore, the
same past as $c$).  Thus, $(\bar P)_0 = P$.

Let $Q$ be an IP in $\bar X$, generated by a future
chain $c$, with an interweaving chain $c_0$ in $X$. 
Then $\overline{Q_0}$ has $c_0$ as generating chain 
in $\bar X$.  Since $c_0$ and $c$ are
interweaving,  they have the same past in $\bar
X$, i.e., $\overline{Q_0} = Q$.

If $P \subset P'$ are IPs in $X$, then $\bar P =
I_{\bar X}^-[P] \subset I_{\bar X}^-[P'] =
\overline{P'}$. 

If $Q \subset Q'$ are IPs in $\bar X$, then $Q_0 = Q
\cap X \subset Q' \cap X = Q'_0$. \qed \enddemo 

Consider a sequence $\sigma = \{x_n\}$ in $X$.  We
will see that for any  $x \in X$, $x \in
L_X(\sigma)$ if and only if $x \in L_{\bar
X}(\sigma)$, i.e., $ L_X(\sigma) = L_{\bar
X}(\sigma) \cap X$.  With $i: X \to \bar X$
denoting the inclusion map, with the respective \h
topologies, that will establish that $i$ is a
homeomorphism onto its image, i.e., that the \h
topology on $X$ is the same as its subspace
topology in $\bar X$.

Suppose $x$ is a \h limit of $\sigma$ with respect 
to $X$, i.e., (1) for all $y \in X$ with $y \ll x$,
eventually $y \ll x_n$, and (2) for any IP $P$ in 
$X$ with $P \supset I_X^-(x)$, if for all $y \in
P$, for all elements of some subsequence,  $y \ll
x_{n_k}$, then $P = I_X^-(x)$.  We want to
establish the same results for $\bar X$.
 
Clause (1):  For any $p \in \bar X$ with $p \ll x$,
there is some $y \in X$ with $p \ll y \ll x$
(hypothesis 2).  We know eventually $y \ll x_n$, 
whence follows $p \ll x_n$.

Clause (2):  Let $Q$ be any IP in $\bar X$ 
containing $I_{\bar X}^-(x)$.  Then $Q_0 = Q \cap
X$ is an IP in $X$ containing $I_{\bar X}^-(x) \cap
X = I_X^-(x)$.  Suppose for all $q \in Q$,
for all $k$, $q \ll x_{n_k}$; then the same is true
for $Q_0$, whence $Q_0 = I^-(x)$.  Then $Q =
\overline{Q_0} = \overline{I_X^-(x)} = I_{\bar
X}^-(x)$  (the last can be seen by noting that
$(I_{\bar X}^-(x))_0 = I_{\bar X}^-(x) \cap X =
I_X^-(x)$).  That finishes showing $x \in L_{\bar
X}(\sigma)$.

Now suppose $x \in X$ is a \h limit of $\sigma$ 
with respect to $\bar X$, i.e., (1) for all $q \ll
x$, eventually $q \ll x_n$, and (2) for any IP $Q$
in $\bar X$, if for all $q \in Q$, for all $k$, $q
\ll x_{n_k}$, then $Q = I_{\bar X}^-(x)$.  We need
to see the same holds with respect to $X$:

Clause (1):  For any $y \in X$ with $y \ll x$, we
clearly have eventually $y \ll x_n$.  Clause (2):  
Let $P$ be any IP in $X$ with $P \supset
I_X^-(x)$.  Then $\bar P$ is an IP in $\bar X$ and
$\bar P \supset \overline{I_X^-(x)} = I_{\bar
X}^-(x)$.  Suppose for all $z \in P$, for some
subsequence,  $z \ll x_{n_k}$.  For any $q \in \bar
P$, there is some $p \in P$ with $q \ll p$; then
there is some $y \in X$ with $q \ll y \ll p$.  We
have $y \in P$, since $y \ll p$ places $y$ in $\bar
P$,  and $y \in X$.  Therefore, for all $k$, $y \ll
x_{n_k}$,  whence $q \ll x_{n_k}$ also.  It follows
that $\bar P = I_{\bar X}^-(x)$, so $P = (\bar
P)_0 = (I_{\bar X}^-(x))_0 = I_X^-(x)$.   Thus is
$x$ shown to be in
$L_X(\sigma)$.

Finally, consider any $p \in \bar X$.  By 
regularity, $I_{\bar X}^-(p)$ is an IP in $\bar X$,
generated by a future chain $c$, so $(I_{\bar
X}^-(p))_0$ is an IP in $X$, generated by an
interweaving future chain
$c_0$ in $X$ (as in the proof of the Lemma).   
Note that $p$ is a future limit of $c_0$ (in $\bar
X$), since interweaving future chains have the
same past.  Then, by Proposition 2.2, $p$ is a \h
limit of $c_0$ (in $\bar X$).  It follows that $p$
is in the closure of
$X$. \qed \enddemo

We can apply this to the future completion
of a regular chronological set:

\proclaim{2.5 Corollary} Let $X$ be a regular
chronological set.  Then the standard future
injection $\hat\iota_X : X \to \hat X$ is a
homeomorphism onto its image, and $X$ is dense in
$\hat X$. \endproclaim

\demo{Proof}  By the definition of $\ll$ in $\hat
X$, its restriction to the elements of $X$ is
precisely the original chronology relation on
$X$.  Since $X$ is regular, so is $\hat X$.  Again
using the definition of $\ll$ in $\hat X$, we see
that for any $P \ll Q$ in $\hat X$, there is some
$x \in X$ with $P \ll x \ll Q$:  For instance, if
$P$ and $Q$ are in $\hat \partial X$, then there is
some $y \in Q$ with $P \subset I^-(y)$; then there
is some $x \in Q$ with $y \ll x$, and we have $P
\ll x \ll Q$.  Thus, Theorem 2.4 applies. \qed
\enddemo 

If $\hat\partial X$ is to have the expected
properties of a boundary for $X$, then not only
should it be in the closure of $X$, but it should
itself be closed.  Now, this cannot be true in
general:  $X$ could, for instance, consist of a
spacetime $M$ plus a portion of
$\hat\partial(M)$, in which case $\hat\partial(X)$ 
is the remainder of $\hat\partial(M)$ and need not
be closed.  (Simple example: $M = \Bbb L^2 -
\{(x,0)  \;|\; x \le 0\}$, so $\hat\partial(M)$
consists of the elements of $\hat\partial(\Bbb
L^2)$ plus $\{(x,0) \;|\; x \le 0\}$ (as elements
attached to the ``underside" of the slit).  If $X$
is $M$ with $\{(0,0)\}$ adjoined, then
$\hat\partial(X) =  \{(x,0) \;|\; x < 0\}$ and has
$(0,0)$ in its closure, though $(0,0)$ is not in
$\hat\partial(X)$.)

But so long as we restrict ourselves to
spacetimes---with no boundary
elements con\-joined---then the Future
Chronological Boundary is, indeed, closed:

\proclaim{2.6 Proposition} Let $M$ be a strongly
causal spacetime.  Then $\hat\partial(M)$ is
closed in $\hat M$. \endproclaim

\demo{Proof} Let $\sigma = \{P_n\}$ be a sequence
of elements of $\hat\partial(M)$; we must show
that it is impossible for an element $x \in M$ to
be in $L(\sigma)$.  (It then follows that
$L^\Omega[\hat\partial(M)] \subset
\hat\partial(M)$.)

For suppose $x \in L(\sigma)$.  By strong causality we
have a relatively compact neighborhood $U$ of $x$,
such that no causal curve exits and re-enters $U$. 
Each $P_n$ is the past of a future-endless curve
$\tau_n$; each $\tau_n$ is eventually outside $U$.  For
all $z \ll x$, eventually $z \ll P_n$ (clause (1) 
of the definition of $L$), i.e., $z \in P_n$; thus,
there is a future-timelike curve from $z$ to a
point on $\tau_n$, which we can choose to be on a
portion of $\tau_n$ which is not in $U$ and which
from that point on never enters $U$.  

We are now in exactly the same position as in the
latter part of the proof of Theorem 2.3:   With
choice of a future chain $\{z_n\}$ approaching
$x$, applying the paragraph above to the points
$\{z_n\}$ yields  future-timelike curves $c^n_k$
from $z_n$ to $\tau_k$ outside of $U$,
intersecting $\partial U$ in $y^n_k$; these
curves have a subsequence converging to a
future-causal $c$ from $x$ to a point $y \in
\partial U$.  Every point in $I^-(y)$ is in the
past of that subsequence, so, by clause (2) of the
definition of $L$, $I^-(y) = I^-(x)$, implying $y =
x$, which is false. \qed \enddemo

Section 5 will give the example of a large class of
spacetimes  $M$ with spacelike boundary (including
interior Schwarzschild) in which the
\h topology gives an entirely reasonable topology 
for $\hat M$ (the intuitively ``right" topology). 
It will also be shown that for spacetimes $M$ with
spacelike boundary, the \h topology agrees with
the topology given to a nice completion $\bar M$ 
formed by embedding $M$ into a larger manifold.  

But the true value of the \h topology construction
is that it leads to a categorical result, an
extension into a topological category of the
categorical results for the Future Chronological
Boundary as a construction among chronological
sets.  The basic strategy is to define a new set
of categories, subcategories of {\bf Chron} and
its allied categories, by restricting the
morphisms to \h continuous functions.  For now, we
will look at (past-)regular chronological sets: 
Let {\bf PregFtopChron} be the category whose
objects are (past-)regular chronological sets and
whose morphisms are \h continuous,
future-continuous functions ({\bf Ftop} denoting
``topology defined with future boundaries in
mind").  Our task is to show that all manner of
future-continuous functions, needed for the
categorical results of
\cite{H}, are \h continuous, also.  Then the same
form of categorical results will apply in the new,
topological, categories; Corollary 2.5 was the 
first of such results.

However, there is a substantial fly in the 
ointment:  It is not true, in general, that the
extension of a \h continuous function to
future completions remains \h continuous---one can
have $f: X \to Y$ future-continuous and \h
continuous, but $\hat f: \hat X \to \hat Y$,
though future-continuous, fail to be
\h continuous.  Essentially, this can happen if $X$
has a timelike boundary.  Here's an example (see
Figure 7):

Let $X = \{(x,t) \in \Bbb L^2 \;|\;x > 0\}$, and 
let $Y = \Bbb L^2$.  Define $f: X \to Y$ by

$$f(x,t) = \left\{
\aligned
&(x,t), \\
&(x, (1 + 1/x)t), \\
&(x, t + 1),
\endaligned \quad \aligned
&t \le 0 \\
&0 \le t \le x \\
&t \ge x 
\endaligned
\right.\;.$$  
It is easy to examine
$f_*(\frac{\partial}{\partial t})$ and
$f_*(\frac{\partial}{\partial x})$ and see thereby
that $f_*$ carries the two future-null vectors
$\frac{\partial}{\partial t} +
\frac{\partial}{\partial x}$ and
$\frac{\partial}{\partial  t} -
\frac{\partial}{\partial x}$ into future-causal
vectors, in each of the regions where $f$ is
differentiable.  Since $f$ is continuous, it 
follows that it is chronological.  The continuity
of $f$ in the obvious sense is equivalent to the
\h continuity of
$f$ (Theorem 2.3).  Thus, from the observation
following Proposition 2.2, it follows that
$f$ is future-continuous (since $Y$ is evidently
past-distinguishing). 

Now consider $\hat f : \hat X \to \hat Y$.  The 
Future Chronological Boundary $\hat\partial(X)$
consists of the right half of future null
infinity, $\frak I_R^+$; future timelike infinity,
$i^+$; and a timelike boundary component that can
be conventionally described as $\{(0,t) \;|\; t
\in \Bbb R\}$; the \h topology is just as is to be
expected.  We have the usual future boundary for
$\Bbb L^2$: $\hat\partial(Y)$ consists of future
null infinity $\frak I^+$ and future timelike
infinity $i^+$, with the expected topology.  On
$\frak I_R^+$ and $i^+$, $\hat f$ does the
expected, essentially the identity map.  But on
the other boundary points, we have 

$$\hat f(0,t) = \left\{ 
\aligned 
&(0,t), \\
&(0, t + 1),
\endaligned \quad \aligned
&t \le 0 \\
&t > 0
\endaligned \right.\;.$$
Thus, $\hat f$ is not \h continuous.

So, what to do?  It appears that what allows 
things to go awry is $\hat f$ carrying boundary
elements into points (including boundary points)
that are timelike-related.  So we will take the
following tack in constructing a subcategory where
this does not happen:  We'll constrain our objects
to have only ``spacelike boundaries", and we'll
insist that $\hat f$ preserve the spacelike nature
of such. 

To this end, let us call a regular point $x$ in a
chronological set $X$ {\it inobservable\/}
(in $X$) if the only IP which contains $I^-(x)$ is
$I^-(x)$ (non-regular points will be considered in
Section 4); we will say a chronological set
$X$ {\it has only spacelike boundaries\/} if (1)
all elements of $\hat\partial(X)$ are inobservable
(in $\hat X$) and (2) the inobservables in $\hat
X$ form a closed subset, or $\hat\partial(X)$ is
closed in $\hat X$ (the reason for (2) is
technical in nature, allowing proofs to go 
through); and a chronological map $f : X
\to Y$ will be said to {\it preserve spacelike
boundaries\/} if $\hat f$ (not just
$f$) preserves inobservables (i.e., for any $p$
inobservable in $\hat X$, whether boundary element 
or otherwise, $\hat f(p)$ is inobservable in $\hat
Y$).  Then we define the category {\bf
Preg\-Spbd\-Ftop\-Chron} to have as objects
regular chronological sets with only spacelike
boundaries, and to have as morphisms
future-continuous, \h continuous maps which
preserve spacelike boundaries.   

What are the objects of {\bf 
Preg\-Spbd\-Ftop\-Chron}?  Suppose $M$ is a
spacetime and $X$ is $M$ with some elements of
$\hat\partial(M)$ conjoined; then
$\hat\partial(X)$ consists of the remaining 
elements of $\hat\partial(M)$, i.e., those not
already in $X$; $\hat X$ is the same as $\hat M$,
and the inobservables there are all in
$\hat\partial(M)$ (since any point of a spacetime
is observable).  Thus, to say $X$ has only
spacelike boundaries is to say (1) the remaining
elements of
$\hat\partial(M)$---those not already in
$X$---are all inobservable, and (2) either those
remaining elements of $\hat\partial(M)$ form a
closed subset, or the entire collection of
inobservables in $\hat M$ (i.e., in
$\hat\partial(M)$) forms a closed set.  Thus, for
instance, $M$ might have a timelike portion in its
boundary, so long as $X$ includes the timelike
components of $\hat\partial(M)$; but either all the
inobservable boundary points are required to form 
a closed set, or the non-included boundary points
must do so.  In the case that $M$ has only
inobservable boundary points to begin with, it
doesn't matter which boundary points, if any, are
included in $X$, as then the set of inobservables
is the same as
$\hat\partial(M)$, and that is always closed
(Proposition 2.6).  In any case, for any
chronological set $X$, $\hat X$ has only spacelike
boundaries, as $\hat\partial(\hat X)$ is empty
(and, hence, closed).

It is fruitful to note that a strongly causal
spacetime $M$ which has only spacelike boundaries
is necessarily past-determined:  For $I^-(x)
\subset I^-(w)$, let $\{x_n\}$ be a future chain
with $x$ as future limit; then each $x_n \ll w$,
so there is a future-directed timelike curve
$\gamma_n$ from $x_n$ to $w$.  The curves
$\{\gamma_n\}$ have a limit curve $\gamma$, which
is a future-directed causal curve beginning at
$x$; either $\gamma$ terminates at $w$ or it has
no future endpoint.  In the latter case, $\gamma$
generates an IP $P$ which is in $\hat M$; but then
$P$ is a proper subset of $I^-(w)$, so $P$ is
observable.  So if $M$ has only spacelike
boundaries, $\gamma$ must reach $w$, whence $x
\prec w$.  Thus, if $w \ll y$, we have $x \prec w
\ll y$, so $x \ll y$.

But for chronological sets in general, having only
spacelike boundaries does not imply
past-determination:  If $M$ is not past-determined,
then neither is $\hat M$, but the latter has only
spacelike boundaries.

What are the morphisms of {\bf
Preg\-Spbd\-Ftop\-Chron}?  For $f: X \to Y$ to be 
in this category, it is required that $\hat f$
preserve inobservables:  This is a restriction not
just on $f$, but also on what $\hat f$ does to
elements of $\hat\partial(X)$.  Thus, for instance,
if $X$ is lower-$\Bbb L^2$, $\{(x,t)
\;|\; t < 0\}$, we cannot have $Y$ as $\Bbb L^2$ 
and $f$ as the inclusion map, as $\hat f$ carries
inobservables of $\hat X$ into observables in $\hat
Y$.  The effect of this restriction
is to require, in essence, that we consider only 
maps which take boundaries to boundaries, not to
interiors.

We need to note that {\bf Preg\-Spbd\-Ftop\-Chron} 
is, indeed, a category:  For any $X$, $\hat 1_X =
1_{\hat X}$, so $\hat 1_X$ preserves
inobservables.  For $f : X
\to Y$ and $g: Y \to Z$, if $\hat f$ and $\hat g$ 
both preserve inobservables, so does $\widehat{g
\circ f} = \hat g \circ \hat f$.  

Next we need to see that future completion is a
functor for the appropriate subcategory;  
in particular, for $f: X \to Y$ in our {\bf
SpbdFtop} category, we need $\hat f : \hat X \to
\hat Y$ also \h continuous.  From Proposition 6 in
\cite{H}, one would expect that we need to have
$Y$ past-determined in order to have $\hat f$
chronological; but, as it turns out, the
assumption of spacelike boundaries makes that an
unnecessary hypothesis.

\proclaim{Proposition 2.7}Let $f: X \to Y$ be a 
map in the category {\bf
Pdis\-Preg\-Spbd\-Ftop\-Chron}.  Then  $\hat
f: \hat X \to \hat Y$ is future-continuous and \h
continuous.\endproclaim

\demo{Proof} First we consider why $\hat f$ is
even chronological:  This is established in
Proposition 6 of \cite{H}, under the additional
hypothesis that $Y$ is past-determined.  That
hypothesis is used only to show that for $P \in
\hat\partial(X)$ and $q \in \hat X$ (whether in
$X$ or in $\hat\partial(X)$), if
$P \ll q$, then $\hat f(P) \ll \hat f(q)$.  But
since $X$ has only spacelike boundaries, for any
$P \in \hat\partial(X)$, there is no $q \in \hat
X$ with $P \ll q$ (that would imply $P \subset
\hat I^-(q)$, whence $P = \hat I^-(q)$, so $P
\ll q$ is false).  That is the only portion of
the proof that uses past-determination. 
Consequently, $\hat f$ is future-continuous. 

To show \h continuity of $\hat f$, consider first a
sequence of points $\sigma = \{x_n\}$ in $X$ with
$x \in X$ in  $L_{\hat X}(\sigma)$.  From the proof
of Theorem 2.4 we know that $L_X(\sigma) = L_{\hat
X}(\sigma) \cap X$ for $\sigma$ in $X$, so this is
equivalent to $x$ being in $L_X(\sigma)$. Since
$f$ is \h continuous, it follows that $f(x) \in
L_Y^\Omega[\,f[\sigma]\,]$. 

From knowing that, in any regular chronological set
$Z$, $L_Z(\sigma) = L_{\hat Z}(\sigma) \cap Z$ for
$\sigma$ in $Z$, one easily derives that for any
subset $A \subset Z$, $L_Z[A] = L_{\hat Z}[A] 
\cap Z$; then an easy transfinite induction
establishes that for all ordinals $\alpha$,
$L_Z^\alpha[A] \subset L_{\hat Z}^\alpha[A] \cap
Z$ (equality fails at $\alpha = 2$: one has
$L^2_Z[A] = L_{\hat Z}[L_{\hat Z}[A] \cap Z] \cap
Z$; in addition to $ L^2_Z[A]$, 
$L^2_{\hat Z}[A] \cap Z$ also contains
$L_{\hat Z}[L_{\hat Z}[A] \cap
\hat\partial(Z)] \cap Z$).  Thus, in particular,
$L^\Omega_Y[\,f[\sigma]\,] \subset L^\Omega_{\hat
Y}[\,f[\sigma]\,] \cap Y$. Therefore, we have
$f(x) \in L_{\hat Y}^\Omega[\,f[\sigma]\,]$.

Now consider the same sequence $\sigma$ with $P 
\in \hat\partial(X)$ in $L_{\hat X}(\sigma)$.  $P$
is generated by some future chain $c = \{z_m\}$ in
$X$; the IP $ Q = I^-[\,f[c]\,]$ in $Y$ is either
in $\hat\partial(Y)$---in which case $\hat f(P)$ =
$Q$---or is $I^-(y)$ for some unique $y \in Y$---in
which case $\hat f(P) = y$.  In either case, we 
have each $z_m$ is eventually in the past of
$x_n$, so the same is true for $f(z_m)$ and
$f(x_n)$; this gives us that for any $q \ll \hat
f(P)$, eventually $q \ll f(x_n)$ (irrespective of
whether $q \in Y$ or $q \in
\hat\partial(Y)$).  Now consider any IP $Q'$ in 
$\hat Y$ containing $\hat I^-(\hat f(P))$ ($\hat
I^-$ denoting past in $\hat Y$):  Since $P$ is in
$\hat\partial(X)$ and $X$ has only spacelike
boundaries, $P$ is inobservable; since $f$ 
preserves spacelike boundaries, $\hat f(P)$ is also
inobservable; therefore, $Q' = \hat I^-(\hat f(P))$. 
This finishes showing that $\hat f(P) \in L_{\hat
Y}(f[\sigma])$. 

Now consider a sequence $\sigma = \{P_n\}$ of 
points in $\hat\partial(X)$.  Since $X$ has only
spacelike boundaries, all the $P_n$ are
inobservable.  Either the inobservables in $\hat
X$ form a closed subset, or
$\hat\partial(X)$ is closed.  Thus, for $p
\in L_{\hat X}(\sigma)$, either $p$ is inobservable
(because the $P_n$ are) or $p$ is in 
$\hat\partial(X)$ (because the $P_n$ are); but in
the latter case, again $p$ is inobservable, since
$X$ has only spacelike boundaries.  Since
$f$ preserves spacelike boundaries, $\hat f(p)$ is
also inobservable (irrespective of whether $p$ is 
in $X$ or in $\hat\partial(X)$).  Now proceed much
as in the last paragraph:  Since $X$ is
regular, there is some future chain $c =
\{z_m\}$ with $p$ as future limit of $c$ (we need
the assumption of regularity only in case
$p$ might be in $X$, not in $\hat\partial(X)$). 
Each $z_m$ is eventually in the past of $P_n$, so
$f(z_m)$ is eventually in the past of $\hat
f(P_n)$ (since $f$ chronological implies that
$\hat f$ is chronological).  As above, this gives
us that for any $q \ll \hat f(p)$, eventually $q
\ll \hat f(P_n)$.  Since $p$ is inobservable, as
before, no IP can properly contain $\hat I^-(\hat
f(p))$.  Therefore, $\hat f(p) \in L_{\hat Y}(\hat
f[\sigma])$.

Finally consider a general sequence $\sigma = 
\{p_n\}$ in $\hat X$ with $p \in L_{\hat
X}(\sigma)$.  If $\sigma$ contains a subsequence
$\sigma'$ lying in $\hat\partial(X)$, then since
$p$ is also in $L_{\hat X}(\sigma')$, it must be
that $p$ is inobservable (as explained in the
previous paragraph).  As before, for all $q \ll
\hat f(p)$, eventually $q \ll \hat f(p_n)$; there
is no IP properly containing $\hat I^-(\hat
f(p))$; and, therefore, $\hat f(p) \in L_{\hat
Y}(\hat f[\sigma])$.  If, on the other hand, there
is no subsequence of $\sigma$ lying in
$\hat\partial(X)$, then, except for a finite
number of points, which can be ignored, $\sigma$
lies in $X$, and the first two cases apply.
\qed \enddemo

\proclaim{2.8 Corollary} Future completion
$\;\widehat{}\;$ is a functor from
{\bf Pdis\-Preg\-Spbd\-Ftop\-Chron}
to {\bf Fcpl\-Pdis\-Preg\-Spbd\-Ftop\-Chron}.
\endproclaim

\demo{Proof} We need to see that if $X$ is
past-distinguishing, regular, and  has only
spacelike boundaries, then the same is true for
$\hat X$.  The only issue is spacelike boundaries;
but this is trivial, since
$\hat\partial(\hat X)$ is empty (and, hence, is
closed).  We also need to see that if
$f$ is future-continuous and \h continuous and
preserves spacelike boundaries, then the same is 
true for $\hat f$.  By Proposition 2.7, the only
issue is preserving spacelike boundaries; but this
also is trivial, as ``$f$ preserves spacelike
boundaries" is a statement about $\hat f$, and
$\Hat{\Hat f} = \hat f$. \qed \enddemo

\proclaim{2.9 Theorem} The future completion 
functor $\;\widehat{}\; :$ {\bf
Pdis\-Preg\-Spbd\-Ftop\-Chron} $\to$ {\bf
Fcpl\-Pdis\-Preg\-Spbd\-Ftop\-Chron} and the
standard future injections $\hat\iota_X: X \to
\hat X$ yield a left adjoint to the forgetful 
functor, hence a categorically unique way of
providing a future completion, in a topological
category respecting the \h topology, for
past-distinguishing, regular
chronological sets with spacelike boundaries.
\endproclaim

\demo{Proof} By Corollary 2.5, the map $\hat 
\iota_X : X \to \hat X$ is \h continuous, and it
clearly preserves spacelike boundaries
(since $\Hat{\Hat\iota}_X :
\hat X \to \Hat{\Hat X} = \hat X$ is the identity). 
Thus, $\hat\iota_X$ is in the category {\bf
Pdis\-Preg\-Spbd\-Ftop\-Chron}, yielding a
natural transformation $\hat{\boldsymbol\iota} :
\bold I \;\dot\to\; \bold U \circ
\;\widehat{}\;$, where {\bf I} is the identity 
functor on {\bf
Pdis\-Preg\-Spbd\-Ftop\-Chron} and {\bf U}
is the forgetful functor from {\bf
Fcpl\-Pdis\-Preg\-Spbd\-Ftop\-Chron} to {\bf
Pdis\-Preg\-Spbd\-Ftop\-Chron}; what this
says is that for any $f: X \to Y$ in {\bf
Pdis\-Preg\-Spbd\-Ftop\-Chron}, $\hat\iota_Y
\circ f = \hat f \circ \hat\iota_X$ (part of
Proposition 6 in \cite{H}).

All that is needed to establish the left 
adjunction is the universality property:  For any
$f: X \to Y$ in {\bf
Pdis\-Preg\-Spbd\-Ftop\-Chron}, with $Y$
also future-complete, $\hat f: \hat X \to Y$ is
the unique future-continuous, \h continuous map,
preserving spacelike boundaries, such that $\hat f
\circ \hat \iota_X = f$.  The unique existence of
$\hat f$ for being future-continuous and
satisfying this condition is Corollary 7 in
\cite{H}; that $\hat f$ is also \h continuous and
preserves spacelike boundaries is Corollary 2.8
just above.  Standard category theory (as in [M])
then gives the categorical uniqueness of any left
adjoint (i.e., any other functor left-adjoint to
the forgetful functor would be naturally
equivalent to $\;\widehat{}\;$; a natural
equivalence is a natural transformation made up of
isomorphisms). \qed
\enddemo

Upon application of this functor to one of the
objects most of interest---a strongly causal
spacetime $M$---the resultant $\hat M$ is actually
$M^+$, i.e., the GKP Future Causal
Boundary construction (defined as the
past-determination of $\hat M$); this is because,
as observed above, a spacetime with only spacelike
boundaries is already past-determined (as is,
therefore, its future chronological completion). 
However, for categorical results, we cannot
identify $\;\widehat{}\;$ with $\boldkey +$, since
the requisite category contains objects which are
not past-determined.  For categorical discussion of
the GKP operation, we must pursue a discussion of
past-determination, the subject of Section 3.

\head 3. Regular Chronological Sets and Past
Determination
\endhead

The past-determination functor {\bf p} : {\bf
PregChron} $\to$ {\bf PdetPregChron} is the 
categorical method for making a regular
chronological set past-determined; it adds some
additional chronology relations, in case the
original object is not past-deter\-mined already. 
Specifically, if $X$ is a chronological set with
chronology relation $\ll$, then
$X^p$ is the same set with chronology relation 
$\ll^p$ holding between $x$ and $y$ if either $x
\ll y$ or if $I^-(x)$ is non-empty and $I^-(x)
\subset I^-(w)$ for some $w \ll y$.  For any $f: X
\to Y$, $f^p : X^p \to Y^p$ is the same
set-function; for any $X$, $\iota^p_X : X \to X^p$
is the identity on the set-level.  

Propositions 10 and 11 and Corollary 12 in
\cite{H} establish these facts:  For any
chronological set $X$, $X^p$ is past-determined; if
$X$ is past-determined, then $X^p = X$; if $X$ is
past-distinguishing or future-complete, so is
$X^p$.  For any future-continuous $f: X \to Y$, so
long as $X$ is regular, $f^p$ is also
future-continuous; thus {\bf p} is a functor as
claimed above.  Within the past-regular
category, the maps $\iota^p_X$ are
future-continuous and form a natural transformation
$\boldsymbol\iota^\bold p : \bold I \;\dot\to\; 
\bold U \circ \bold p$, where {\bf I} is the
identity functor on {\bf PregChron} and {\bf U} :
{\bf PdetPregChron}
$\to$ {\bf PregChron} is the forgetful functor.  
The functor {\bf p} and the natural transformation
$\boldsymbol \iota^{\bold p}$ form a left adjoint 
to the forgetful functor {\bf U} in virtue of the
universality property:  For any $f: X \to Y$ in 
{\bf PregChron} with $Y$ past-determined, $f^p :
X^p \to Y$ is the unique future-continuous map
with $f^p \circ \iota^p_X = f$.

Our first task is to see that all this 
machinery carries over to the topological
categories:  We need that past-determination
preserves objects in the right categories, that
functions stay continuous (with respect to the \h
topologies) upon past-determination, and that the
maps $\iota^p_X$ are continuous.  This all falls
in line quite nicely; the crucial fact is that
past-determination has no effect on the \h
topology.

We need a lemma to identify the IPs in $X^p$ with
those in $X$, analogous to the lemma used in 
Theorem 2.4:  Let $\I(X)$ denote the set of all IPs
of a chronological set $X$.  Given a chronological 
set $X$, for any $P \in \I(X)$, let $P^p$ denote
$I^{-p}[P]$, where $I^{-p}$ denotes the past using
the $\ll^p$ relation; and for any $Q \in \I(X^p)$,
let $Q_0$ denote $I^-[Q]$ ($I^-$ denoting past
using $\ll$).

\proclaim{Lemma 3.1} For any chronological set $X$,
the maps $P \mapsto P^p$ and $Q \mapsto Q_0$ 
establish an isomorphism, as partially ordered
sets under inclusion, between $\I(X)$ and
$\I(X^p)$.
\endproclaim

\demo{Proof} First note that for any $x, y, z$ in 
$X$, $x \ll y \ll^p z$ implies $x \ll z$ (because
$I^-(y) \subset I^-(w)$ for some $w \ll z$ and $x
\in I^-(y)$, so $x \in I^-(w)$, i.e., $x \ll w$);
and $ x \ll^p y \ll z$ implies $x \ll^p z$ (because
$I^-(x)$ is non-empty and $I^-(x) \subset I^-(w)$
for some $w \ll y$, hence $w \ll z$).  Conversely,
$x \ll z$ implies there is some $y$ with $x \ll y
\ll^p z$ (there is some
$y$ with $x \ll y \ll z$, and $y \ll z$ implies $y
\ll^p z$); and $x \ll^p z$ implies there is some 
$y$ with $x \ll^p y \ll z$ ($I^-(x) \subset
I^-(w)$ for some $w \ll z$; pick $y$ with $w \ll y
\ll z$). It follows that for any $z \in X$,
$I^-[I^{-p}(z)] = I^-(z)$ and $I^{-p}[I^-(z)] =
I^{-p}(z)$; therefore, for any $A \subset X$, 
$I^-[\,I^{-p}[A]\,] = I^-[A]$ and 
$I^{-p}[\,I^-[A]\,] = I^{-p}[A]$.

In particular, for any $P \in \I(X)$, let $c$ be a 
future chain generating $P$; then $P^p = I^{-p}[P] 
= I^{-p}[\,I^-[c]\,] = I^{-p}[c]$.  Therefore,
since $c$ is also a future chain in $X^p$, we have
that $P^p$ is  in $\I(X^p)$.  Also, for any $Q \in
\I(X^p)$, let $c$ be a future chain in $X^p$
generating $Q$, and let
$c'$ be an associated chain in $X$; then $Q_0 =
I^-[Q] = I^-[\,I^{-p}[c]\,] = I^-[\,I^{-p}[c']\,]
= I^-[c']$.  Therefore, $Q_0$ is in $\I(X)$. 
Thus, the two maps at least have targets as
advertised.

We have, for any $P \in \I(X)$, $(P^p)_0 =
I^-[\,I^{-p}[P]\,] = I^-[P] = P$; and for any $Q\in
\I(X^p)$, $(Q_0)^p = I^{-p}[\,I^-[Q]\,] = I^{-p}[Q] 
= Q$.  Thus, the the two maps yield a bijection of 
the respective sets of IPs.  The preservation of
the subset relation is evident. \qed \enddemo

Note that $X$ is regular if and only if $X^p$ is
regular:  If $I^-(x) = P \cup P'$ for $P$ and 
$P'$ in $\I(X)$, neither equal to $I^-(x)$, then
$I^{-p}(x) = I^{-p}[I^-(x)] = I^{-p}[P \cup P'] =
I^{-p}[P] \cup I^{-p}[P'] = P^p \cup (P')^p$,
neither equal to $I^{-p}(x)$ (if, e.g., $P^p =
I^{-p}(x)$, then $P = (P^p)_0 = I^-[P^p] =
I^-[I^{-p}(x)] = I^-(x)$). The other direction
follows in exactly analogous fashion.

\proclaim{Proposition 3.2} For any regular
chronological set $X$, $\iota^p_X : X \to X^p$ is a
homeomorphism in the respective \h topologies.
\endproclaim

\demo{Proof} Let $L$ and $L^p$ denote the
limit-operators for, respectively, $X$ and $X^p$.  
Let $\sigma$ be any sequence in $X$ with
$ x \in L(\sigma)$.  For all $y \ll^p x$, there is 
some $z$ with $y \ll^p z \ll x$.  For all $n$
sufficiently large, $z \ll \sigma(n)$; then $y
\ll^p  z \ll \sigma(n)$, so $y \ll^p \sigma(n)$. 
For any IP $Q$ of $X^p$ with $Q \supset I^{-p}(x)$,
Lemma 3.1 gives us $Q_0 \supset (I^{-p}(x))_0 = 
I^-(x)$.  Suppose for every $y \in Q$, there is
some subsequence $\tau \subset \sigma$ with, for
all $n$, $y \ll^p \tau(n)$.  For any $z \in Q_0$,
there is some $y \in Q$ with $z \ll y$; then for 
some subsequence $\tau \subset \sigma$, for all
$n$, $y \ll^p \tau(n)$.  This gives us $z \ll y
\ll^p \tau(n)$, so $z \ll \tau(n)$ for all $n$. 
Therefore, $Q_0 = I^-(x)$.  By Lemma 3.1, $Q =
I^{-p}(x)$.  This shows $x \in L^p(\sigma)$.

Now let $\sigma$ be any sequence with $x \in
L^p(\sigma)$.  A formally identical proof, with 
$\ll$, $I^-$, and the $(\;)^p$ map swapped,
respectively, with $\ll^p$, $I^{-p}$, and the
$(\;)_0$ map, establishes that $x \in L(\sigma)$.

With identical limit-operators, $X$ and $X^p$ have 
the same \h topologies; specifically, the identity
map ($\iota^p_X$) is a homeomorphism. \qed \enddemo

\proclaim{Theorem 3.3} Past determination is a 
functor {\bf p} $:$ {\bf Preg\-Ftop\-Chron} $\to$
{\bf Pdet\-Preg\-Ftop\-Chron}; the maps $\iota^p_X$
form a natural transformation 
$\boldsymbol\iota^{\bold p}$ in the {\bf
Preg\-Ftop} categories; and {\bf p} and
$\boldsymbol\iota^{\bold p}$ form a left adjoint 
for the forgetful functor {\bf U} $:$ {\bf
Pdet\-Preg\-Ftop\-Chron} $\to$ {\bf 
Preg\-Ftop\-Chron}.

All this is also true in the {\bf SpbdFtop}
categories.
\endproclaim  

\demo{Proof} To show that {\bf p} is a functor 
for the {\bf Ftop} categories, we need only show
that for $f: X \to Y$ future-continuous and \h
continuous, that $f^p : X^p \to Y^p$ is \h
continuous.  This follows immediately from
Proposition 3.2 and the commutative diagram $f^p
\circ \iota^p_X = \iota^p_Y \circ f$.  To show
$\boldsymbol\iota^{\bold p}$ is a natural
transformation for the topological categories, we
just need each $\iota^p_X$ \h continuous, i.e.,
Proposition 3.2.  Then the universality property
for the chronological categories translates
immediately into the appropriate universality
property for the topological categories---for any
future-continuous, \h continuous $f: X \to Y$ with
$X$ regular and $Y$ regular and past-determined,
$f^p : X^p \to Y$ is the unique future-continuous
and \h continuous map satisfying $f^p \circ
\iota^p_X = f$.

To extend these results to the {\bf SpbdFtop}
categories, we need to show that 
$\widehat{\iota_X^p}$ preserves inobservables, and
if $\hat f$ preserves inobservables, then so does
$\widehat{f^p}$:

Since the maps of Lemma 3.1 between $\I(X)$ and
$\I(X^p)$ preserve inclusion among IPs, they
also preserve the property of being inobservable or
not.  For $P \in
\hat\partial(X)$,
$\widehat{\iota^p_X}(P) = P^p$
(reason: $I^{-p}[\,\iota^p_X[P]\,] = I^{-p}[P] =
P^p$,  which is in $\hat\partial(X^p)$, so that is
$\widehat{\iota_X^p}(P)$).  Thus,
$\widehat{\iota^p_X}$ preserves inobservables in
$\hat\partial(X)$.  For $x \in X$, if
$I^{-p}(\iota^p_X(x)) = I^{-p}(x)
\subset Q$ for some $Q \in \I(X^p)$, then
$I^-(x) = I^-[I^{-p}(x)] 
\subset I^-[Q] = Q_0 \in \I(X)$; therefore, 
$\iota^p_X$ preserves inobservables in $X$.

For $f: X \to Y$, with $\hat f: \hat X \to \hat Y$
preserving inobservables, consider $\widehat{f^p}:
\widehat{X^p} \to \widehat{Y^p}$.  First note that
any $x \in X$ is inobservable in $X$ if and only
if it is inobservable in $X^p$ (if $I^-(x)
\subset P$, then $I^{-p}(x) = I^{-p}[I^-(x)]
\subset I^{-p}[P] = P^p$; other directions
similarly).  Thus, if $x \in X^p$ is inobservable
in $X^p$, then it's also inobservable in $X$, so
$f(x)$ is inobservable in $Y$, so $f^p(x) =f(x)$
is inobservable in $Y^p$.  Second, consider $Q
\in \hat\partial(X^p)$:  If $Q$ is inobservable,
then so is $Q_0$, so $\hat f(Q_0)$ is also
inobservable.  Now suppose $\hat f(Q_0) = y \in
Y$, i.e., that
$I^-[\,f[Q_0]\,] = I^-(y)$.  Let $Q_0$ be
generated by a future chain $c$, which then also
generates $Q$ (in $X^p$); then
$I^{-p}[\,f^p[Q]\,] = I^{-p}[\,f^p[c]\,] =
I^{-p}[\,f[c]\,] = I^{-p}[\,f[Q_0]\,] =
I^{-p}[\,I^-\,[\,f[Q_0]\,]\,] = I^{-p}[I^-(y)] =
I^{-p}(y)$, so $\widehat{f^p}(Q) = y$ also.  Thus,
$\widehat{f^p}(Q)$ is inobservable.  If, on the 
other hand, $\hat f(Q_0) \in \hat\partial(Y)$,
i.e., $I^-[\,f[Q_0]\,]$ is not any $I^-(y)$, then
for $Q_0$ generated by $c$, $\hat f(Q_0) =
I^-[\,f[c]\,]$.  Then
$I^{-p}[\,f^p[Q]\,] =I^{-p}[\,f[Q]\,] =
I^{-p}[\,f[c]\,]$ is no $I^{-p}(y)$, for then 
$I^-(y) = I^-[I^{-p}(y)] =
I^-[\,I^{-p}[\,f[c]\,]\,] = I^-[\,f[c]\,] = \hat
f(Q_0)$, which we are supposing cannot be any
$I^-(y)$.  Therefore,
$I^{-p}[\,f^p[Q]\,]$ is some $P \in 
\hat\partial(Y^p)$ (specifically, $P =
I^{-p}[\,f[c]\,]$) and
$\widehat{f^p}(Q) = P$.  If
$P \subset P'$ for some $P' \in
\hat\partial(Y^p)$, then $P_0  \subset P'_0$, and
$P_0 = I^-[\,I^{-p}[\,f[c]\,]\,] = I^-[\,f[c]\,] =
\hat f(Q_0)$.  Since this last is inobservable, we
must have $P'_0 = P_0$, so $P' = P$.  This says,
again, that $\widehat{f^p}(Q)$ is inobservable.
\qed \enddemo

We thus have a functor $\;\widehat{}\; \circ 
{\bold p} :$ {\bf Pdis\-Preg\-Spbd\-Ftop\-Chron}
$\to$ {\bf
Fcpl\-Pdet\-Pdis\-Preg\-Spbd\-Ftop\-Chron} and a
natural transformation $\hat{\boldsymbol \iota} 
\circ \boldsymbol \iota^{\bold p}$, forming a left
adjoint to the forgetful functor.  But to reach
the actual GKP construction we need the functor
$\boldkey +$, where $X^+ = (\hat X)^p$.  Recall
that this functor is constructed using $j_X :
\widehat{X^p} \to (\hat X)^p$ via $f^+ = j_Y \circ
\widehat{f^p} \circ  (j_X)^{-1} : X^+ \to Y^+$
(for $f: X \to Y$); and the same maps are also
used to form the natural transformation
$\boldsymbol \iota^{\boldkey +}$ via
$\iota^+_X = j_X \circ \hat\iota_{X^p} \circ 
\iota^p_X : X \to X^+$.  The action of $j_X$ on
the elements of $\hat\partial(X^p)$ is just the map
$Q \mapsto Q_0$ of Lemma 3.1.

We need to show that $j_X$ and $(j_X)^{-1}$ are \h
continuous.  But this follows automatically, since
$j_X$ is an isomorphism of the chronological sets
(Proposition 13 of \cite{H}), and the \h topologies 
are constructed directly from the respective
chronology relations.  We also need that
$\widehat{j_X}$ and  its inverse preserve
inobservables; this follows automatically for the
same reason.  Thus, all the elements are in order:

\proclaim{Theorem 3.4} There is a functor 
$\boldkey +$ from
{\bf Pdis\-Preg\-Spbd\-Ftop\-Chron} to {\bf
Fcpl\-Pdet\-Pdis\-Preg\-Spbd\-Ftop\-Chron}, with a
natural transformation $\boldsymbol
\iota^{\boldkey +}$ forming a left adjoint to the
forgetful functor. \qed
\endproclaim 

This completes the categorical formulation of the
Future Chronological Boundary in the topological,
spacelike-boundary category.  But we can also 
state results that are not strictly categorical,
such as those that involve objects in different
categories (i.e., obeying different hypotheses). 
For instance,  From Theorem 14 in \cite{H} we obtain
this result:

\proclaim{Theorem 3.5} Let $X$ and $Y$ be 
regular chronological sets with only spacelike
boundaries, with $Y$ past-distinguishing.  For any
future-continuous, \h continuous map $f: X \to Y$,
preserving spacelike boundaries, there is a unique
future-continuous, \h continuous map $f^+ : X^+
\to Y^+$, preserving spacelike boundaries,
that commutes with the natural transformation
$\boldsymbol\iota^{\boldkey +}$.
\endproclaim

\demo{Proof} Theorem 14 of \cite{H} gives the 
unique existence of $f^+$ with $f^+ \circ
\iota_X^+ = \iota_Y^+ \circ f$, just for $f$
future-continuous.  $Y$ does not need to be regular
for that, but it appears in the hypotheses here,
as we have not yet defined \h topology for
non-regular chronological sets.

Theorems 2.7 and 3.3, respectively, give us \h
continuity for the the future completion and
past-determination of future-continuous functions. 
Theorem 3.3 requires no more than the hypotheses 
given here, but Theorem 2.7 is stated for the
domain being past-distinguishing, as well as
regular.  However, examination of the proof there
shows that past-distinguishment is never
employed:  It is in the hypotheses for the
target-space only so as to ensure that $\hat f:
\hat X \to \hat Y$ can be defined, and is in the
hypotheses of the domain only so that the theorem
is stated in terms of a functor on a category. 
Thus, we already have all we need to know that
$f^+$ here is \h continuous. \qed \enddemo

More along these lines can be accomplished once we
admit non-regular chronological sets into our
consideration in Section 4 (see Proposition 4.6).

\smallpagebreak

Theorem 1.1 yields an obvious topological result,
not requiring spacelike boundaries:  

\proclaim{Theorem 3.6} Let $X$ be any regular
chronological set; then any future-completing,
past-distinguishing boundary on $X$ is \h
homeomorphic to $\hat\partial(X)$.
\endproclaim

\demo{Proof} If $Y$ is a past-distinguishing
future completion of $X$ in the sense of Theorem 
1.1 (with $i: X \to Y$), then that theorem yields
a chronological isomorphism $i^+$ of $X^+ = (\hat
X)^p$ with $Y^+ = Y^p$.  This map carries $X$ in
$\hat X$ onto $Y_0 = i[X]$ in $Y$; hence, it
carries $\hat\partial(X)$ onto $\partial(Y) = Y -
Y_0$. Since the
\h topologies are determined solely by the
chronology relations (in this case, the
past-determined chronology relations), this is
also a \h homeomorphism.  In particular, we get a
homeomorphism of $\hat\partial(X)$ with
$\partial(Y)$ in the \h topologies determined by
$\ll^p$ in $\hat X$ and $Y$.  But by
Proposition 3.2, there is no difference in the \h
topologies determined by $\ll^p$ and by $\ll$. 
Thus, we have a homeomorphism of the two
boundaries in the original spaces, $\hat X$ and
$Y$. \qed \enddemo  

Again, this strong rigidity will be relaxed once 
we consider non-regular chronological sets,
allowing for a generalized sense of
future completion (see Corollary 4.9).

If we don't insist on the future boundary actually
future-completing the chronological set we start
out with, then we must have a portion of the
Future Chronological Boundary, at least
\h topologically (the chronology can be
different, but it is the same in the
past-determination).  This can be couched as a
semi-rigid version of Theorem 2.4, with the
addition of past-distinguishing:

\proclaim{Theorem 3.7}  Let $\bar X$ be a
past-distinguishing, regular chronological set with
$X$ a subset of
$\bar X$ satisfying the following:
\roster
\item The restriction of $\ll$ to $X$ yields 
another regular chronological set; and
\item for any $p \ll q$ in $\bar X$, there is some $x
\in X$ so that $p \ll x \ll q$.
\endroster
Then there is a topological embedding of $\bar X$
into $\hat X$, which is the identity on $X$.
\endproclaim

\demo{Proof} Let $I^-$ denote the past in $X$,
$\bar I^-$ the past in $\bar X$, and $\hat I^-$
the past in $\hat X$.  Let $\bar\partial(X)$ denote
$\bar X - X$, our (partial) future boundary on
$X$. 

Consider $p \in \bar\partial(X)$, and let $Q
= \bar I^-(p)$.  Apply the operation from the Lemma
in Theorem 2.4 to obtain $Q_0 = Q \cap X$. 
Suppose there were $x \in X$ such that $Q_0 =
I^-(x)$; then $\bar I^-(x) = \bar I^-[Q_0] =
\overline{(Q_0)} = Q = \bar I^-(p)$; then
past-distinguishing yields $x = p$, which cannot
be.  Therefore, $Q_0 \in \hat\partial(X)$.  

Thus, we can define a map $i : \bar X \to \hat X$
by $i(x) = x$ for $x \in X$ and $i(p) = (\bar
I^-(p))_0$ for $p \in \bar\partial(X)$, and this
is the identity on $X$, taking boundary points to
boundary points.  This map is injective, since
$i(p) = i(p')$ means $(\bar I^-(p))_0 = (\bar
I^-(p'))_0$, hence  $\overline{(\bar I^-(p))_0} =
\overline{(\bar I^-(p'))_0}$, hence $\bar I^-(p) =
\bar I^-(p')$, so $p = p'$ by past-distinguishing.

That $i$ is chronological is very easy to
establish.  But we need more:  We want a
chronological isomorphism onto the image.  This,
however, is not necessarily true for $i$; but it
is true for the past-determination, $i^p : \bar
X^p \to (\hat X)^p$.  The only real difficulty
comes with $i(p) \ll^p x$ for $p \in
\bar\partial(X)$ and $x \in X$; we need to obtain
$p \ll^p x$:

Knowing $i(p) \ll^p x$, we can find (as in the
proof of Lemma 3.1) $q \in \hat X$ with $i(p) \ll^p
q \ll x$.  Then, whether $q$ is in $X$ or in
$\hat\partial(X)$, there is $x' \in X$ with $q
\ll x' \ll x$ (since for $q \in \hat\partial(X)$,
$q \ll x$ means there is $w \ll x$ in $X$ with
$I^-(q) \subset I^-(w)$).  Then we have $i(p)
\ll^p x'$, so for some $w \ll x'$ in $\hat X$,
$\hat I^-(i(p)) \subset \hat I^-(w)$.  Let $Q =
\bar I^-(p)$, so $i(p) = Q_0$.  Then $\hat
I^-(Q_0)$ includes all the elements of the IP
$Q_0$.  Since $\hat I^-(Q_0) \subset \hat I^-(w)$,
it follows that $Q_0 \subset \hat I^-(w) \cap X
\subset I^-(x')$.  Therefore, $Q =
\overline{(Q_0)} \subset \overline{I^-(x)} = \bar
I^-(x')$.  That, together with $x' \ll x$, gives
us $p \ll^p x$. 

All the other implications follow readily,
yielding that $i^p$ is a chronological isomorphism
onto its image.  This gives us $i^p$ as a
\h homeomorphism onto its image.  Finally,
Proposition 3.2 gives us that $i$ is also a \h
homeomorphism onto its image. \qed \enddemo

\head 4. Topology for Non-Regular Chronological 
Sets
\endhead 

The example cited at the beginning of Section 2
illustrates both the desirability and the
challenge of non-regular boundary elements for a
spacetime:  In that example, the spacetime $M$,
consisting of $\Bbb L^2$ with the negative
time-axis deleted, has, as Future Chronological 
Boundary, two closed half-lines, representing
either side of the missing axis.  This yields the
rather eccentric situation of a pair of
non-Hausdorff elements in the boundary:  The two
future-ends of those half-lines,
$P^+_0$ and $P^-_0$, each trying to enact the
role of the missing point $(0,0)$, are both limits
of the sequence $\sigma(n) = (0, 1/n)$, i.e., both
are in $L(\sigma)$.  The solution to this somewhat
unnatural construction, as provided in the full
Causal Boundary of \cite{GKP}, is to meld the two
points together, into the single boundary point
$B_0$; the future of $B_0$ is defined to be
$I^+((0,0))$ (plus the elements of the Future 
Causal Boundary ``at infinity"), and its past is
defined to be $I^-((0,0))$ except for the deleted
semi-axis (plus the other boundary elements
$P^+_s$ and $P^-_s$ for $s < 0$; and, in the full
Causal Boundary, the points ``at infinity" of the
Past Causal Boundary).  This produces a
chronological set, but $I^-(B_0)$ is decomposable
as $\bar P^+_0 \cup \bar P^-_0$ (the bar denoting
inclusion of the other boundary elements, as
appropriate).  This is, in some sense, a much more
natural object to use as a boundary of $M$ than is
$\hat\partial(M)$.  The challenge is to topologize
it in a way that makes $B_0$ the (unique) limit of
the sequence $\sigma$; $B_0$ should also be the
limit of such sequences as $\tau^+(n) = (1/n,0)$
and $\tau^-(n) = (-1/n,0)$, and even of $\tau^0$
defined by $\tau^0(2n) =
\tau^+(n)$ and $\tau^0(2n+1) = \tau^-(n)$ (see
Figure 8).

To explore this topology, we must make explicit
use of those portions of the past of a point $x$
which make $I^-(x)$ decomposable:  For any point
$x$ in a chronological set $X$, consider $\I_x =
\{P \in \I(X) \;|\; P \subset I^-(x)\}$ as a
partially ordered set under inclusion.  Call any
$P \in \I_x$ a {\it past component\/} of $x$
if $P$ is a maximal element of $\I_x$.  IPs
are required to be non-empty, so $I^-(x)$ must be
non-empty if $x$ is to have any past components. 
So long as that condition is met, there will be
a past component containing every point in its
past:

\proclaim{Proposition 4.1} Let $X$ be a
chronological set and $x$ a point of $X$.  For
every $y \ll x$, there is a past component $Q$ of
$x$ containing $y$.
\endproclaim

\demo{Proof}  First note that the union of any
chain of IPs is an IP:  Recall that
a past set $P$ is an IP if and only if for any
$x_1$ and $x_2$ in $P$, there is some $z \in P$
with $x_1 \ll z$ and $x_2 \ll z$.  If $\{P_\alpha
\;|\; \alpha < \Gamma\}$ is a chain of IPs, ordered
by inclusion ($\Gamma$ an ordinal number), then $P
= \bigcup_{\alpha < \Gamma} P_\alpha$ is clearly a
past set, and by the criterion just mentioned, it
is an IP (for $x_1 \in P_\alpha$ and $x_2 \in
P_\beta$, both $x_1$ and $x_2$ are in $P_\gamma$
for $\gamma = \text{max}\{\alpha, \beta\}$, and we
then locate $z \in P_\gamma$).  

Therefore, every chain in $\I_x$ has an upper
bound. By Zorn's Lemma, every element of $\I_x$ is
contained in a maximal element of $\I_x$,
i.e., a past component of $x$.

For $y \ll x$, there is some $y_1$ with $y \ll
y_1 \ll x$.  We can define a future chain
$\{y_n\}$ thus, by choosing $y_{n+1}$ such that
$y_n \ll y_{n+1} \ll x$.  Then $P =
I^-[\{y_n\}]$ is an element of $\I_x$
containing $y$.  $P$ sits in a past-component $Q$
of $x$, and we have $y \in P \subset Q$. \qed
\enddemo

We next define a notion of a sequence
``converging" to an IP, expressed in terms of a
function $\Cal L: \Cal S(X) \to \frak P(\I(X))$
from sequences in $X$ to the power set of IPs in
$X$: 

\definition{Limit-operator for IPs in a
chronological set} For $X$ any chronological
set, the limit-operator $\Cal L$ for IPs  is
defined thus:

For a sequence $\sigma$ in $X$ and $P
\in \I(X)$, $P \in \Cal L(\sigma)$ if and only if

\roster 
\item for all $x \in P$, eventually $x \ll
\sigma(n)$, and
\item for any IP $Q \supset P$, if for all $x
\in Q$, there is some subsequence $\tau \subset
\sigma$ such that for all $k$, $x \ll \tau(k)$,
then $Q = P$.
\endroster
\enddefinition

(Clause (2) is equivalent to this: for any IP $Q$
properly containing $P$, for some $x \in Q$,
eventually $x \not\ll \sigma(n)$.)

Thus, $\Cal L$ plays the same role for IPs that
the limit-operator $L$ played before for points. 
In the example above, we have $\bar P^+_0$ is
in $\Cal L(\sigma)$ and in $\Cal L(\tau^+)$, but
not in $\Cal L(\tau^-)$ or in $\Cal L(\tau^0)$;
and dually for $\bar P^-_0$.  Our goal is to have
all four sequences converge to $B_0$, whose past
components are $\bar P^+_0$ and $\bar P^-_0$.  It
is especially the sequence $\tau^0$, bouncing back
and forth from $\tau^+$ to $\tau^-$, that
motivates this definition:

\definition{Limit-operator for points in a 
chronological set} For $X$ any chronological
set, the limit-operator $L$ for points is
defined thus: 

For $\sigma$ a sequence in $X$ and $x \in X$, $x
\in L(\sigma)$ if and only if for every
subsequence $\tau \subset \sigma$, there is a
subsubsequence $\rho \subset \tau$ and a past
component $P$ of $x$, such that $P \in \Cal
L(\rho)$.  
\enddefinition

(We shall have frequent occasion to
mention the family of past components of a point
$x$; call this $\frak P_x$.)

Thus, in the example at hand, we have $B_0 \in
L(\tau^0)$ because for any subsequence $\tau'
\subset \tau^0$, either $\tau'$ has infinitely many
points in $\tau^+$ or it has infinitely many in
$\tau^-$ (or both).  If the former, then it has a
subsubsequence $\rho^+$ which lies in $\tau^+$,
and $\bar P^+_0 \in \Cal L(\rho^+)$ (because
$\bar P^+_0 \in \Cal L(\tau^+)$); and if the
latter, then we find a subsubsequence
$\rho^- \subset \tau^-$ and $\bar P^-_0 \in \Cal
L(\rho^-)$.

Note first of all that this definition of
limit-operator reduces to the original one given
in Section 2---call it $L_0$--- for those points of
$X$ which are regular:  For such $x$, this
definition says that $x \in L(\sigma)$ if and only
if, for every $\tau \subset \sigma$, there is a
$\rho \subset \tau$ such that $I^-(x) \in \Cal
L(\rho)$; since $\sigma_1 \subset \sigma_2$
implies $\Cal L(\sigma_2) \subset \Cal
L(\sigma_1)$, this is equivalent to saying that
for all $\sigma' \subset \sigma$, $I^-(x) \in \Cal
L(\sigma')$. That is equivalent to saying that for
all $\sigma' \subset \sigma$, $x \in
L_0(\sigma')$.  But that last is equivalent to
saying $x \in L_0(\sigma)$. 

Next note that this definition of limit-operator
does, indeed, define a topology:  If $\sigma'
\subset \sigma$ and $x \in L(\sigma)$, then for
any $\tau \subset \sigma'$, we also have $\tau
\subset \sigma$, so there is $\rho \subset \tau$
and a past component $P$ of $x$ with $P \in \Cal
L(\rho)$; thus, $x \in L(\sigma')$ also. We will
again call this the \h topology. 

Note that for any $x \in X$, with $\hat x$
denoting the constant sequence $\hat x(n) = x$,
$\Cal L(\hat x) = \frak P_x$, the set of past
components of $x$:  To say that for all $y \in P$,
$y \ll \hat x(n)$ (whether for all $n$ large
enough or for some subsequence of integers) is
just to say that $P \subset I^-(x)$.  Thus, the
first clause in the definition of $\Cal L(\hat x)$
says $P \subset I^-(x)$, and the second says that
for any IP $Q \supset P$, if $Q \subset I^-(x)$
also, then $Q = P$; and that is precisely the
definition of a past component of $x$.  

What we're aiming for is an analogue of
Proposition 2.1, to show that points are closed. 
But we need the generalization of
past-distinguishing from \cite{H} for this: 
Call a chronological set $X$ {\it generalized
past-distinguishing\/} if for all $x$ and $y$ in
$X$, if $x$ and $y$ share a past
component---i.e., if $\frak P_x \cap \frak P_y
\neq \emptyset$---then $x = y$.

\proclaim{Proposition 4.2} For any 
generalized past-distinguishing chronological
set $X$, for any $x \in X$, $L(\hat x) =
\{x\}$; thus, all points are closed in the \h
topology.
\endproclaim

\demo{Proof} We have $y \in L(\hat x)$ if and only
if there is a past component of $y$ in $\Cal
L(\hat x)$, i.e., $\frak P_y \cap \frak P_x \neq
\emptyset$. \qed \enddemo

There are other notions that need to be
generalized for non-regular chronological sets. 
For instance, in the example above, we have $B_0$
is the \h limit of the future chain
$c = \{(1/n,-2/n)\}$, but it's not the future
limit of $c$, since $I^-(B_0)$ includes both
$\bar P^+_0$ and $\bar P^-_0$, while $I^-[c]$ is
just the one past component $\bar P^+_0$.  Thus,
we are led to this definition (also from
\cite{H}):  For $c$ a future chain in a
chronological set $X$, a point
$x$ is a {\it generalized future limit\/} of $x$ if
$I^-[c]$ is a past component of $x$.

Then we have an analogue of Proposition 2.2:

\proclaim{Proposition 4.3} Let $c = \{x_n\}$ be a
future chain in a chronological set $X$.  Then a
point
$x$ is a generalized future limit for $c$ if and
only if it is a \h limit for $c$; furthermore,
if $X$ is generalized past-distinguishing, then 
$L^\Omega[c] = L[c]$. 
\endproclaim

\demo{Proof} Suppose $x$ is a generalized future
limit of $c$, so $P = I^-[c] \in \frak P_x$.  Then
for all $y \in P$, eventually $y \ll x_n$. Also,
for any IP $Q \supset P$, if for all $y \in Q$,
$y \ll x_{n_k}$, then we also have for all $y
\in Q$, $y \in I^-[c] = P$, so $Q = P$. 
Therefore, $P \in \Cal L(c)$.  For any subchain
$c' \subset c$, we also have $P \in \Cal L(c')$. 
Therefore, $x \in L(c)$.

Suppose $x \in L(c)$.  For any subchain $c'
\subset c$, there is a subsubchain $c'' \subset
c'$ and a past component $P$ of $x$ with $P \in
\Cal L(c'')$.  But for a future chain $c$, the
elements of $\Cal L(c)$ are the same as those of
$\Cal L(c')$ for any subchain $c'$; so the
previous statement amounts just to saying that
there is a past component $P$ of $x$ with $P \in
\Cal L(c)$.  And that says that for all $y \in P$,
eventually $y \ll x_n$, and for any IP $Q \supset
P$, if for all $y \in Q$, $y \ll x_{n_k}$, then $Q
= P$.  But that just says that $P \subset I^-[c]$
and that $P$ is a maximal IP for that property. 
Since $I^-[c]$ is an IP, that means $P = I^-[c]$,
so $x$ is a generalized future limit of $c$.

To show $L^\Omega[c] = L[c]$, we need only show
that $L^2[c] = L[c]$ (which is $c \cup L(c)$). 
Consider any sequence of points $\sigma$
in $L(c)$.  From what we've just seen, each
$\sigma(n)$ is a generalized future limit of $c$,
i.e., has a past component which is $I^-[c]$. 
But by generalized past-distinguishment, there
can be only a single generalized future limit of
$c$, so this is a constant sequence $\hat x$, and
$L(\hat x) = \{x\}$.
\qed \enddemo

We can have an analogue of Theorem 2.4, showing
how any sort of future boundary on a chronological
set does, indeed, form a boundary in a reasonable
topological sense.  In Theorem 2.4, it was the
insistence on (past) regularity that justified the
word ``future" as applied to the boundary; here,
we will be content merely to have pasts of points
non-empty:

\proclaim{Theorem 4.4}Let $\bar X$ be a 
chronological set with every point having
a non-empty past, and having a subset $X$  
satisfying the following:
\roster
\item The restriction of $\ll$ to $X$ yields 
another chronological set with a non-empty past
for each point; and
\item for any $p \ll q$ in $\bar X$, there is some $x
\in X$ so that $p \ll x \ll q$.
\endroster
Then the \h topology on $X$ (as a
chronological  set in its own right) is the
same as the subspace topology it inherits from the
\h topology on $\bar X$, and $X$ is dense in $\bar
X$. \endproclaim 

\demo{Proof}  Let $\bar I^-$, $\bar L$, $\bar\Cal
L$, and $\bar\frak P$ denote the past operator,
limit-operators, and past components in $\bar X$,
the unbarred versions denoting the same in $X$. 
The Lemma of Theorem 2.4 is still valid, as it
made no use of regularity: For $P \in \I (X)$,
$\bar P \in \I (\bar X)$, and for $Q \in \I (\bar
X)$, $Q_0 \in \I (X)$.  In fact, for any $x \in X$,
these maps yield a bijection between $\frak P_x$
and $\bar\frak P_x$:  For $P$ a past component of
$x$, $\bar P$ is clearly a subset of $\bar
I^-(x)$, and its maximality as such is provided by
the preservation of inclusion by these maps; the
other direction is similar.

We need to show that for any sequence
$\sigma$ in $X$ and $x \in X$, $x \in L(\sigma)$
if and only if $x \in \bar L(\sigma)$.

Suppose $x \in L(\sigma)$.  We must show that
for every $\tau \subset \sigma$, there is a $\rho
\subset \tau$ and a $Q \in \bar\frak P_x$ with $Q
\in \bar\Cal L(\rho)$.  We have $P \in \frak P_x$
with $P \in \Cal L(\rho)$.  Then $Q = \bar P$
works: We already know $\bar P \in \bar\frak P_x$. 
For any $p \in \bar P = \bar I^-[P]$, we can find
$x \in P$ with $p \ll x$; then eventually $x \ll
\rho(n)$, so $p \ll \rho(n)$.  For any $Q' \in
\I (\bar X)$ with $Q' \supset \bar P$, if for all
$p \in Q'$, $p \ll \rho(n_k)$, then the same is
true for $Q'_0$, and $Q'_0 \supset P$.  Hence,
$Q'_0 = P$, whence $Q' = \bar P$.  Therefore,
$\bar P \in \bar\Cal L(\rho)$.

A formally identical proof, with barred and
unbarred reversed, shows that if $x \in \bar
L(\sigma)$, then $x \in L(\sigma)$ (note that
for $Q \in \I (\bar X)$, $Q_0 = I^-[Q]$). 

For density of $X$ in $\bar X$, consider any $p
\in \bar X$:  It has a non-empty past, so it has a
past component $Q$ generated by a future chain
$c$ (Proposition 4.1).  There is an interweaving
chain $c_0$ which lies in $X$, generating
$Q_0$.  Then $p$ is a generalized future limit
of $c_0$ (in $\bar X$), so by Proposition 4.3,
$p \in \bar L(c_0)$. \qed \enddemo

We have the immediate application to future
completion, as in Corollary 2.5:

\proclaim{Corollary 4.5} Let $X$ be a
chronological set with non-empty pasts for its
points.  Then the standard future injection
$\hat\iota_X : X \to
\hat X$ is a homeomorphism onto its image, and $X$ is dense in
$\hat X$. \endproclaim 

\demo{Proof} All we need do is observe that for
all $P \in \hat\partial(X)$, the past of $P$ in
$\hat X$ is non-empty. \qed \enddemo

Applying future-completion to a non-regular
chronological set is not actually a very
satisfactory process, as the result is never
generalized past-distinguishing, even if the
original object is:  Suppose
$X$ is generalized past-distinguishing and has a
point $x$ with distinct past components $P$ and
$Q$.  Note that $P \in \hat\partial(X)$:  If $P =
I^-(y)$, then $x$ and $y$ share the past
component $P$, so $x = y$, but we're assuming that
$I^-(x) \neq P$.  Thus, $P$ is a point in its own
right in $\hat X$.  If $P$ is generated by the
future chain $c$ in $X$, then $\bar P = \hat
I^-[c]$ is the past of $P$ in $\hat X$; but $\bar
P$ is also a past component of $x$ in $\hat X$,
and $x \neq \bar P$.  Therefore, $\hat X$ is not
generalized past-distinguishing.  In short: Forming
$\hat X$ creates non-Hausdorff clusters of points
associated with all the past components of any
non-regular point in $X$.

A way around this can be found by instituting a
new functor, generalized future completion:  For
any chronological set $X$, let $\hat\partial^g(X)
= \{P \in \I(X) \;|\; P$ is not a past
component of any point in $X\}$, the {\it
Generalized Future Chronological Boundary\/} of
$X$; and let $\hat  X^g = X \cup
\hat\partial^g(X)$, with chronology relation
defined as for $\hat X$, the {\it generalized
future completion\/} of $X$; among generalized
past-distinguishing chronological sets, this
amounts to a functor (into a category of generalized
future-complete objects, a concept explained below),
with $f: X \to Y$ giving rise to $\hat f^g : \hat
X^g \to \hat Y^g$.  In essence, $\hat X^g$
dispenses with the extraneous points that appear in
$\hat X$, associated with each non-regular point in
$X$, and interrelationships between the two functors
$\;\widehat{}\;$ and $\;\widehat{}\,\,^\bold{g}$ can
be explored (for example, there are maps $\pi_X :
\hat X \to \hat X^g$---sending the past components
of a non-regular point to that point---forming a
natural transformation $\boldsymbol \pi:
\;\widehat{}\;\; \dot\to
\;\;\widehat{}\,\,^\bold{g}$).  But there seems to
be little point in pursuing a full categorical
treatment of generalized future completion, as
non-regular chronological sets are seldom the
object upon which one wishes to construct a
completing boundary; rather, the more usual course
is to form a non-regular object as the intended
completion of a starting object which is regular
(e.g., a spacetime).

Probably the most common usage of non-regular
chronological sets is the formation of boundaries
for spacetimes like that of the example $M$ above,
where the Future Chronological Boundary yields an
unsatisfying result: the non-Hausdorff pair of
$P^+_0$ and $P^-_0$; in this case, the full GKP
Causal Boundary results in the somewhat more
natural boundary (with $B_0$)---at any rate, it's a
Hausdorff boundary.  It is not apparent what
purpose is served by looking at the categorical
operations applied to such constructs; for
instance, if $\bar M = M \cup \bar\partial(M)$ is
formed from $\hat M$ by replacing  $P^+_0$ and
$P^-_0$ by $B_0$, then $\widehat{\bar
M}$ gives us all three boundary points: $\bar B_0$
as part of $\bar M$, then $\bar P^+_0$ and $\bar
P^-_0$ as IPs in $\bar M$, and it is unclear what
use this object has.

None the less, it may be of use to pursue the
categorical approach for non-regular chronological
sets, if for no other purpose than to convince
ourselves that the \h topology in the non-regular
case really does have appropriate properties.  

Proceeding in this spirit, we have to generalize
our sense of appropriate morphism in {\bf Chron}: 
If we look at the projection map $\pi: \hat M \to
\bar M$ ($\pi$ is the identity on $\bar M$, save
that $\pi(P^+_0) = \pi(P^-_0) = B_0$), we see it is
not future-continuous: 
$P^+_0$ is the future limit of the future chain
$c = \{(1/n, -2/n)\}$, but $B_0$ is not the
future limit of $c$, as the past of $B_0$ contains
both past components, not just the one that $c$
is in.  But $B_0$ {\it is\/} a generalized future
limit of $c$.  That is the key to the morphisms we
need to be considering between non-regular
chronological sets:  As in
\cite{H}, a chronological map $f: X \to Y$ between
chronological sets will be called {\it generalized
future-continuous\/} if for any future chain $c$
in $X$ with generalized future limit $x$, $f(x)$
is a generalized future limit of $f[c]$.  Let
{\bf GChron} be the category whose objects are
chronological sets with all points having a
non-empty past, and with morphisms being
generalized future-continuous maps.  As in
\cite{H}, we will call a chronological set {\it
generalized future-complete\/} if every future
chain has a generalized future limit.  Finally,
$X$ will be called {\it generalized
past-determined\/} if whenever a past component of
$x$ is contained in
$I^-(w)$ for some $w \ll y$, we have $x \ll y$.

It is easy to see that the notions of generalized
past-distinguishing and generalized
past-determined are stronger notions than,
respectively, past-distinguishing and
past-determined, while generalized future-complete
is a weaker notion than future-complete.  (Note
that a spacetime with only spacelike
boundaries---or its future completion---is
generalized past-determined:  It is
past-determined and also regular.) But generalized
future-continuous is neither weaker nor stronger
than future-continuous.

We also must generalize our notion of
inobservability; in fact, we will redefine it, in
a manner which includes the previous definition
when applied to regular points:  A point $x$ in a
chronological set is {\it inobservable\/} if for
every past component $P$ of $x$, the only IP which
contains $P$ is $P$ itself.  (Notions of spacelike
boundaries and functions preserving spacelike
boundaries are unchanged.)

Here is a collection of categorical and
quasi-categorical results in the generalized
categories.  Generalized past-determination is
utilized only in the purely chronological (i.e.,
not topological) category.  For the topological
category, the assumption of spacelike boundaries
(necessary for the application of the techniques
from Proposition 2.7) suffices in place of
generalized past-determination.  The second
statement here leads to some further general
categorical results. The third statement  is notable
for leading to the quasi-rigidity results in the
immediately succeeding theorems, and its
quasi-categorical nature will prove important for
proving Theorem 5.3. 

\proclaim{Proposition 4.6} Let $f : X \to Y$ be a
morphism in {\bf GChron}.  
\roster
\item Suppose $Y$ is generalized past-determined
and generalized past-distingui\-shing; then $\hat
f :
\hat X \to \hat Y$ is also in {\bf GChron} and is
the unique generalized future-continuous map such
that $\hat f \circ \hat\iota_X = \hat\iota_Y \circ
f$.
\item Suppose that $f$, $X$, and $Y$ are in
{\bf Spbd\-Ftop\-Gchron} with $Y$ generalized
past-distinguishing; then $\hat f$ is also in {\bf
Spbd\-Ftop\-Gchron}. 
\item Suppose, in addition to the above, that $Y$
is generalized future-complete; then there is a
unique generalized future-continuous and \h
continuous map $\tilde f^g: \hat X \to Y$ such that
$\tilde f^g \circ \hat\iota_X = f$. 
\endroster \endproclaim

\demo{Proof} (1) We only need $Y$ to be
past-determined and past-distinguishing to obtain
the chronological map $\hat f$ (Proposition 6 in
\cite{H}). We need to show here that
$\hat f$ is generalized future-continuous.  But
this is easy:  If $x \in X$ is a generalized
future limit of a future chain $c$ in $\hat X$,
then we can interpolate elements of $X$ to obtain
a chain $c_0$ in $X$, for which $x$ is a
generalized future limit; then $f(x)$ is a
generalized future limit of $f[c_0]$ in $Y$, so it
also is in $\hat Y$.  If $P \in \hat\partial(X)$ is
a generalized future limit of $c$, then $P$ is
generated by the chain $c_0$ (and, actually, $P$
is an ordinary future limit of $c$), so $\hat f(P)$
is generated by $f[c_0]$ (and, actually, $\hat
f(P)$ is an ordinary future limit of $f[c]$). And
$\hat f$ is unique for a generalized
future-continuous function obeying $\hat f \circ 
\hat\iota_X = \hat\iota_Y \circ f$, since $\hat
f(P)$ is then forced to be a generalized future
limit of $f[c_0]$, and there is only one such,
since $Y$ is generalized past-distinguishing
(which implies the same for $\hat Y$).

(2) Now assume we're in the spacelike boundaries
category:  As in the proof of Proposition 2.7, to
have $\hat f$ future-continuous, we don't need $Y$
to be past-determined, since $X$ has only
spacelike boundaries; but we need to show
$\hat f$ is \h continuous.  The proof of Theorem
4.4 demonstrates that for any
$Z$ in {\bf GChron}, for any sequence $\sigma$ in
$X$, $L_Z(\sigma) = L_{\hat Z}(\sigma) \cap Z$,
just as was needed in the proof of Proposition
2.7.  Then the formal properties of the
limit-operator complete the portion of the proof
showing that if $x \in X$ is in $L_{\hat
X}(\sigma)$ for a sequence $\sigma$ lying in $X$,
then $f(x) \in L^\Omega_{\hat Y}[\,f[\sigma]\,]$.

Now consider the same sequence $\sigma$ with $P
\in \hat\partial(X)$ in $L_{\hat X}(\sigma)$. 
Then $P$ is regular, as is $\hat f(P)$ (reason: 
Let $c$ be a chain in $X$ generating $P$, and let
$Q = I^-[\,f[c]\,]$.  If $Q \in \hat\partial(Y)$,
then $\hat f(P) = Q$ and $Q$ is regular (because
it's in $\hat\partial(Y)$); while if $Q = I^-(y)$
for some $y \in Y$, then $\hat f(P) = y$ and $y$
is regular (because its past is the IP $Q$).).
Therefore, the same proof as in Proposition 2.7
applies, showing $\hat f(P)
\in L_{\hat Y}(f[\sigma])$.

Now consider the case of a sequence $\sigma =
\{P_n\}$ lying in $\hat\partial(X)$, with $x \in
X$ in $L_{\hat X}(\sigma)$.  Let $\sigma' =
\hat f[\sigma]$; we need to show that $f(x) \in
L_{\hat Y}(\sigma')$, i.e., that for any $\tau'
\subset \sigma'$, there is a $\rho' \subset
\tau'$ and a past component $Q$ of $f(x)$ with
$Q \in \Cal L_{\hat Y}(\rho')$.  For any such
subsequence $\tau'$ of $\sigma'$, let $\tau$ be
the corresponding subsequence of $\sigma$, i.e.,
so that $\hat f[\tau] = \tau'$; then there is a
subsubsequence $\rho \subset \tau$ and a past
component $P$ of $x$ such that $P \in \Cal L_{\hat
X}(\rho)$.  Let $c$ be a chain in $X$ generating
$P$, and let $Q = I^-[\,f[c]\,]$.  We have $x$ a
generalized future limit of $c$, so $f(x)$ must be
a generalized future limit of $f[c]$ ($f$ being
generalized future continuous); that says
precisely that $Q$ is a past component of $f(x)$. 
We know that for all $m$, eventually $c(m) \ll
\rho(n)$; therefore, for all $m$, eventually
$f(c(m)) \ll \hat f(\rho(n))$.  Since $x$ is a
limit of boundary elements, which must be
inobservable, $x$ is also inobservable
(inobservables in $\hat X$ being closed);
therefore, $f(x)$ is also inobservable.  Hence, we
need not worry about any IP $Q' \supset Q$, as
necessarily then $Q' = Q$.  Therefore, $Q \in \Cal
L_{\hat Y}(\rho')$ for $\rho' = \hat f[\rho]$:  We
have $f(x) \in L_{\hat Y}(\sigma)$.

Next consider $P \in \hat\partial(X)$ in $L_{\hat
X}(\sigma)$ for the same $\sigma = \{P_n\}$. 
Then $P$ and $\hat f(P)$ are regular, and the
proof of Proposition 2.7 applies.

For a general sequence $\sigma$, the same
stratagem as used in Proposition 2.7 works here:  If
there is some subsequence $\sigma' \subset
\sigma$ lying in $\hat\partial(X)$, then any $p
\in L_{\hat X}(\sigma)$ is inobservable, and the
proof above applies.  And if
there is no such subsequence, then we can ignore
the finite number of elements of $\sigma$ lying in
$\hat\partial(X)$ and apply the other parts of the
proof.

That $\hat f$ preserves spacelike boundaries
follows precisely as in Corollary 2.8, so $\hat f$
is a morphism of {\bf SpbdFtopGChron}.

(3) Finally, consider $Y$ to be generalized
future-complete.  Theorem 15 of \cite{H} yields a
unique generalized future-continuous $\tilde f^g :
\hat X \to Y$ with $\tilde f^g
\circ \hat\iota_X = f$ ($\tilde f^g$ is defined
thus: for  $P \in \hat\partial(X)$ generated by a
future chain $c$, $\tilde f^g(P)$ is the unique
generalized future limit of $f[c]$); we just need
to show it \h continuous.  (In \cite{H}, this map
was called $\hat f^g$, disrespecting the potential
categorical usage of that symbolism.  Using that
symbolism properly, we can identify
$\tilde f^g$ in terms of the functor and natural
transformation alluded to above:
$\tilde f^g = \hat f ^g \circ \pi_X$.)

For a sequence $\sigma$ lying in
$X$, consider a point
$x \in X$ within $L_{\hat X}(\sigma)$:  As before,
this means $x \in
L_X(\sigma)$, whence, by
\h continuity of $f$, $f(x) \in
L^\Omega_Y(f[\sigma])$, as needed.  Now consider
$P \in \hat\partial(X)$ in $L_{\hat X}(\sigma)$: 
Then, $P$ being regular, the proof in Proposition
2.7 applies with minor modification (for $P$
generated by a chain $c$, $\tilde f^g(P)$ is in $Y$,
even if $I^-[\,f[c]\,]$ is not $I^-(y)$ for any $y
\in Y$; $I^-[\,f[c]\,]$ is a past component of a
unique $y$, however, and that is $\tilde f^g(P)$). 
For a sequence $\sigma$ lying in $\hat\partial(X)$
or, more generally, in $\hat X$, the same proof
applies as in (2) above.
\qed \enddemo

It may be instructive to examine a couple of
examples, illustrating how the spacelike nature of
the boundary is important for continuity of the
induced map in statement (3), even for cases of
simple inclusion, with an isomorphism onto the
image:

First consider a case with spacelike boundary (see
Figure 9):  Let $X$ be lower Minkowski half-space,
$\Bbb L^3_- = \{(x,y,t) \;|\; t < 0\}$; then $\hat
X$ is clearly $\{(x,y,t) \;|\; t \le 0\}$ (actually,
$(x,y,0)$ refers to the obvious IP in $X$), and
$X$ manifestly has only spacelike boundaries and is
past-distinguishing.  We will let $Y$ be derived
from $\hat X$ by identification of some points:  Fix
attention on two curves in $\hat X$, $L^- =
\{(x,-1,0) \;|\; x > 0\}$ and $L^+ = \{(x,1,0)
\;|\; x > 0\}$. For each $x \ge 0$, let $p^-_x =
(x,-1,0)$ and $p^+_x = (x,1,0)$.  Define the set
$Y$ as $\hat X/\!\!\sim\,$, where $\sim$ is the
equivalence relation given by $p^-_x \sim p^+_x$
for each $x >0$, and no other points identified
together; thus, $\sim$ identifies the two
open half-lines $L^-$ and $L^+$ into a single
line $L$.  For each $x > 0$, let $p_x$ denote the
equivalence class of $p^+_x$ (or of $p^-_x$).  Make
$Y$ into a chronological set by giving it the same
chronology relation as on $\hat X$, with the
proviso that $I_Y^-(p_x) = I_{\hat X}^-(p^-_x) \cup
I_{\hat X}^-(p^+_x)$.  Then $Y$ also has only
spacelike boundaries ($\hat Y$ is just $Y$ with
the two lines $L^-$ and $L^+$ added back in) and is
past-distinguishing; it is also generalized
future-complete.  Let $i: X \to Y$ just be
inclusion; this clearly is continuous and
generalized future-continuous and preserves
spacelike boundaries (and it is a chronological
isomorphism onto its image).  Note that $\tilde i^g:
\hat X \to Y$ takes each of the lines $L^-$ and
$L^+$ onto $L$.  Since in $\hat X$,  $p_0^-$ is the
limit of the curve $L^-$, as is $p_0^+$ for $L^+$,
we'd better have both $p_0^-$ and $p_0^+$ limits
of the curve $L$ in $Y$.  And this is, indeed, the
case: \{$p_0^-, p_0^+\}$ is a non-Hausdorff pair in
$Y$, with any neighborhood of either point
necessarily containing an end of the curve $L$,
i.e., all $p_x$ with $x < x_0$ for some $x_0 > 0$
(this can be seen by noting, for instance, that
for any sequence $\sigma = \{p_{x_n}\}$ with $x_n$
going to 0, $p_0^- \in L_Y(\sigma)$---since
$p_0^-$ is regular, just use the original
definition of \h limit---so any closed set
containing such a $\sigma$ contains
$p_0^-$, so any closed set excluding $p_0^-$
excludes any such sequence, so any neighborhood of
$p_0^-$ contains an end of $L$).  The \h topology
on $Y$ is precisely the same as the quotient
topology on $\hat X/\!\!\sim\,$.

Now consider a case with the
boundary being null:  Let $\Pi$ be the null plane
$\{z = y\}$ in $\Bbb L^3$, and let $X = I^-(\Pi)$.
Clearly $\hat\partial(X)$ can be identified with
$\Pi$ and $\hat X$ with the closure of $X$ in $\Bbb
L^3$.  Do the same thing as above:   For $x
\ge 0$, let $p_x^- = (x,-1,-1)$ and $p_x^+ =
(x,0,0)$; let $Y = \hat X/\!\!\sim\,$, where $\sim$ 
is the equivalence relation given by $p^-_x \sim
p^+_x$ for each $x >0$, and no other points
identified together (let $p_x$ name the
corresponding equivalence class).  With the half
lines $L^-$ and $L^+$ in $\hat X$ defined
analogously as above, this equivalence relation
identifies $L^-$ and $L^+$ into a single half-line
$L$.  As before, define  $I_Y^-(p_x) = I_{\hat
X}^-(p_x^-) \cup I_{\hat X}^-(p_x^+)$.  We have $Y$
past-distinguishing and generalized future-complete.
Let $i : X \to Y$ be inclusion: continuous and
generalized future-continuous, and also a
chronological isomorphism onto its image.

The interesting fact is that $\tilde i^g: \hat X \to
Y$ is not continuous:  $p_0^-$ is not a limit of
$L$ (though $p_0^+$ is); more specifically, for
any sequence $\sigma$ lying in $L$ with
$x$-coordinate approaching 0, $p_0^- \notin
L_Y(\sigma)$.  The reason is that although every
point in $I^-(p_0^-)$ is eventually in the past of
$\sigma$, the same is also true for an IP which
properly contains $I^-(p_0^-)$, namely,
$I^-(p_0^+)$ (see Figure 10).  The neighborhoods of
$p_0^+$ are analogous to those in the previous
example, but the neighborhoods of $p_0^-$ are quite
different:  They include sets formed by starting
with an open set in $\hat X$, and then deleting all
points of
$L^-$ (or $L^+$).  This gives $Y$ some
decidedly odd-looking open sets (though $Y$ is still
non-Hausdorff, as neighborhoods of $p_0^-$ must
contain points in a deleted neighborhood of an end
of $L^-$, as must neighborhoods of $p_0^+$).  The
\h topology here is not at all the quotient topology
on $\hat X$, but takes special cognizance of the
null nature of the boundary and its relation to the
identified points.  (Note that
$\hat i : \hat X \to \hat Y$ {\it is\/}
continuous, even though $X$ and $Y$ are not in {\bf
Spbd}: $\hat Y$ consists of $Y$ with the curves
$L^-$ and $L^+$ added back in, not
Hausdorff-separated from $L$; $\hat i$ takes $L^-$
to $L^-$---not to $L$---and $p_0^-$ is the limit
of $L^-$ in $\hat Y$.  Theorem 3.6 is not
applicable because $Y$ has points---those in
$L$---which are not future limits of chains in
$i[X]$.)

The sense in which the boundary in this example is
null is evident from the embedding into Minkowski
space; but it can also be formally addressed in the
manner of \cite{GKP}:  Starting with a set $X$ with
both a chronology relation $\ll$ and a causality
relation $\prec$, obeying appropriate properties, one
can extend $\prec$ (as well as $\ll$) to $\hat X$ by
defining, for $P$ and
$Q$ in $\hat\partial(X)$ and $x$ in $X$, $x \prec
P$ for $I^-(x) \subset P$, $P \prec x$ for $P
\subset I^-(x)$, and $P \prec Q$ for $P \subsetneqq
Q$.  Then the boundary above possesses
points $P \prec Q$, but with $P \ll Q$ failing.

\medpagebreak

Statement (3) of Proposition 4.6 looks rather like
a universality principle, but it isn't quite, since
we're not making use of the appropriate functor
(generalized future completion).  But it does mean
that we have a strong characterization of any
generalized future-completing, generalized
past-distinguishing boundary on a chronological
set with spacelike boundaries:

\proclaim{Proposition 4.7} Let $Y$ be a
chronological set with only spacelike boundaries,
pasts of points non-empty, generalized
past-distinguishing, and generalized
future-complete \rom(i.e., in {\bf
GFcpl\-GPdis\-Spbd\-Ftop\-GChron}\rom).  Then, in
the \h topology, $Y$ is a topological quotient of
its future completion, $\hat Y$.
\endproclaim

\demo{Proof} Apply Proposition 4.6(3) to $1_Y: Y
\to Y$, the identity map; this yields 
$\widetilde{1_Y}^g : \hat Y \to Y$, with
$\widetilde{1_Y}^g \circ \hat
\iota_Y = 1_Y$.  Now make use of the categorical
fact that if we have $\alpha : A \to B$ and $\beta
: B \to A$ with $\beta \circ \alpha = 1_A$, then
$A \cong B/\!\!\sim\,$, where $\sim$ is the
equivalence relation defined by $b_1 \sim b_2$ if
and only if $\beta(b_1) = \beta(b_2)$.  (This has
an easy categorical proof:  For any $\phi: B \to
C$ which respects the $\sim$ relation, there is a
unique $\tilde \phi : B/\!\!\sim\; \to C$ with
$\tilde\phi \circ \pi = \phi$, where $\pi : B \to
B/\!\!\sim$ is the projection onto the quotient. 
Then apply this to $\beta$, obtaining $\tilde\beta
: B/\!\!\sim\; \to A$ with $\tilde\beta \circ \pi =
\beta$.  We also have $\pi \circ \alpha : A \to
B/\!\!\sim\,$.  Then $\tilde\beta \circ \pi \circ
\alpha = \beta \circ \alpha = 1_A$, and $\pi
\circ \alpha \circ \tilde\beta = 1_{B/\sim}\,$, as
can be directly calculated.  Thus, $A \cong
B/\!\!\sim\,$, with $\beta$ the projection map.) 
Result: $Y \cong \hat Y/\!\!\sim$ with projection
map $\widetilde{1_Y}^g$. \qed
\enddemo

Proposition 4.7 identifies a generalized
future-complete chronological set (in the
spacelike boundaries category) as necessarily
being derived, as a quotient, from a standard
future completion; but it doesn't help us
determine {\it which\/} future completion if we
start from an incomplete object.  What would be
better would be an analogue of Theorem 3.6, giving a
rigidity result for any general future-completing
boundary, starting with a given chronological set. 
This, too, can be done, though spacelike boundaries
must be assumed (the example above with a null
boundary illustrates how things can
go wrong for a generalized future completion,
absent that assumption):

We'll model this the same way as in the proof of
Theorem 3.6:  Start with $X$, a generalized
past-distinguishing chronological set with only
spacelike boundaries and no points with empty
pasts.  To effect a generalized future completion
of $X$, we need to embed $X$ in a generalized
future-complete chronological set, $i : X \to Y$
in {\bf GPdis\-Spbd\-Ftop\-GChron}; we require that
$Y$ be generalized future-complete, and that with
$Y_0 = i[X]$ and $i_0: X \to Y_0$ the
restriction of $i$, that $i_0$ be a chronological
isomorphism.  We want $Y$ to consist only of the
image of $X$ and of necessary boundary points: 
So we also require that every point in
$\partial(Y) = Y - Y_0$ be the generalized future
limit of some future chain in $Y_0$.  What we want
is to show that $Y$ is a topological
quotient of $\hat X$ (not just of
$\hat Y$) and also to identify $\partial(Y)$.

The proofs involved are perhaps the most intricate in
this paper, but the results are among the most
interesting; important applications arise in Theorem
5.3.

\proclaim{Theorem 4.8} Let $X$ be a generalized
past-distinguishing chronological set with spacelike
boundaries and no points with empty pasts. Let
$Y$ be any generalized past-distinguishing,
generalized future completion of $X$ (also with
spacelike boundaries and no points with empty
pasts); then, in the \h topology, $Y$ is a
topological quotient of $\hat X$. 
\endproclaim 

\demo{Proof} We have $i : X \to Y$ in {\bf
GPdis\-Spbd\-Ftop\-GChron}, with $Y$ generalized
future-complete.  Consequently, by Proposition
4.6(3), we have $\tilde i^g: \hat X \to Y$ with
$\tilde i^g \circ \hat\iota_X = i$.  (The continuity of
this map is the only place where the spacelike
boundaries play an essential role; but the example
above, following Proposition 4.6, demonstrates the
crucial aspect of that assumption.)  We will  establish
a homeomorphism between
$Y$ and the quotient of $\hat X$ by the equivalence
relationship $\sim$ defined by $p \sim q$ if and
only if $p$ and $q$ have the same image under
$\tilde i^g$.  Let $\pi : \hat X \to \hat
X/\!\!\sim$ denote the projection map.  We
automatically have a (unique) continuous map $j:
\hat X/\!\!\sim\; \to Y$ such that $j \circ \pi =
\tilde i^g$, and $j$ is clearly injective.  We need
to see that $j$ is onto and bicontinuous.

It will help to be clear about the action of
the map $\tilde i^g$:  For elements of $X$, $\tilde
 i^g$ is just the same as $i$.  For $P \in
\hat\partial(X)$, we look at any future chain $c$
in $X$ generating $P$ and consider $i[c]$:  As $Y$
is generalized future-complete, this future chain
must have a generalized future limit $y$; and as
$Y$ is generalized past-distinguishing, it has
only one such. In other words, for some unique $y$,
$P' = I^-[\,i[P]\,] \in \frak P_y$; and  $\tilde 
i^g(P) = y$.  (Note that an image of an element of
$\hat\partial(X)$ need not be in $\partial(Y)$: 
Suppose  $X$ has a non-regular point $x$; let $P$
be a past component.  Then $P$ is in
$\hat\partial(X)$---$P$ cannot be $I^-(x')$, for
then $x'$ and $x$ would share a past component,
but $X$ is assumed to be generalized
past-distinguishing, and $x' \neq x$ since $x$ is
non-regular.  Then $\tilde i^g(P) = i(x)$.) 

Clearly, $j$ is onto $Y_0$, since $j(\pi(x))
= i(x)$.  To show $j$ is also onto
$\partial(Y)$, we need the fact that for any $z \in \partial(Y)$, there
is a future chain $c$ in $X$ with $z$ the
generalized future limit of $i[c]$.  Let $P =
I^-[c]$.  $P$ must be in
$\hat\partial(X)$:  If $P = I^-(x)$,
then with $P' =I^-[\,i[c]\,]$, we have $P' =
I^-(i(x))$ (since $i_0 : X
\to Y_0$ is a chronological isomorphism); but then
$P'$ is a common past component for $i(x)$ and
$z$, so $i(x) = z$, so $z \in Y_0$.  Knowing that
$P \in \hat\partial(X)$, we have $\tilde i^g(P) =
z$.  Thus $\tilde i^g$ is onto $Y$, and $j$ must be
as well.

To show $j : \hat X/\!\!\sim\; \to Y$ is
bicontinuous, we have to show that if a sequence
in $Y$ has a limit, then the same holds for the
images under $j^{-1}$.   By the same process used
in Section 2 (before Proposition 2.1) to
characterize the continuity of a map between
topological spaces defined by limit-operators, we
can characterize the current need this way:  For any
sequence $\sigma$ and point $y$ in $Y$, if $y \in
L_Y(\sigma)$, then $j^{-1}(y)$ is in the closure
of $j^{-1}[\sigma]$ (this uses the fact that a
point is a limit of a sequence if and only if it
is in the closure of all subsequences).

The quotient topology has, as open sets in $\hat
X/\!\!\sim\,$, sets of the form $\pi[U]$ for $U
\subset X$ open and full (i.e., if $p \in U$ and
$q \sim p$, then $q \in U$ also); similarly, a
closed set in $\hat X/\!\!\sim$ is anything of the
form $\pi[A]$, where $A$ is a closed and full
subset of $\hat X$.  Thus, to show $\bar p \in
\hat X/\!\!\sim$ in the closure of a sequence
$\bar\sigma$, it suffices to show that for any
subsequence $\bar\tau \subset \bar\sigma$, there
is a subsubsequence $\bar\rho \subset \bar\tau$, a
sequence $\rho'$ in $\hat X$ with $\pi[\rho'] =
\bar\rho$, and some point $p \in L_{\hat
X}(\rho')$ with $\pi(p) = \bar p$---for then any
closed and full set in $\hat X$ containing
$\pi^{-1}[\bar\sigma]$ must contain
$\pi^{-1}(\bar p)$. 

Thus, our objective comes down to the following
(eschewing $\hat X/\!\!\sim$ altogether): to show
that for any sequence $\sigma$ in $Y$ and
$y \in L_Y(\sigma)$, for any subsequence $\tau
\subset \sigma$, there is a sequence $\rho'$ in
$\hat X$ and  $p \in L_{\hat X}(\rho')$ with $\tilde 
i^g[\rho'] \subset \tau$ and  $\tilde i^g(p) = y$.

We will need a Lemma much like that in Proposition
2.4:  For any IP $P$ in $Y_0$, let $\bar P =
I^-[P]$, where  $I^-$ denotes the past in $Y$
(we'll use $I_0^-$ for the past in
$Y_0$); for any IP $Q$ in $Y$, let
$Q_0 = Q \cap Y_0$.

\proclaim{Lemma} The maps $P \mapsto \bar P$ and
$Q \mapsto Q_0$ establish an isomorphism of
partially ordered sets (under inclusion) between
$\I(Y_0)$ and $\I(Y)$. \endproclaim

The proof of this Lemma is surprisingly
complicated, though essentially technical; it
relies on the assumption of spacelike boundaries,
absent a separate hypothesis that $Y_0$ be
chronologically dense in $Y$.  The Lemma's proof
will be delayed till after the other elements of
the proof of this theorem.

\medpagebreak

Given a sequence $\sigma$ in $Y$ with $y \in
L_Y(\sigma)$, consider any subsequence $\tau
\subset \sigma$:  We know there is a
subsubsequence $\rho \subset \tau$ and a past
component $Q$ of $y$ (in $Y$) with $Q \in \Cal
L_Y(\rho)$.  Then $Q_0$ is generated by a
future chain $c_0$ in $Y_0$, i.e., there is a
chain $c$ in $X$ with $c_0 = i[c]$, so that $Q_0 =
I_0^-[\,i[c]\,]$.  Let $P = I^-[c]$, an IP in $X$. 
If for some $x \in X$, $P \in \frak P_x$, then let
$p = x$; otherwise, let $p = P$.  In either case,
we have $P$ is a past component of $p$ in $\hat
X$ and $\tilde i^g(p) = y$:  In the first case, $x$
is a generalized future limit of $c$, so $i(x)$
must be a generalized future limit of $i[c] = c_0$
in $Y_0$; then $i(x)$ is also a generalized future
limit of $c_0$ in $Y$ (as is seen by applying the
Lemma: $Q_0$ is maximal in $I_0^-(i(x))$ if and
only if $Q$ is maximal in $I^-(i(x))$), so, $y$
being the unique generalized future limit of $c_0$
in $Y$, $i(x) = y$.  In the second case, $P$ {\it
qua\/} subset of $X$ is the only past component in
$\hat X$ of $P$ {\it qua\/} element of $\hat X$
(since $X$ has spacelike boundaries, the past in
$\hat X$ of $P$ can contain no elements of
$\hat\partial(X)$, for any such would then be
properly contained by $P$; thus, $I^-_{\hat X}(P) =
P$); since $i[c] = c_0$ generates, in $Y$, an IP
which is a past component of $y$ (namely, $Q$), we
have $\tilde i^g(P) = y$. 
 
So we have $P \in \frak P_p$ where $\tilde i^g(p) =
y$.  What we need now is a sequence $\rho'$ in
$\hat X$ with $\tilde i^g[\rho'] = \rho$ and $P \in
\Cal L_{\hat X}(\rho')$ (for then $p \in L_{\hat
X}(\rho')$).  We will construct this now:

We know that $Q = I^-[\,i[c]\,] \in \Cal L_Y(\rho)$;
in other words, for each $m$, eventually $i(c(m))
\ll \rho(n)$: Specifically, for each $m$, there is
some $n_m$ such that for all $n > n_m$, $i(c(m))
\ll \rho(n)$; we may safely take the sequence of
numbers $\{n_m\}$ to be monotonic increasing. 
Then whenever $n > n_m$, there is a past component
$Q_n^m$ of $\rho(n)$ containing $i(c(m))$.  It is
these past components in $Y$ that will be
used to generate the proper sequence in $\hat X$, but
a diagonal process is needed to reduce from a
doubly-indexed family to a singly-indexed one.  

First we invert the roles of $m$ and $n$ above: For
each $n$, let $M_n = \max\{m \;|\; n_m < n\}$;
conceivably, some $M_n = \infty$ (and then all
$M_{n'} = \infty$ for $n' > n$), though we don't
expect that to be the case usually.  As $n$
increases, $n$ eventually surpasses any one $n_m$,
so if the $\{M_n\}$ are finite, they grow to
infinity; thus, in any event, we can pick a
monotonically increasing sequence $\{m_n\}$ such
that $\{m_n\}$ goes to infinity and for all $n$,
$m_n \le M_n$; then for all $n$, for all $m \le
m_n$, $i(c(m)) \in Q_n^m$. 

Now we diagonalize:  For each $n$, let $Q_n =
Q_n^{m_n}$; then for all
$n$, for all $m \le m_n$, $i(c(m)) \in Q_n \in
\frak P^Y_{\rho(n)}$.  As before, for each $n$ we
can find a future chain $c^n$ in $X$ such that
$Q_n = I^-[\,i[c^n]\,]$.  We have for all $m \le
m_n$, there is some $j$ with $i(c(m)) \ll
i(c^n(j))$; therefore, $c(m) \ll c^n(j)$. 
Thus, if we let $P_n = I^-[c^n]$, we have for
all $m \le m_n$, $c(m) \in P_n$.  These
$\{P_n\}$ will be the basis for our sequence
$\rho'$:  For each $n$, if there is some $x_n \in
X$ with $P_n$ a past component of $x_n$, then let
$\rho'(n) = x_n$; otherwise, let $\rho'(n) =
P_n$, an element of $\hat\partial(X)$.  Note that
in the first case, $x_n$ is a generalized future
limit of $c^n$, so $i(x^n)$ is the generalized
future limit in $Y_0$ of $i[c^n]$, and the same is
true in $Y$ (the Lemma guaranteeing the preservation
of relations among past components); therefore,
$\rho(n)$ also being the generalized future limit
of $i[c^n]$, we have $i(x_n) = \rho(n)$.  In the
second case, $\tilde i^g(P_n)$ is the generalized
future limit of $i[c^n]$---again, $\rho(n)$.  Thus,
in either case,  $\tilde i^g[\rho'] = \rho$.

To show $P \in \Cal L_X(\rho')$, we will need to
have, for each $m$, eventually $c(m) \in P_n$.
What we have is that for each $n$, for all $m \le
m_n$, $c(m) \in P_n$.  Concentrate attention on any
fixed $m$:  Since the sequence $\{m_n\}$ is increasing
to infinity, for all $n$ sufficiently large, $m$ will be
less than $m_n$, placing $c(m)$ in $P_n$. 

To check for maximality of $P$, suppose we have $P'
\in \I(\hat X)$ with $P' \supset P$ and for all $q \in
P'$, for some $\{n_k\}$, for all $k$, $q \ll
\rho'(n_k)$.  Actually, since $X$ has only
spacelike boundaries, for all $q \in P'$, $q$ is 
in $X$, not in $\hat\partial(X)$, so we can write
$P'$ simply as $I^-_X[c']$ for some chain $c'$ in
$X$ (though we could always use the Lemma in Theorem
2.4 to write any IP in any $\hat X$ as the $\hat
X$-past of a chain in $X$).  Let $Q' =
I^-_0[\,i[c']\,]$; from $P' \supset P$, {\it
i.e.}, $I^-[c'] \supset I^-[c]$ in $X$, we have
$I_0^-[c'] \supset I_0^-[c]$ {\it i.e.}, $Q'
\supset Q_0$, since
$Y_0$ is just a reflection of $X$.  Then we also
have $\overline{Q'} = I^-[\,i[c']\,] \supset Q$. 

We have for all $n$ and $k$, $c'(n) \ll
\rho'(n_k)$.  In case $\rho'(n_k) \in X$, this
gives us $i(c'(n)) \ll i(\rho'(n_k)) = \rho(n_k)$. 
In case $\rho'(n_k) \in \hat\partial(X)$,
$\rho'(n_k) = I^-[c^{n_k}]$, and we have for all
$j$ sufficiently high, $c'(n) \ll c^{n_k}(j)$,
whence $i(c'(n)) \ll i(c^{n_k}(j))$; thus,
$i(c'(n)) \in I^-[\,i[c^{n_k}]\,] = Q_{n_k}$, so
$i(c'(n)) \ll \rho(n_k)$.  Therefore, we have that
for all $q \in \overline{Q'}$, $q \ll \rho(n_k)$. 
It follows that $\overline{Q'} = Q$, whence $Q' =
Q_0$, whence $P' = P$.  Thus, $P \in \Cal
L(\rho')$. \qed \enddemo
 
\demo{Proof of Lemma} If $P \in \I(Y_0)$ is
generated by a chain $c$, then $\bar P = 
I^-[c]$, so
$\bar P \in \I(Y)$; thus $\overline{(\;)} :
\I(Y_0) \to
\I(Y)$.  It takes a bit of doing to go the other
way around:  We must first establish that for
any future chain $c$ in $Y$, there is an
interweaving chain $c_0$ in $Y_0$; this is where
the spacelike character of the boundaries of $Y$
comes into play.  Once we know that, then for
any $Q \in \I(Y)$, we take a chain $c$
generating $Q$ and then an interweaving chain
$c_0$ in $Y_0$; then $Q$ is also generated by
$c_0$, so $Q_0 = I^-[c_0] \cap Y_0 =
I^-[c_0] \in \I(Y)$ and $(\;)_0 : \I(Y) \to
\I(Y_0)$.  Then these maps are inverses of one
another:  For $P = I_0^-[c]$, we have
$(\bar P)_0 = (I^-[c])_0 = I^-[c] \cap Y_0 = P$
(since $c$ is already in $Y_0$); and for $Q = 
I^-[c]$, with interweaving chain $c_0$ in $Y_0$, we
have $\overline{Q_0} = I^-[c_0] = Q$.  The
preservation of inclusion by each map is clear.

So we need to establish the existence of
interweaving chains:  Given a chain $c = \{z_n\}$
in $Y$, we want to find a chain $c_0 = \{y_n\}$ in
$Y_0$ with $z_{n-1} \ll y_n \ll z_n$.  We might as
well assume all $z_n$ are in $\partial(Y)$:  If
only a finite number aren't we can just drop them,
and if a subsequence lies in $\partial(Y)$, we can
concentrate our attention on that subsequence. 
Then each $z_n$ is the generalized future limit of
a chain $c^n = \{y^n_m\}_{m \ge 1}$ in $Y_0$.  For
each $n$, let $P_n = I^-[c^n]$, and let $Q_n$ be
the past component of $z_n$ containing $z_{n-1}$.
We will show that for each $n$, $P_n$ and $Q_n$
must be the same; then, since $Q_n$ is generated
by $c^n$, there must be some $m$ with $z_{n-1} \ll
y^n_m \ll z_n$, so letting $y_n = y^n_m$ finishes
the job.

Fix $n$; we will show $z_n$ must be regular (see
Figure 11).   Suppose $z_n$ is non-regular; then
$P_n$ must be in $\hat\partial(Y)$:  If $P_n = 
I^-(w)$ for some $w$, then, $w$ and $z_n$
share a past component (namely, $P_n$), so, by
generalized past-distinguishing, $w = z_n$; but
that can't be, because $w$ is regular and $z_n$
isn't.  We have for all $m$, $y^n_m \ll z_n \ll
z_{n+1}$; therefore, $P_n \subset Q_{n+1}$.  But
since all elements of $\hat\partial(Y)$ are
inobservable, this implies $P_n = Q_{n+1}$. 
Therefore, since $z_n \in Q_{n+1}$, for some $m$,
$z_n \ll y^n_m$; but we also know that for all
$m$, $y^n_m \ll z_n$, so we have a contradiction. 
Ergo, $z_n$ must be regular.

Once we know $z_n$ is regular, all its past
components are the same: $P_n = Q_n$, and we are
done. \qed \enddemo

As a result of Theorem 4.8, to identify any
generalized future-completing boundary for $X$, we
need only look in $\hat X$ and its quotients by
equivalence relations (actually, quotients by
continuous maps).  However, although $X$ sits
nicely in $\hat X$, once we apply the equivalence
relation to get $\hat X/\!\!\sim\,$, the image of
$X$ is no longer sharply separated from the
boundary, since non-regular points of $X$ get
identified with their past components in the
projection to $\hat X/\!\!\sim\,$.  Thus, a cleaner
picture is available in $\hat X^g$, which omits
from consideration those IPs of $X$ which are past
components, so that nothing in $\hat\partial^g(X)$
becomes identified with a point of $X$ under an
analogous equivalence relation.  This gives us a
clean statement for $\partial(Y)$:

\proclaim{Corollary 4.9} Let $X$ be a generalized
past-distinguishing chronological set with spacelike
boundaries and no points with empty pasts. Let
$Y$ be any generalized past-distinguishing,
generalized future completion of $X$ (also with
spacelike boundaries and no points with empty
pasts); then $\partial(Y)$ (the generalized
future-completing boundary) is a topological
quotient of $\hat\partial^g(X)$, the Generalized
Future Chronological Boundary of $X$. 
\endproclaim

\demo{Proof} Since, by Theorem 4.8, $Y$ is
homeomorphic to $\hat X/\!\!\sim\,$ (for $\sim$ the
equivalence relation defined by $\tilde i^g: \hat X
\to Y$), we need concern ourselves only with
identifying the boundary placed on $X$ in $\hat
X/\!\!\sim\,$.  But, as related above, we really
want, instead, to look at $\hat X^g/\!\!\sim\,$,
where the $\sim$ relation is essentially the same: 
$\hat X^g$ is a subset of $\hat X$, just omitting
those elements of $\hat\partial(X)$ which are past
components of non-regular points in $X$.  Since
$\tilde i^g$ maps the past components of a
non-regular point $x$ to $i(x)$, all the
equivalence classes in $\hat X$ have a
representative in $\hat X^g$.  

Also note that the chronological sets $\hat X^g$
and $\hat X$ bear the relation of $X$ and $\bar X$
in Theorem 4.4, since no elements of
$\hat\partial(X)$ are chronologically related;
thus, $\hat X^g$, with the induced chronology
relation, has the subspace topology from $\hat X$.

We will put this all together to show that $\hat
X^g/\!\!\sim\,$ is homeomorphic to $\hat
X/\!\!\sim\,$:

We have the situation of topological spaces $A
\subset B$ with an equivalence relation $\sim$ on
$B$, and the equivalence relation passing to the
subspace $A$.  With $\pi : B \to B/\!\!\sim\,$ 
being the projection to the quotient space and $j:
A \to B$ the inclusion map, we have $\pi \circ j :
A \to B/\!\!\sim\,$ is continuous.  Since this map
respects the induced equivalence relation on $A$,
we have the continuous map $\tilde j: A/\!\!\sim\;
\to B/\!\!\sim\,$, i.e., $\tilde j: \hat
X^g/\!\!\sim\; \to \hat X/\!\!\sim\,$.  Now we need to
construct a continuous inverse for $\tilde j$: 

Since $\hat X^g$ is generalized
past-distinguishing and generalized
future-complete, we can apply Proposition 4.6(3) to
$\hat\iota^g_X : X  \to \hat X^g$ (where
$\hat\iota^g_X$ is the obvious inclusion), yielding
a map $\widetilde{\hat\iota^g_X}^g : \hat
X \to \hat X^g$; this is the map $\pi_X$ forming
the natural transformation $\boldsymbol \pi$.  To
understand its action, it is best to think of
$\hat\partial(X)$ as broken into two parts: $\bigcup
\{\frak P_x \;|\; x \text{ is non-regular}\}$ and
$\hat\partial^g(X)$.  The action of $\pi_X$ is to take
$x \in X$ to $x$; for non-regular $x$, to take $P \in
\frak P_x$ to $x$ (since $x$ is then the generalized
future limit of a chain generating $P$); and to take
$P \in \hat\partial^g(X)$ to $P$ (since $P$ is then
itself the future limit of its generating chain). 
Since the relation $\sim$ from Theorem 4.8 always
identifies, for non-regular $x$, all elements of $\frak
P_x$ with $x$, we see that $\pi_X$ respects
$\sim\,$; thus, as above, $\pi_X$ induces a map
$\tilde \pi_X: \hat X/\!\!\sim\; \to \hat
X^g/\!\!\sim\,$.  

The map $\pi_X \circ j: \hat X^g \to \hat X^g$ is
just the identity, so $\tilde\pi_X \circ \tilde j:
\hat X^g/\!\!\sim\,\; \to \hat X^g/\!\!\sim\,$ is
also the identity.  The map $j \circ \pi_X: \hat X
\to \hat X$ is the identity on $\hat X^g$; for the
remainder---$P \in \frak P_x$ for $x$
non-regular---it takes $P$ to $x$.  However, note 
that $P \sim x$ in that situation (essentially
because the map $\tilde i^g$ itself factors through
$\pi_X$: $\tilde i^g = \hat i^g \circ \pi_X$, as
alluded to in the proof of Proposition 4.6(3)); thus,
we have $\widetilde{j \circ \pi_X} = \tilde j \circ
\tilde\pi_X$ is the identity on $\hat
X/\!\!\sim\,$.  Therefore, we have a homeomorphism
$\hat X^g/\!\!\sim\; \cong \hat X/\!\!\sim\,$.

Finally, we note that $X$ is nicely embedded in
$\hat X^g/\!\!\sim\,$:  If $\tilde i^g(P) = x$ for
$P \in \hat\partial(X)$, then $i(x)$ is a generalized
future limit for the image of a chain generating
$P$; but this reflects in $X$, implying that $P$ is
a past component of $x$.  If $x$ is regular, then
$P = I^-(x)$, and $P$ is not in $\hat\partial(X)$;
and if $x$ is non-regular, then $P$ is not in
$\hat\partial^g(X)$.  Therefore, no element of $X$
can be identified with any element of
$\hat\partial^g(X)$ by $\sim$.  Thus, the boundary
attached to $X$ by $\hat X^g/\!\!\sim\,$ (ergo, by
$Y$) is just what is left over:
$\hat\partial^g(X)/\!\!\sim\,$.
\qed \enddemo 

We close with a medley of categorical observations:

In virtue of statement (2) of Proposition 4.6, we
have a categorical formulation of future
completion, including universality, without the
assumption of regularity, so long as we strengthen
past-distinguishing to generalized
past-distinguishing ({\bf GPdis}):

\proclaim{Theorem 4.10} Future completion is a
functor $\;\widehat{}\;$ \rom: {\bf
GPdis\-Spbd\-Ftop\-GChron} $\to$ {\bf
Fcpl\-GPdis\-Spbd\-Ftop\-GChron}\rom; together with
the natural transformation
$\hat{\boldsymbol\iota}$, this forms a left
adjoint to the forgetful functor.
\endproclaim

\demo{Proof}  Proposition 4.6 provides the
functoriality of future completion for these
categories.  All we need to observe is that the
maps $\hat\iota_X: X \to \hat X$ are generalized
future-continuous (obvious, since any generalized
future limit in $X$ is also one in $\hat X$) and
\h continuous (Corollary 4.5). \qed \enddemo

We can also establish categorical results for
past-determination without the regularity
assumption.  First we need an analogue of
Proposition 3.2:

\proclaim{Proposition 4.11} For any chronological
set $X$ with pasts of all points non-empty,
$\iota^p_X: X \to X^p$ is a \h homeomorphism.
\endproclaim

\demo{Proof}  Let the maps $P \mapsto P^p$ and
$Q \mapsto Q_0$ be as in Lemma 3.1.  Let
$L$, $\Cal L$, $L^p$, and $\Cal L^p$ denote the
various limit-operators in $X$ and $X^p$; let
$\frak P_x$ and $\frak P^p_x$ denote the families
of past components of $x$ in $X$ and in $X^p$. 

We will see that for any sequence $\sigma$, $L(\sigma)
= L^p(\sigma)$.

Let $\sigma$ be any sequence in $X$ with $x \in
L(\sigma)$.  For any $\tau \subset
\sigma$, there are $\rho \subset \tau$ and $P \in
\frak P_x$ with $P \in \Cal L(\rho)$.  Then $P^p \in
\frak P^p_x$; to have $x \in L^p(\sigma)$, we just need
$P^p \in \Cal L^p(\rho)$:  For any $y \in P^p$, there is
some $y_0 \in P$ with $y \ll^p y_0$, and $y_0 \ll
\rho(n)$ (eventually) implies $y \ll^p \rho(n)$
(eventually).  For any $Q \in \I (X^p)$ with $Q
\supset P^p$, suppose for all $y \in Q$, $y \ll^p
\rho(n_k)$ (for some subsequence); then for all $y
\in Q_0$, we can find $y' \in Q$ with $y \ll y'$,
and $y' \ll \rho(n_k)$ implies $y \ll \rho(n_k)$. 
We also know $Q_0 \supset P$, so $Q_0 = P$;
therefore, $Q = P^p$.  This finishes showing $P^p 
\in \Cal L^p(\rho)$.  Thus, $x \in L^p(\sigma)$.

A formally identical proof, with the roles of $X$
and $X^p$ reversed, establishes that $x \in
L^p(\sigma)$ implies $x \in L(\sigma)$.  With
identical limit-operators, $X$ and $X^p$ have the
same \h topologies, with $\iota^p_X$ a
homeomorphism. \qed \enddemo

With past determination well in hand, we obtain
results for the GKP Future Causal Boundary
construction, much as in Theorems 3.4 and 3.5:

\proclaim{Theorem 4.12} 
\roster

\item 
Past determination is a functor {\bf p} $:$ {\bf
Ftop\-GChron} $\to$ {\bf Pdet\-Ftop\-GChron} and
also {\bf p} $:$ {\bf Spbd\-Ftop\-GChron} $\to$ {\bf
Pdet\-Spbd\-Ftop\-GChron}; together with the
natural transformation $\boldsymbol\iota^{\bold
p}$, these are left adjoints for the respective
forgetful functors.  In particular, for any map
$f: X \to Y$ in {\bf FtopGChron} and
$Y$ past-determined, there is a unique generalized
future-continuous and \h continuous map $f^p : X^p
\to Y$ with $f^p \circ \iota^p_X = f$; and if $f$
is also in {\bf Spbd}, so is $f^p$.

\item 
There is a functor $\boldkey + :$ {\bf
GPdis\-Spbd\-Ftop\-GChron} $\to$ {\bf
Fcpl\-Pdet\-GPdis\-Spbd\-Ftop\-Gchron} and a
natural transformation $\boldsymbol
\iota^{\boldkey +}$, forming a left adjoint to
the forgetful functor.

\item 
For any map $f: X \to Y$ in {\bf
Spbd\-Ftop\-GChron} with $Y$ generalized
future-complete, generalized
past-distin\-gui\-shing, and past-determined, there
is a unique generalized future-continuous and \h
continuous map $\tilde f^{+g} : X^+ \to Y$ with
$\tilde f^{+g} \circ \iota^+_X = f$. 
\endroster
\endproclaim

\demo{Proof} (1) The same proof as in Theorem 3.3
applies for the functoriality of {\bf p}.  The
statements about $f^p$ when $Y$
is past-determined just amount to the universality
property.  (They are included here for comparison
with statement (3) here and statement (3) of
Proposition 4.6; these are  more general than the
analogous statements with $Y$ being generalized
past-determined.)

(2) The same proof applies as in Theorem 3.4.

(3) This follows the pattern of the universality
properties, even though it is not categorical: 
With $f$ in {\bf SpbdFtopGChron} and
$Y$ also generalized future-complete and
generalized past-distinguishing, Proposition
4.6(3) gives us
$\tilde f^g: \hat X \to Y$ (unique with $\tilde f^g
\circ \hat\iota_X = f$).  Then with $\tilde f^g$ in
{\bf SpbdFtopGChron} and $Y$ also past-determined,
statement (1) above gives us $(\tilde f^g)^p : (\hat
X)^p \to Y$ (unique with $(\tilde f^g)^p \circ
\iota_{\hat X}^p = \tilde f^g$).  Then
$\tilde f^{+g}$ is just $(\tilde f^g)^p \circ
(j_X)^{-1}$ (unique for
$\tilde f^{+g} \circ \iota^+_X = f$, where
$\iota^+_X = j_X \circ \iota^p_{\hat X} \circ \hat
\iota_X$).
\qed \enddemo

\head 5. Examples
\endhead

This Section will largely be devoted to examining
in detail the Future Chronological Boundary of a
class of spacetimes with spacelike boundary,
comparing the \h topology with what might be
expected; and to examining how the
\h topology works for boundaries derived in a simple
manner from embedding spacetimes with spacelike
boundaries into larger manifolds, where non-regular
points come into play. (All the explicit spacetimes
examined in this Section are globally hyperbolic, so
there is no need to be concerned about
past-determination:  The $\;\widehat{}\;$ and
$\boldkey +$ functors are the same for these spaces,
and what is I am calling the Future Chronological
Boundary here could just as easily be called the GKP
Future Causal Boundary.) In a nutshell:  The
\h topology gives the ``right" results in a variety of
situations.  However, it does give different results
from the embedding topology for some embeddings which
might be deemed questionable.

But we will begin with the most elementary of 
examples: Minkowski n-space, $\Bbb L^n$. 

\subhead
5.1 Minkowski Space
\endsubhead

The causal structure of Minkowski space is
well-known, with explication in such sources as
\cite{HE}.  But a small amount of detail here is not
out of place.  Let us split $\lo^n$ orthogonally as
$\r^{n-1} \times \lo^1$; inside $\r^{n-1}$ we'll
locate the unit sphere about the origin,
$\s^{n-2}$.  It is evident that any null line
$\beta$ in $\lo^n$ gives rise to an IP in the form of
$I^-[\beta]$, and that this is the same as the past
of the null hyperplane determined by $\beta$: 
Specifically, if $\beta$ is given by $\beta(s) =
(z,0) + s(p,1)$ for $z \in \r^{n-1}$ and $p \in
\s^{n-2}$, the null hyperplane bounding the 
corresponding IP is $\Pi =
\{(x,t) \;|\; t = \<x,p\> - \<z,p\>\}$
(where $\<\,,\,\>$ denotes the Euclidean inner
product in $\r^{n-1}$).  Thus, the collection of IPs
definable this way corresponds to the set of all
null hyperplanes, parametrized by
$\s^{n-2} \times \r^1$ as $\Pi_{p,a} = \{(x,t) \;|\;
t = \<x,p\> + a\}$; let $P_{p,a} = I^-[\Pi_{p,a}] =
\{(x,t) \;|\; t < \<x,p\> + a\}$. (These are the IPs
constituting $\frak I^+$, future null infinity.)

In fact, these are all the IPs of $\lo^n$, aside
from those which are pasts of a single point and
the IP $P_\infty$, often known as $i^+$ (or future
timelike infinity), which consists of the entire
spacetime.  Sketch of a proof: Let $\gamma$ be any
future-endless timelike curve; we can represent
$\gamma$ by $\gamma(s) = (c(s), s)$ for $c: \r \to
\r^{n-1}$ a curve with
$|\dot c| < 1$ (where $|\;|$ denotes Euclidean
length).  It's not hard to see that $I^-[\gamma] =
\{(x,t) \;|\; t < \sup_s(s - |c(s) - x|)\}$, i.e.,
the past of the graph of the function $b: x
\mapsto \sup_s(s - |c(s) - x|)$.  For any
$x \in \r^{n-1}$, let $b_x : \r \to \r$ be the
function given by $b_x(s) =  s - |c(s) - x|$;
this is monotonic increasing, so $b(x) = \sup{b_x} =
\lim_{s \to \infty}{b_x(s)}$.  Since for all $s$, $x
\mapsto b_x(s)$ is Lipschitz-1, either for all
$x \in \r^{n-1}$, $b(x) = \infty$; or for all $x$,
$b(x)$ is finite, and $b$ is Lipschitz-1.  In the
first case, $I^-[\gamma] = P_\infty$; in the
second case, $I^-[\gamma] = P_{p,a}$ where $p =
\lim_{s\to\infty}{c(s)/|c(s)|}$ and $a =
b(0)$ (the fact that for all $x$, $b_x$ has a
limit, is what shows that $c(s)/|c(s)|$ has a
limit; then calculation shows $b(x) = \<x,p\>+a$).

Thus we know $\hat\partial(\lo^n) = \{P_{p,a}
\;|\; (p,a) \in \s^{n-2}\times\r^1\} \cup
\{P_\infty\}$, giving us a nice parametrization
of the boundary as a cone over the
$(n-2)$-sphere.  Indeed, that is the topology of
the future boundary of $\lo^n$ in its conformal
embedding into the Einstein static spacetime,
$\s^{n-1} \times \lo^1$:  This embedding can be
described as follows (taken from \cite{HE}, Section
5.1) :  Express $\r^{n-1}$ as $\s^{n-2} \times
\r^+$ via $x \mapsto (x/|x|, |x|)$, so that $\lo^n
= \s^{n-2}\times\r^+\times\lo^1$; express
$\s^{n-1}$ as a subset of $\r^n = \r^{n-1} \times
\r^1$, so that $\s^{n-1} \times \lo^1$ is a subset
of $\r^{n-1} \times \r^1 \times \lo^1$.  Then we
define the map $\phi : \s^{n-2} \times \r^+ \times
\lo^1 \to \r^{n-1} \times \r^1 \times \lo^1$ by
$$\phi(p,r,t) = (p\sin\theta, \cos\theta, \tau),$$
where
$$\align
\theta &= \tan^{-1}(t+r) - \tan^{-1}(t-r) \\
\text{and  }\tau &= \tan^{-1}(t+r) +
\tan^{-1}(t-r).
\endalign$$ 
Since this is a conformal map, the causal
structure is preserved.  For a fixed value of
$\tau$ with $0 <\tau < \pi$, the image of $\phi$  
has $|\theta| < \pi - \tau$.  This yields, for the
boundary of the image of
$\phi$ in the $\tau > 0$ region, a copy of
$\s^{n-2}$ as $(\s^{n-2}\sin(\pi-\tau),
\cos(\pi-\tau), \tau)$.  Thus, the future boundary
of the embedding (i.e., with $0<\tau<\pi$) is
$\s^{n-2}\times(0,\pi)$ together with the
boundary-element at $\tau=\pi$, the point
$(\boldkey 0,1,\pi)$: in other words, a cone on
$\s^{n-2}$.

To explore the \h topology of
$\hat\partial(\lo^n)$, we need to know when it is
that one IP can contain another.  This is quite
simple:  Every $P_{p,a}$ is contained in
$P_\infty$, and $P_{p,a} \subset P_{q,b}$ if and
only if $p = q$ and $a \le b$ (any two hyperplanes
will intersect if they aren't exactly parallel). 
For questions of convergence, it's easiest to work
with a generating chain for each IP:  $P_{p,a}$ is
generated by $c_{p,a}(m) = (mp,m+a-1/m)$.  Then
a sequence of IPs $\sigma = \{P_{p_n,a_n}\}$
approaches an IP $P_{p,a}$ (i.e.,$P_{p,a} \in
L(\sigma)$) if and only if
\roster
\item
for each $m$, eventually $c_{p,a}(m) \in
\sigma(n)$, and
\item
for any $\epsilon > 0$, for all $m$ sufficiently
large, eventually $c_{p,a+\epsilon}(m) \notin
\sigma(n)$.
\endroster
(The second clause is the contrapositive of saying
that any IP containing $P_{p,a}$ and having each
element of its generating chain eventually in the
past of $\sigma$, must actually be $P_{p,a}$.)  It
is easy to work this out and show it equivalent to
$\lim_{n\to\infty} (p_n,a_n) = (p,a)$ in the usual
topology of $\s^{n-2}\times\r^1$.  (For example: If
for some sequence $\{n_k\}$, $\<p,p_{n_k}\> <
1-\delta$, then by clause (1), for all $m$,
eventually $m+a-1/m < m\<p,p_{n_k}\> +
a_{n_k}$, so eventually $a_{n_k} > m\delta + a -
1/m$. This means $\lim a_{n_k} = \infty$; but
that contradicts clause (2).)  It is also easy to
show $\sigma$ approaches $P_\infty$ if and only if
$\lim a_n = \infty$.  Thus, the \h topology for the
boundary is precisely that of a cone on $\s^{n-2}$,
just as in the conformal embedding into
$\s^{n-1}\times\lo^1$.

We also need to be concerned about how the
boundary is topologically connected to the
spacetime:  When is it that $\sigma$ has $P_{p,a}$
(or $P_\infty$) as a limit for $\sigma(n) =
(x_n,t_n) \in \lo^n$?  This is considerably
messier to work out, and the answer is perhaps a
bit surprising: $P_{p,a} \in L(\sigma)$ if and
only if
\roster
\item $\lim t_n = \infty$,
\item $\lim x_n/|x_n| = p$, and
\item $\lim (t_n - |x_n|) = a$. 
\endroster

One might have expected condition (3) to be,
instead, $\lim (t_n - \<x_n,p\>) = a$, in light of
the defining equation for $P_{p,a}$; but this
turns out not to be a sufficient condition. 
However, the condition as given is precisely that
needed for $\phi[\sigma]$ to have as limit the
point $(p\sin(\pi-\tau), \cos(\pi-\tau),
\tau)$ in the boundary of the image of $\phi$,
where $\tau = \tan^{-1}a$.  

Just a word on how to establish this result:  The
two clauses needed for convergence to $P_{p,a}$ are
\roster
\item for all $m$, eventually, $t_n - |x_n - mp| -
m > a - 1/m$, and
\item for any $\epsilon>0$, for $m$ sufficiently
large, eventually $t_n - |x_n - mp| - m \le a +
\epsilon - 1/m$.
\endroster
The key is to replace the expression on the left
of the inequalities by one that is independent of
$m$, that being $t_n - |x_n|$.  The difference,
$\delta^m_n = |x_n|-|x_n-mp|-m$, is always
non-positive, but must be shown to go to 0 for
fixed $m$.  This turns out to be the case so long
as $\{\lambda_n\}$ goes to infinity and
$\{\mu_n/\lambda_n\}$ goes to 0, where $x_n =
\lambda_np + \mu_nq_n$ for $q_n$ unit-length and
perpendicular to $p$.

It is easy to show $\sigma$ approaches $P_\infty$
if and only if $\lim a_n = \infty$. 
  
This is all summarized thus:

\proclaim{Proposition 5.1} Let $M = \lo^n$ and $E
= \s^{n-1}\times\lo^1$, and let $\phi: M \to E$ be
the standard conformal embedding of Minkowski
space in the Einstein static spacetime.  Then
$\hat\phi : \hat M \to \hat E$ is a
homeomorphism onto its image, with the \h topology
for $\hat M$; in particular, $\hat\partial(M)$ is
a cone on $\s^{n-2}$. \qed
\endproclaim 

It is worth noting the contrast between this
result and those in Proposition 2.7 and in
Theorem 3.6:  Theorem 3.6 states that any
future-completing boundary on $M$ must be
homeomorphic to $\hat\partial(M)$---but that
assumes the use of the \h topology for the
boundary, and the point of Proposition 5.1 is to
use a boundary coming from an embedding in
another spacetime.  Proposition 2.7 fares better
in this regard, in respect of the continuity of
$\hat\phi$, as in this case, the image of
$\hat\phi$ lies entirely within $E$, so that
there is no concern over which topology to use
for $\hat\partial(E)$---but that Proposition
applies only in the case of spacelike boundaries,
and $M$ has a null boundary.

It is also worth contrasting the \h topology for
$\hat\partial(\lo^n)$ with that given in the
original GKP topology for the Causal Boundary. 
This is explicated in \cite{HE}, Section 6.8: 
$M^\#$ is defined to be $\hat M \cup \check M$
(where $\check M$ is $M$ plus the Past
Chronological Boundary, formed of IFs).  For any
IF $F$ in $M$, the sets $F^{\text{int}}$ and
$F^{\text{ext}}$ are considered to be open sets
in $M^\#$, where $F^{\text{int}} = F \cup \{P \in
\hat\partial(M) \;|\; P \cap F \neq \emptyset\}$
and $F^{\text{ext}} = (M - \slanted{closure}(F))
\cup \{P \in \hat\partial(M) \;|\; \text{for
any } A \text{ with } P = I^-[A], I^+[A]
\not\subset F\}$; and dually for
$P^{\text{int}}$ and $P^{\text{ext}}$ for any IP
$P$.  The collection of all of these four types of
sets provides a sub-basis for the GKP topology in
$M^\#$. 

Apply this to $M = \lo^n$, examining
$(F_{p,a})^{\text{ext}}$ for any $(p,a) \in
\s^{n-2}\times\r^1$:  The portion in $\lo^n$ is
unexceptional, making up $P_{p,a}$.  But for which
IPs $P_{q,b}$ (or $P_\infty$) is it true that
when expressed as $I^-[A]$, the $A$ must have
points in its future not in $F_{p,a}$?  The only
possibilities are those with $q=p$ and $b<a$, as
anything else will intersect $F_{p,a}$ and
will clearly be expressible as $I^-[A]$ for some
$A \subset F_{p,a}$ (more precisely: $P_{p,a}$
is ruled out because it is expressible as
$I^-[l]$ for $l$ a null half-line lying in
$\Pi_{p,a}$, the boundary of $F_{p,a}$, and
$I^+[l] \subset F_{p,a}$).  Thus, the GKP topology
on $\lo^n \cup \hat\partial(\lo^n) \cup
\check\partial(\lo^n)$ has, as an open set,
anything of the form $P_{p,a} \cup \{P_{p,b} \;|\;
b<a\}$ (where $P_{p,a}$ is to be read as a subset
of $\lo^n$).  Then the GKP topology on
$\hat\partial(\lo^n)$ has these as open sets: for
any $(p,a) \in \s^{n-2}\times\r^1$, $\{P_{p,b}
\;|\; b<a\}$.  This is not at all the topology of
a cone on $\s^{n-2}$. (The other GKP-open sets in
$\hat\partial(\lo^n)$ come from
$(I^+((x,t)))^{\text{ext}}$, yielding $\{P_{p,b}
\;|\; \<x,p\>+b < t\}$---but these add nothing
new, as they wholly contain the open sets already
mentioned.) This topology would have
$\{P_{p_n,a_n}\}$ approach $P_{p,a}$ if and only
if all $p_n = p$ and those elements of $\{a_n\}$
which are greater than $a$, if infinite in number,
approach $a$ in the usual sense; and every sequence
$\{P_{p_n,a_n}\}$ approaches $P_\infty$ (since
the only neighborhood of $P_\infty$ is all of
$M^\#$).

(The GKP construction of the Causal Boundary
includes identifications on $M^\#$ to produce a
Hausdorff $M^*$, but this has no bearing on
$\lo^n$, as no identifications are made.  Other
methods of identification within $M^\#$ have been
proposed, such as by Szabados in \cite{S} and by Budic
and Sachs in \cite{BS}; but these also make no
identifications for $M = \lo^n$.)

It is possible to generalize the techniques
employed for $\lo^n$ to any standard static
spacetime.  The results for topology of the Future
Chronological Boundary are what one would hope for,
though the causal structure can show some curious
anomalies.  But it is sufficiently complicated as to
warrant appearance in a separate paper. 

\subhead
5.2 Multiply Warped Products
\endsubhead

Perhaps the next most obvious example to look at is
Schwarzschild space, especially interior
Schwarz\-schild, inside the event horizon, as the
boundary---the Schwarz\-schild singu\-larity---is
spacelike.  Using $r$ and $t$ as the standard
Schwarzschild coordinates (see, e.g, \cite{HE},
Section 5.5), which switch timelike/spacelike roles
inside the event horizon, we have, as the metric $g$
for the $r < 2m$ region,
$$g \;=\; -\frac1{\frac{2m}r-1}(dr)^2 \,+\,
(\frac{2m}r-1)(dt)^2 \,+\, r^2h_{\s^2},$$
where $m$ is the mass and $h_{\s^2}$ is the
standard metric on the unit 2-sphere.  We can get all
the causal information we need by looking at a
conformal metric, $$\bar g \;=\; -(dr)^2 \,+\,
(\frac{2m}r-1)^2(dt)^2
\,+\, r^2(\frac{2m}r-1)h_{\s^2}.$$

In other words, interior Schwarzschild is conformal to
the spacetime $(0,2m) \times \r^1 \times \s^2$ with
metric $-(dr)^2 + f_1(r)h_{\r^1} + f_2(r)h_{\s^2}$
(where $h_{\r^1}$ is the standard Riemannian metric on
$\r^1$) for some specific functions $f_1$ and $f_2$. 
The natural expectation is that the
Schwarzschild singularity---the Future Chronological
Boundary of internal Schwarzschild---should be
$\r^1\times\s^2$; and we will see that this is
precisely so, in the \h topology.  But rather than look
just at interior Schwarzschild, we will examine a large
class of spacetimes of this same form.  Let us call a
spacetime a {\it multiply warped product spacetime\/}
if it has the following form:  The manifold is $M =
(a,b) \times K_1 \times \cdots \times K_m$ for some
$a<b$ (possibly infinite) and for some manifolds $K_1$
through $K_m$.  Each $K_i$ has a Riemannian metric
$h_i$, and for each $i$ there is a positive function
$f_i: (a,b) \to \r^+$.  The spacetime metric is  $g =
-(dt)^2 + f_1(t)h_1 + \cdots + f_m(t)h_m$ (more
precisely: $g = -(dt)^2 + \sum_i(f_i\circ t)
\pi_i^*h_i$, where $t: M \to (a,b)$ and $\pi_i : M
\to K_i$ are projection).  The Riemannian manifolds
$(K_i,h_i)$ will be called the spacelike factors,
the functions
$f_i$ the warping functions.

Interior Schwarzschild (a spherically symmetric vacuum
spacetime) is conformal to a multiply warped  product
spacetime with two spacelike factors, the standard
$\r^1$ and $\s^2$.  Robertson-Walker spacetimes
(homogeneous and isotropic, perfect fluid; see, e.g,
\cite{HE}, Section 5.3) are multiply
warped product spacetimes, each with a single spacelike
factor: A Robertson-Walker spacetime is $((\alpha,
\omega) \times K,\, -(dt)^2 + r(t)^2h)$, where
$(\alpha, \omega)$ is some interval in $\r^1$ (finite,
infinite, or half-infinite),  $(K,h)$ is a
constant-curvature Riemannian manifold,  and $r: \r^1
\to \r^+$ is some positive function, scaling the size
of the universe.  The Kasner spacetimes (spatially
homogenous and vacuum; see, e.g, \cite{W}, Section 7.2)
are multiply warped product spacetimes with three
spacelike factors:  These have the form of
$((0,\infty) \times \r^3,\, -(dt)^2 + t^{2p_1}(dx)^2 +
t^{2p_2}(dy)^2 + t^{2p_3}(dz)^2)$ for some constants
$p_1$, $p_2$, and $p_3$ satisfying $\sum_i p_i = \sum
(p_i)^2 = 1$; this has three spacelike factors of
standard $\r^1$.

Not all multiply warped product spacetimes have
spacelike boundaries:  If any of the spacelike factors
is incomplete as a Riemannian manifold, then the
Future Chronological Boundary has timelike regions
(i.e., chronology relations between boundary points);
and if any of the warping functions goes to 0 too
quickly (for a finite future end of the interval
$(a,b)$) or to infinity too slowly (for an infinite
future end), then the Future Chronological Boundary has
null or timelike relations (i.e., inclusion of some
boundary point in another).  But here we will consider
only the spacelike case.

The matter addressed in the following proposition is
not akin to that of Proposition 2.7, concerning the
continuity of $\hat f: \hat X \to \hat Y$ (say, with a
multiply warped product spacetime for $X$,
$(a,b]\times K$ for $Y = \hat Y$, and inclusion for
$f$), because in that proposition it is the \h
topology that is used for $Y$, and here it is the
product topology on $(a,b]\times K$ that is of
interest.  In fact, we will see how to put a
chronology relation on $(a,b]\times K$, and the point
of the proposition is that the \h topology from that
is the same as the product topology.  In similar vein,
Theorem 3.6---concerned with any future-completing and
past-distinguishing boundary---addresses a different
concern than this proposition, as it, also, looks at
the \h topology of such a boundary, and we are
concerned now with a ``naturally occurring" topology
for a boundary.

\proclaim{Proposition 5.2} Let $M$ be conformal to a
multiply warped product spacetime with timelike factor
$(a,b)$ \rom(assume $b$ to be the future-end of the
interval\rom), spacelike factors $(K_i, h_i)$, and
warping functions $f_i: (a,b) \to \r^+$ \rom($1 \le i
\le m$\rom).  Suppose that for each $i$,
\roster
\item $h_i$ is a complete Riemannian metric, and
\item for some finite $c \in (a,b)$, $\int_c^b
f_i^{-\frac12} < \infty$.
\endroster
Then $M$ has only spacelike boundaries; $\hat M$ is
\h homeomorphic to $(a,b]\times K$, where $K =
K_1\times\cdots\times K_m$; and $\hat\partial(M)$ is \h
homeomorphic to $K$, included in $\hat M$ as
$\{b\}\times K$.
\endproclaim

\demo{Proof} The proof is somewhat lengthy, though
largely uncomplicated:  First we must identify all the
IPs in $M$; then we have to show
$\hat\partial(M)$ is spacelike; and finally we have
to show the \h topology reflects the topology of
$(a,b]\times K$.

\medpagebreak

\flushpar Step (1): Identifying $\hat\partial(M)$.
\medpagebreak

First we will construct a new chronological set (which
will turn out, essentially, to be $\hat M$):  Let the
set be $\bar M = (a,b]\times K$, and extend the
chronology relation $\ll$ from $M$ to $\bar M$
by defining, for any $t$ with $a<t<b$ and any $x$ and
$y$ in $K$, $(t,x) \ll (b,y)$ if and only if there
is a timelike curve $\gamma: [t,b) \to M$ with
$\gamma(t) = (t,x)$ and $\lim_{s\to b}\gamma(s) =
(b,y)$.  Since
$\gamma$ can always be expressed as $\gamma(s) =
(s,c(s))$ for some curve $c$ in $K$, this is
equivalent to there being a curve $c: [t,b) \to K$ with
$c(t) = x$, $\lim_{s\to b}c(s) = y$, and for all $s$,
$\sum_if_i(s)^{\frac12}|\dot c_i(s)|_i < 1$, where
$c_i = \pi_i\circ c$ ($\pi_i$ being projection to the
$i$th spacelike factor) and $|\;|_i$ denotes norm
with respect to $h_i$.  No other chronology relations
(aside from those in $M$) are defined in $\bar M$; it
should be evident that $(\bar M, \ll)$ is a
chronological set. We will use $\bar I^-$ and $\bar
I^+$ to denote past and future within $\bar M$, while
$I^-$ and $I^+$ denote the same in $M$.

The crucial step is to note that, just as in a
spacetime, $\bar I^+$ yields an open set in $\bar M$
(using the product topology for $(a,b]\times K$):

\proclaim{Lemma} For any $(t,x) \in M$, $\bar
I^+((t,x))$ is open in $\bar M$.
\endproclaim

\demo{Proof of Lemma} All we need show is that $\{y
\in K \;|\; (t,x) \ll (b,y)\}$ is open in $K$, i.e.,
that for $(b,y) \gg (t,x)$, we can find a neighborhood
$U$ of $y$ in $K$ such that for all $z \in U$, $(b,z)
\gg (t,x)$.  Since we are dealing only with small
neighborhoods of $y$, the differences in the metrics
$h_i$ become irrelevant, and the question reduces to
a Euclidean one:  Given continuous curves $c_i : [t,b]
\to\r^{k_i}$, C$^1$ on $[t,b)$ with
$\sum_if_i(s)^{\frac12}\|\dot c_i(s)\| < 1$ for
$s<b$ ($\|\;\|$ denoting Euclidean norm), is there a
neighborhood $U$ of $c(b)$ in $\r^n$ ($n = \sum_i k_i$
and $c = (c_1, ..., c_m)$) such that for all
$\bar y \in U$, there are continuous curves
$\bar c_i : [t,b] \to \r^{k_i}$, C$^1$ on
$[t,b)$ with $\sum_if_i(s)^{\frac12}\|\Dot{\Bar
c}_i(s)\| < 1$ for $s<b$, satisfying $\bar c_i(t) =
c_i(t)$ and $\bar c_i(b) = \bar y_i$?

We need only consider each factor separately, i.e.,
deal with $m = 1$:  Given continuous $c: [t,b] \to
\r^k$, C$^1$ on $[t,b)$ with $f(s)^{\frac12}\|\dot
c(s)\| < 1$ for $s<b$, we need to find a family of
variations $\bar c$ of $c$, starting at the same point
and satisfying the same differential inequality, with
endpoints at $b$ forming a neighborhood of $c(b)$. 
Let $\epsilon(s) = f(s)^{-\frac12} - \|\dot c(s)\|$;
$\epsilon$ is positive and continuous on $[t,b)$.  For
any continuous $\theta: [t,b] \to \r^k$,
C$^1$ on $[t,b)$ with $\|\dot\theta(s)\| < 
\epsilon(s)$ for $s<b$ and obeying $\theta(t) = 0$,
the curve $\bar c = c+\theta$ is an acceptable
variation of $c$.  If we let $\Theta$ be the set of
all such $\theta$, then it is clear that
$\{\theta(b) \;|\; \theta \in
\Theta\}$ contains a neighborhood of $0$ in $\r^k$
(just consider separately each one-dimensional
projection). \qed \enddemo

For any $x \in K$, define $Q_x = \bar I^-((b,x))$.  We
will show that $\hat\partial(M) = \{Q_x \;|\; x \in
K\}$.

First it is clear that any $Q_x$ is an IP in $M$, as
$Q_x = I^-[\gamma_x]$, where $\gamma_x$ is the timelike
curve given by $\gamma_x(s) = (s,x)$ for all $s$: 
Surely anything in the past of $\gamma_x$ lies in
$Q_x$.  Conversely, for any $(t,y) \in Q_x$, i.e., 
$(t,y) \ll (b,x)$, since $\bar I^+((t,y))$ is an open
neighborhood of $(b,x)$ in $\bar M$, $\bar I^+((t,y))$
contains some $(s,x)$ for $s<b$, i.e., $(t,y) \ll
(s,x)$; thus, $(t,y) \in I^-[\gamma_x]$.  Furthermore,
$Q_x \in \hat\partial(M)$, as it is clear that any
$I^-((t,y))$ contains no $(s,z)$ with $s \ge t$, while
$Q_x$ contains $(s,x)$ for $s$ arbitrarily close to
$b$.

Next, we show that every IP in $\hat\partial(M)$ is a
$Q_x$ for some $x \in K$:  Let $P = I^-[\gamma]$ for
$\gamma$ a future-endless timelike curve in $M$, i.e.,
$\gamma(s) = (s,c(s))$ for $c : [t_0,b) \to K$ 
C$^1$ and satisfying $\sum_if_i(s)^{\frac12}|\dot
c_i(s)|_i < 1$.  In particular, for each $i$, $|\dot
c_i(s)|_i < f_i(s)^{-\frac12}$.  Let $L_i$ denote the
Riemannian length functional in $(K_i,h_i)$; then for
each $i$, $L_i(c_i) = \int_{t_0}^b |\dot c_i(s)|\,ds <
\int_{t_0}^b f_i(s)^{-\frac12}\,ds$, which is finite. 
Therefore, since $K_i$ is complete, $c_i$ must have an
endpoint $x_i = \lim_{s\to b} c_i(s)$.  Let $x =
(x_1,...,x_m)$.  Then $P = Q_x$:  

Since $c$ approaches $x$, $\gamma$ approaches $(b,x)$;
thus, for any $(t,y) \in I^-[\gamma]$, there is a
timelike curve from $(t,y)$ to $(b,x)$: $P \subset
Q_x$.  For any $(t,y) \in \bar I^-((b,x))$, we have
$(b,x) \in \bar I^+((t,y))$, which is an open
neighborhood in $\bar M$ of $(b,x)$.  Since $\gamma$
approaches $(b,x)$, eventually it enters $\bar
I^+((t,y))$, so $(t,y) \in I^-[\gamma]$: $Q_x \subset
P$.

Finally, we need to know that the $\{Q_x \;|\; x \in
K\}$ are all distinct: Given $x \in K$, for each
$j<m$, let $Q_x^j$ be the intersection of $Q_x$ with
$N_x^j = \{(s,z) \;|\; z_i = x_i \text{ for all } i \neq
j\}$.  Let $M_j$ be the warped product spacetime
$((a,b)\times K_j,\, -(dt)^2 \,+\, f_j(t)h_j)$. Then
$Q_x^j = \{(s,x_1,...,x_{j-1},y,x_{j+1},...,x_m) \;|\;
(s,y) \ll (b,x_j) \text{ in } \bar M_j\}$ (the idea
being that if there is a timelike curve in $M$ from
$(t,z)$ to $(b,x)$ and $(t,z) \in N_x^j$, then there is a
timelike curve wholly within $N_x^j$ between those
points).

Now, in $\bar M_j$, we can identify $\bar I^-((b,x_j))$
as $\{(s,y) \;|\; d_j(y,x_j) < \int_s^b
f_j^{-\frac12}\}$, where $d_j$ is the Riemannian
distance function in $K_j$.  For $s$ sufficiently close
to $b$, $\int_s^b f_j^{-\frac12} < \text{diam}(K_j)$,
where diam denotes diameter (in case this is not
infinite).  Therefore, if we let $S_s$ denote the
$s$-slice of $\bar I^-((b,x_j))$, i.e., $S_s = \bar
I^-((b,x_j)) \cap (\{s\}\times K_j)$, then $S_s$
changes, depending on what $x_j$ is---in fact, $S_s$
narrows down to $x_j$ as $s$ approaches $b$.  Then we
can express $Q^j_x$ as 
$\{(s,x_1,...,x_{j-1},y,x_{j+1},...,x_m) \;|\; (s,y)
\in P_j\}$, where $P_j = \bar I^-((b,x_j))$ is a region
in $M_j$ that uniquely identifies $x_j$.  

Suppose $Q_x = Q_{x'}$; then for each $j$, $Q_x^j =
Q_{x'}^j$, so $P_j$ and $P'_j$ must be the same region
in $M_j$; thus, $x_j = x'_j$ for each $j$.  Therefore,
$x = x'$. 

Thus, we have fully identified $\hat\partial(M)$
with $\{b\}\times K$ (and, hence, $\hat M$ with
$(a,b]\times K$).

\medpagebreak

\flushpar Step (2): $\hat\partial(M)$ is spacelike.

\medpagebreak

This is now quite easy: If $Q_x \subset Q_{x'}$, then,
for each $j$,  use the same $s$-slices through $\bar
I^-((b,x_j))$ as above, yielding for all
$s < b$, $S_s \subset S'_s$.  Let $B_r(p)$ denote the
ball of radius $r$ around $p$ in $K_j$ in the
Riemannian distance function; then $B_{r(s)}(x_j)
\subset B_{r(s)}(x'_j)$, where $r(s) = \int_s^b
f_j^{-\frac12}$.  Since $r(s)$ goes to 0 as $s$ goes to
$b$, if $x_j \neq x'_j$, we can find $s<b$ with $r(s) <
d(x_j,x'_j)$, yielding a contradiction.  Thus, for all
$j$, $x _j = x'_j$, and $x = x'$.  

Clearly, we cannot have $Q_x \subset I^-(t,y)$, so
all elements of $\hat\partial(M)$ are inobservable. 
Since, by Proposition 2.6, $\hat\partial(M)$ is
necessarily closed in $\hat M$ (i.e., in the \h
topology; we have yet to verify that is the same as
the product topology on $\bar M$), this is all we need
to have $M$ in the {\bf Spbd} category.

\medpagebreak

\flushpar Step (3): The topology on $\hat M$.

\medpagebreak

First consider a sequence $\sigma$ lying in
$\hat\partial(M)$, so $\sigma(n) = Q_{x_n}$ for some
$x_n \in K$.  We want to show that for any $y \in K$,
$Q_x \in L(\sigma)$ if and only if $\lim x_n = y$. 
Since $Q_x$ is inobservable, $Q_x \in L(\sigma)$ if and
only if for all $(t,y)$ with $t<b$, eventually $(t,y)
\ll (b,x_n)$.  If $\{x_n\}$ approaches $y$, then this
is clearly true: $\bar I^+((t,y))$ is an open
neighborhood of $(b,y)$, so eventually $(b,x_n) \in
\bar I^+((t,y))$.  Conversely, suppose for all $t<b$,
eventually $(t,y) \ll (b,x_n)$.   Then, for each $j$,
looking at the region in $M_j$ as above, we
have, eventually, $d_j(y_j,(x_n)_j) < \int_t^b
f_j^{-\frac12}$.  For any integer $k$, we can find a
$t_k<b$ such that for all $j$, $\int_{t_k}^b
f_j^{-\frac12} < 1/k$.  Thus, for each $j$, for any
$k$, eventually $d_j(y_j,(x_n)_j) < 1/k$: 
$\{(x_n)_j\}$ must approach $y_j$ for each
$j$, so $\{x_n\}$ approaches $y$.

Since $\hat\partial(M)$ is closed in $\hat M$ ($M$
being a spacetime), we don't have to consider the
possibility of a sequence of boundary elements
approaching a point of $M$.

The last thing to consider is a sequence in $M$,
$\sigma(n) = (t_n,x_n)$; we need to show that for any
$y \in K$, $Q_y \in L(\sigma)$ if and only if $\lim
t_n = b$ and $\lim x_n = y$; this is very similar to
the first case.  That $Q_y$ be in $L(\sigma)$ amounts to
having that for all $t<b$, eventually
$(t,y) \ll (t_n,x_n)$.  First suppose $\{t_n\}$
approaches $b$ and $\{x_n\}$ approaches $y$.  Then
eventually $(t_n,x_n) \in \bar I^+(t,y)$ (being an
open neighborhood of $(b,y)$), so $(t,y) \ll
(t_n,x_n)$, as required.  Conversely, suppose for
all $t<b$, eventually $(t,y) \ll (t_n,x_n)$.  For
all $t<b$, eventually $t_n > t$, so $\{t_n\}$
approaches $b$. Then for each $j$, an examination of
$M_j$ similar to above shows we must have, for any
$t<b$, eventually $d_j(y_j,(x_n)_j) <
\int_t^{t_n} f_j^{-\frac12} < \int_t^b
f_j^{-\frac12}$.  Then the same argument as above
shows $\{x_n\}$ approaches $y$. \qed \enddemo

So how do our three examples of classical multiply
warped product spacetimes fare in respect of
Proposition 5.2?  

First interior Schwarzschild (which is actually only
conformal to a multiply warped product, but that's
close enough):  The spacelike factors are standard
$\r^1$ and $\s^2$, complete.  The warping factors are
$f_1(r) = (2m/r-1)^2$ and $f_2(r) = r^2(2m/r-1)$.  The
future-endpoint of the interval in question, $(0,2m)$,
is at 0; thus, we need to examine $\int_0^c
f_i(r)^{-\frac12}\,dr$.  For $i=1$, this is $-c
+ 2m\ln(2m/(2m-c))$; for $i=2$, this is $\pi/2
-\sin^{-1}(1-c/m)$.  Both are manifestly finite. 
Thus, the  Schwarzschild singularity
is spacelike with \h topology of $\r^1\times\s^2$.

For Robertson-Walker spaces the only spacelike factor
is $K$, a Riemannian manifold of constant curvature;
usually this is taken to be $\r^3$, $\s^3$, or $\Bbb
H^3$ (hyperbolic 3-space), but any quotient of these
by a discrete group of isometries will do as well and
may serve just as appropriately for a cosmological
model.  Any such quotient will be complete.  The
warping factor is
$r(t)^2$, where $r(t)$ scales the universe at time
$t$.  The integral condition to be satisfied is that
$\int_c^\omega r(t)^{-1}\,dt < \infty$; this is
precisely the condition that the spacetime be
conformal to a finite (in the future) portion of the
standard static spacetime $\lo^1 \times K$ (i.e.,
respectively Minkowski space or the Einstein static
space, in case $K$ has 0 or positive constant
curvature, assuming no identifications by isometries). 
For example, if the future-endpoint of time is
$\infty$, then $r(t) = t$ fails, but $r(t) =
t^{1+\epsilon}$ works for any
$\epsilon > 0$; and if the future-endpoint of time is
a finite $\omega$, then $r(t) = \omega-t$ fails, but
$r(t) = (\omega-t)^{1-\epsilon}$ works for any
$\epsilon > 0$.  When the integral condition is met,
the boundary is spacelike and \h homeomorphic to $K$.

For the Kasner spacetimes, the three spacelike
factors are all standard $\r^1$, complete.  The
warping functions are $f_i(t) = t^{2p_i}$, with
$\infty$ as the future-endpoint of the time interval. 
The integral condition is met for all cases except the
``exceptional" ones where one of the numbers $p_i$
is 1, and the other two are perforce 0; in that case,
the spacetime is Rindler space, a portion of
Minkowski space (with null boundary).  Aside from the
exceptional cases, the boundary is spacelike and \h
homeomorphic to $\r^3$.

\subhead 5.3 Non-regular Boundaries from Embeddings
\endsubhead

Up to now, we have looked at examples of spacetimes
with regular boundaries.  But one wants to know how
well the \h topology measures up to non-regular
boundaries, as well.  One way that non-regular
boundaries could conceivably come about (though
perhaps not most naturally) is through embeddings: 
If $\phi : M \to N$ topologically embeds the
spacetime $M$ into the manifold $N$ (of the same
dimension), then $M$ may acquire a boundary in the
form of the boundary of $\phi[M]$ in $N$.  In some
circumstances this $\phi$-boundary carries a
natural extension of the chronology relation,
defined for $x \in M$ and $p$ in the
$\phi$-boundary by $x \ll p$ if there is a timelike
curve in $M$ with past endpoint at $x$ and with
$\phi$-image approaching $p$ in the future; and
this might provide a future-completion for $M$. 
But there is no {\it a priori\/} reason to assume
this boundary is regular.

For example:  If $M$ is a multiply warped product
spacetime $(a,b) \times K$ ($K$ the product of the
spacelike factors), then one can easily embed $M$ 
into the manifold $N = \r \times K$ via the obvious
inclusion map; if $b$ is finite, then this yields an
obvious boundary for $M$ of $\{b\} \times K$, 
agreeing with the Future Chronological Boundary in
the case of a complete $K$ and warping functions
obeying the integral conditions---which is to say,
in the case of a spacelike boundary.  But consider
some other embedding $\phi: M \to \r^1 \times K$;
if, for instance, $\phi$ extends continuously to
$\{b\}\times K$ but is not a homeomorphism
there---say, $\phi(b,x) = \phi(b,y)$ for some $x
\neq y$---then we obtain a new boundary for $M$,
and we may wish to know how the
\h topology compares with the topology induced by
$\phi$.

The answer is that the two topologies are the same
for a fairly wide class of ``reasonable" embeddings
$\phi$; but the proof is not really dependent on
the form of the spacetime, just on the fact that
the it has only spacelike boundaries and $\phi$
extends continuously to the Future Chronological
Boundary (in the \h topology---which,
for these spacetimes of Section 5.2, is $\{b\}\times
K$).  Accordingly, we will examine a general
setting:

Let $M$ be a strongly causal spacetime.  Consider
a topological embedding $\phi: M \to N$, i.e., $N$
is a manifold of the same dimension and $\phi$ is
a homeomorphism onto its image.  We can think of the
boundary in $N$ of $\phi[M]$ as a sort of boundary
for $M$:  Identify $M$ with its image under $\phi$
and just use the subspace topology for
$\slanted{closure}(\phi[M])$.  However, this could
well include points that are not in any sense
causally connected to the spacetime (example: $M =
\{(x,t) \in \lo^2 \;|\; |t| < x^2 \text{ and }
0 < x < 1\}$, $\phi: M \to \lo^2$ is inclusion; the
point $(0,0)$ is in the closure of $M$ but not
causally related to any point of $M$).  To insure
something that has the character of a future
boundary, we will restrict ourselves to looking at
boundary points with what amounts to a past in
$M$:  For any $p \in N$, let $I_M^-(p) = \{x \in M
\;|\; \text{for some timelike curve }\gamma :
[0,\infty) \to M, \gamma(0) = x \text{ and }
\lim_{t\to\infty}\phi(\gamma(t)) = p\}$.  Then the
{\it future $\phi$-boundary\/} of $M$, denoted
$\bdm$, is defined as $\{p \in N -
\phi[M] \;|\; I_M^-(p) \neq \emptyset\}$; and the
{\it future $\phi$-completion\/} of $M$, denoted
$\bm$, is defined as $M \cup
\hat\bdm$, given the subspace
topology from $N$ by identifying $M$ with its
image under $\phi$ (this will sometimes be thought of
as a subset of $N$, using $\phi[M]$ in place of $M$).

At the moment, $\bm$ is just a
topological space, albeit one that may have a
future (topological) limit for various
future-endless timelike curves in $M$.  (If some
timelike curves fail to have future endpoints in
$M_\phi^+$, then ``future $\phi$-completion" is
something of a misnomer; but we will be considering
only instances where this does not happen.)  But we
can also put a chronology relation on $\bm$,
extending the one on $M$:  For $x$ in $M$ and $p$
and $q$ in $\partial_\phi^+(M)$, set $x \ll p$ if
$x \in I_M^-(p)$; $p \ll x$ if for some $y \in M$
with $y \ll x$, $I_M^-(p) \subset I^-(y)$; and $p
\ll q$ if for some $y \in I_M^-(q)$, $I_M^-(p)
\subset I^-(y)$.  Then it is not hard to check that
the extended $\ll$ is a chronology relation on
$\bm$, with $M$ chronologically dense in $\bm$. 
(It therefore follows from Theorem 4.4 that in the
\h topology on $\bm$, as well as the subspace
topology from $N$, $M$ is topologically dense in
$\bm$, and that $M$, in its own topology, is
homeomorphic to its image in $\bm$, as a subspace
of $\bm$ with the \h topology.)

Note that if we assume $M$ has only spacelike
boundaries, then $p \ll a$ cannot occur for $p \in
\bdm$ (and $a$ anything in $\bm$): 
If for some $y \in M$, $I^-_M(p)$ were in $I^-(y)$,
then consider any $x \in I_M^-(p)$:  There must be
a timelike curve $\gamma : [0,\infty) \to M$ with
$\gamma(0) = x$ and
$\lim_{t\to\infty}\phi(\gamma(t)) = p$.  Note that
$P =I^-[\gamma]$ is an IP and that $P$ is not any
$I^-(z)$:  For if it were, then $z$ would be the
future limit of $\gamma$ (i.e., the future limit
of any sequence of points on $\gamma$), hence its
topological limit (Proposition 2.2); that would
put $p = \phi(z)$, contradicting $p$ being in
$\bdm$. Therefore, $P \in
\hat\partial(M)$ and must be inobservable.  But $P
\subset I_M^-(p)$, so if $I_M^-(p) \subset I^-(y)$,
this contradicts the inobservability of $P$.  

In fact, virtually the same argument now shows that
all the elements of $\bdm$ are
inobservable (if $M$ has only spacelike
boundaries):  With $\bar I^-$ denoting the past in
$\bm$, we now know that for any $x \in M$, $\bar
I^-(x) = I^-(x)$, and for any $p \in
\bdm$, $\bar I^-(p) = I^-_M(p)$;
also, we know that $\I(\bm)$ is the same as
$\I(M)$, as they have the same future chains.  Now
let $p$ be any element of $\bdm$,
and let $P$ be any past component of $p$ in
$\bm$.  Then $P$ contains some element $x$ which
has a timelike curve $\gamma$ connecting it to
$p$; in fact, $P = I^-[\gamma]$.  As above, $P$ is
in $\hat\partial(M)$ and must be inobservable in
$\hat M$.  Therefore, $P$ cannot be properly
contained in any IP of $M$, hence, not in any IP
of $\bm$.

For $M$ a spacetime with only spacelike
boundaries, we will be considering that class of
embeddings $\phi : M \to N$ which extend
continuously to $\hat\partial(M)$, yielding 
$\bar\phi : \hat M \to N$ (which will thus include
embeddings of multiply warped product spacetimes
continuously extending to $\{b\}\times K$).  Note
that in such a case, $\phi[M]$ is necessarily
disjoint from $\bar\phi[\hat\partial(M)]$:  For $P
\in \hat\partial(M)$ generated by a timelike curve
$\gamma$, $P$ is the future limit of $\gamma$,
hence, the topological limit of $\gamma$ in  $\hat
M$; therefore, $\bar\phi(P) = \lim
\phi(\gamma(t))$. If
$\bar\phi(P) = \phi(x)$, then $\phi\circ\gamma$
eventually enters every neighborhood of $\phi(x)$, 
so the same is true in $M$: $\gamma$ eventually
enters every neighborhood of $x$, i.e., $x$ is a
future endpoint of $\gamma$ in $M$.  But that makes
$P = I^-[\gamma]$ equal to $I^-(x)$, contradicting
the assumption that $P$ is in $\hat\partial(M)$.
(This does not depend on $M$ having spacelike
boundaries.)

Note that we can readily identify the past 
components of any element $p$ of $\bdm$:  If $P$ is
in $\hat\partial(M)$ and $\bar\phi(P) = p$, then let
$\gamma$ be a generating timelike curve for $P$.  As
$P$ is the future limit of $\gamma$, it is also its
topological limit, so $\bar\phi(P)$ must be $\lim
\phi(\gamma(t))$; therefore, $p = \lim
\phi(\gamma(t))$, so all of $\gamma$ lies in
$I_M^-(p)$.  Thus, $P \subset \bar I^-(p)$; as no IP
can properly contain $P$, it must therefore be a 
past component of $p$.  Furthermore, all past
components  of $p$ arise in this way:  The only IPs
available are those of $M$, and only elements of
$\hat\partial(M)$ can be maximal for lying in
$\bar I^-(p)$.  (Reason (see Figure 12): For any
$I^-(x)
\subset I^-_M(p)$, consider a future chain $\{x_n\}$
approaching $x$; for each $n$, there is a timelike
curve $c_n$ going, in essence, from $x_n$ to $p$. 
These have a causal limit curve $c$ beginning at
$x$.  Let $Q = I^-[c]$; this is an IP containing
$I^-(x)$.  For any $y \in Q$, there is some $t$ with
$y \ll c(t)$, so there is a sequence of numbers
$\{t_n\}$ with $\{c_n(t_n)\}$ approaching $c(t)$. 
Since $c(t) \in I^+(y)$, eventually $c_n(t_n) \in
I^+(y)$, i.e., $y \ll c_n(t_n)$.  This puts $y$ in
$I^-_M(p)$.  Thus, $Q \subset \bar I^-(p)$, so
$I^-(x)$ is not a maximal IP in $\bar I^-(p)$.)  If
$P = I^-[\gamma]$ is one such past component, then
for every $t$ there is a timelike curve $c_t$ from
$\gamma(t)$ to (essentially) $p$.  Let $P_t =
I^-[c_t]$.  For any $x \in P$, for $t$ sufficiently
large, $x \ll \gamma(t) = c_t(0)$, so $x \in P_t$,
i.e., $x \ll P_t$; therefore $P$ is the limit of
$\{P_t\}$ in the \h topology (making use of the
inobservability of $P$).  It follows that
$\bar\phi(P)$ is the limit of $\{\bar\phi(P_t)\}$. 
As shown above, for each $t$, $\bar\phi(P_t) = p$;
this means $\bar\phi(P)$ is the limit of the
constant sequence $\{p\}$, i.e., $\bar\phi(P) = p$.

Net result:  For any $p \in \bdm$, the past
components of $p$ are precisely the elements of
$\bar\phi^{-1}(p)$. 

Not every such embedding produces a
future completion in which the \h topology agrees
with the $\phi$-induced topology on $\bm$. 
Consider lower Minkowski  2-space for $M$, i.e.,
$\{(x,t) \in \lo^2 \;|\; t<0\}$, and $\phi : M \to
\lo^2$ given by $\phi(x,t) = (t\tan^{-1}x,t)$. 
Then the only element of $\bdm$ is
$(0,0)$ ($I_M^-((0,0))$ is all of $M$, but
$I_M^-(t\pi/2,t) = \emptyset$ for $t < 0$).  The
past components of $(0,0)$ are all the IPs of
the form $P_a = \{(x,t) \;|\; |x-a| < -t\}$ for $a
\in \r$.  

A sequence $\sigma$ approaches
the boundary point (0,0) in the \h topology if
for every $\tau \subset \sigma$ there is a $\rho
\subset \tau$ and some $a \in \r$ such that $P_a
\in \Cal L(\rho)$, i.e., $\{\rho(i)\}$ approaches
$(a,0)$ in the ordinary sense.  For $\sigma(n) =
(x_n,t_n)$, this is equivalent to $\{t_n\}$
approaching 0 and $\{x_n\}$ being bounded:  If a
set of numbers is bounded, then every subsequence
has a subsubsequence with a limit; and if a set of
numbers is unbounded, then it has a subsequence
with no such subsubsequence.  But in the
$\phi$-topology, $\{t_n\}$ going to 0 is all
that's needed for convergence to (0,0); the two
topologies differ on the convergence of 
$\{(n,-1/n)\}$.

It turns out that the crucial difference is
whether or not the extension of $\phi$ to
$\hat\partial(M)$ is proper onto its image.  (A
continuous map $f: X \to Y$ is defined to be proper 
if and only if for every compact $K \subset Y$,
$f^{-1}(K)$ is compact.  For second countable
spaces---such as everything considered in this
paper---a continuous map
$f: X \to Y$ is proper onto its image if and only if
every sequence in $X$ whose image under $f$ 
converges to something in $f[X]$, has a convergent
subsequence; thus, if $f$ is injective and
continuous, it is proper onto its image if and only
if it is homeomorphic onto its image.)  

If the entire extension $\bar\phi : \hat M \to N$ is
proper onto its image (which can be identified with
$\bm$), then its restriction to the
boundary, $\bar\phi_0 : \hat\partial(M) \to N$, is
also proper onto its image (which is $\bdm$); this
is a simple consequence of $\hat\partial(M)$ being
closed in $\hat M$ (Proposition 2.6).  But the
converse fails:  Consider $M$ to be lower Minkowski
2-space as above, $N$ the cylinder $\r^2
/\!\!\sim\,$ for $(x,t)\sim(x,t+\pi/2)$, and $\phi
: M \to N$ the map $\phi : (x,t) \mapsto
[x,\tan^{-1}t]$ (the equivalence class):  $\phi$ is
a homeomorphism onto its image (which just barely
fails to wrap once around the cylinder), and $\phi$
extends to a continuous $\bar\phi : \hat M \to N$
($\bar\phi(P_x) = [x,0]$ with notation as before)
for which the restriction $\bar\phi_0 :
\hat\partial(M) = \r \to N$ is proper onto its
image; but the full $\bar\phi :\hat M \to N$ is not
proper onto its image, as
$\{\bar\phi(x,-n)\} = \{[x,-\tan^{-1}n]\}$ converges
to $\bar\phi(P_x) = [x,0]$, but $\{(x,-n)\}$ does 
not converge to $P_x$. And note that the
$\phi$-topology for $\bm$ is that of the cylinder,
while the \h topology is that of the closed
half-plane (but $\bdm$ has the topology of $\r$ in
either case). 

Thus, we really have two separate theorems: 
One for the assumption that $\bar\phi$ is proper 
onto its image, informing the topology of
$\bm$; and one with the weaker assumption that
$\bar\phi_0$ is proper onto its image, informing 
just the topology of $\bdm$.  But we can handle them
simultaneously.

\proclaim{Theorem 5.3} Let $M$ be a strongly causal
spacetime with only spacelike boundaries, and let
$\phi : M \to N$ be a topological embedding of $M$
into a manifold of the same dimension, i.e., $\phi$ is
homeomorphic onto its image. 
\roster
\item
Suppose $\phi$ extends continuously (in the \h 
topology) to $\bar\phi: \hat M \to N$ so that
$\bar\phi$ is proper onto its image; then the \h
topology on $\bm$ is the same as the $\phi$-induced
topology.  
\item
Suppose that  $\phi$ extends continuously to
$\bar\phi: \hat M \to N$ so that the restriction
$\bar\phi_0$ of $\bar\phi$ to $\hat\partial(M)$ is
proper onto its image; then the \h topology on 
$\bdm$ is the same as the $\phi$-induced topology.
\endroster
\endproclaim

\demo{Proof} The method of proof is to show that 
$\bm$ (or $\bdm$),
in either topology, is homeomorphic to a quotient of
$\hat M$ (or of $\hat\partial(M)$), as given in 
Theorem 4.8 (or in Corollary 4.9).

Note first that for any $P \in \hat\partial(M)$,
$\bar\phi(P)$ is in $\bm$:  The IP $P$ is generated 
by some timelike curve  which can be parametrized as
$\gamma : [0,\infty) \to M$, and $P$, being the 
future limit of $\gamma$, is also the topological
limit of $\gamma$; thus, $\bar\phi(P) =
\lim_{t\to\infty} \phi(\gamma(t))$, so $\gamma(0) \in
I^-_M(\bar\phi(P))$.  Furthermore, $\bar\phi(P) \in
\partial^+_\phi(M)$, as was shown above.

Thus we have the continuous maps $\bar\phi : \hat M 
\to \bm$ and $\bar\phi_0 : \hat\partial(M) \to
\bdm$, with the \h topology on
the domains and the topology induced by $N$ (the
$\phi$-topology) on the targets.  In fact, these 
maps are onto:  Given a timelike curve $\gamma :
[0,\infty) \to M$ with $\phi\circ\gamma$ converging
to $p \in N$ and $p \notin \phi[M]$, let $P =
I^-[\gamma]$.  Then $P$ must be in
$\hat\partial(M)$, as otherwise
$I^-[\gamma] = I^-(x)$ for some $x$, and that makes 
$x$ the future endpoint of $\gamma$; then $\phi(x) =
\lim_{t\to\infty} \phi(\gamma(t)) = p$,
contradicting $p$ being in $\bdm$.  That allows us
to consider $\bar\phi(P)$, which must be
$\lim_{t\to\infty}
\phi(\gamma(t)) = p$.

This gives us the continuous bijections 
$\widetilde\phi : \hat M/\!\!\sim\; \to \bm$ and
$\widetilde\phi_0 :
\hat\partial(M)/\!\!\sim\; \to \bdm$, where
$\,\sim\,$ is essentially the same equivalence 
relation for both maps: $P \sim Q$ for $P$ and $Q$
in $\hat\partial(M)$ and $\bar\phi(P) =
\bar\phi(Q)$.  Let $\pi : \hat M \to \hat
M/\!\!\sim\,$ and $\pi_0 :
\hat\partial(M) \to \hat\partial(M)/\!\!\sim\,$ 
denote the projections to equivalence classes. 
Then for any compact $K$ in $\bm$,
$\widetilde\phi^{-1}[K] = \{\pi(z)
\;|\; z \in \hat M \text{ and } \bar\phi(z) \in K\} 
= \pi[\,\bar\phi^{-1}[K]\,]$.  We know
$\bar\phi^{-1}[K]$ is compact if $\bar\phi$ is 
proper, whence its image under $\pi$ is compact;
thus, if $\bar\phi$ is proper, so is
$\widetilde\phi$, and a proper, continuous
bijection is a homeomorphism.  Similarly, if
$\bar\phi_0$ is proper, then
$\widetilde\phi_0$ is a homeomorphism.  

We now want to employ the theorems of Section 4.  
Note that $\bm$ is generalized past-distinguishing: 
First, $M$ is past-distinguishing.  Second, if $x 
\in M$ and $p \in \bdm$ share a past component $P$,
then 
$P$ must be $I^-(x)$ and also $P$ must be in
$\bar\phi^{-1}(p)$; but that is impossible, as
$\bar\phi^{-1}(p)$ consists of elements of
$\hat\partial(M)$.  And last, if $p$ and $q$ in 
$\bdm$ share a past component $P$ then $\bar\phi(P)
= p$ and
$\bar\phi(P) = q$, so $p = q$.  Also note that $\bm$ is
generalized future-complete:  For any future chain
$c$ (which must lie wholly in $M$), let $P =
I^-[c]$; if $P \in \hat\partial(M)$, then we must 
find a generalized future limit for $c$.  Let $p =
\bar\phi(P)$; then $P \in \bar\phi^{-1}(p)$, so $P$ 
is a past component of $p$. 

For case (1) we can now apply Theorem 4.8, which 
tells us that $\bm$, in the \h topology, is a
topological quotient of $\hat M$; more
specifically, it says that
$\bm$ is $\hat M/\!\!\sim\,$ where $\sim$ identifies
elements of $\hat\partial(M)$ if they have the same
image under $\widetilde\phi^g$, where 
$\widetilde\phi^g : \hat M
\to \bm$ is the unique generalized future-continuous
map such that $\widetilde\phi^g \circ \hat\iota_M = 
\phi$ (from Proposition 4.6(3)).  This must be
$\bar\phi$:  Note that $\bar\phi \circ
\hat\iota_M = \phi$; also, $\bar\phi$ is generalized
future-continuous:  For any future chain $c$, if 
its future limit is $x \in M$, then the same is
true in $\bm$ (same chronology relation); and if
its future limit is $P = I^-[c] \in
\hat\partial(M)$, then $P$ is a past component of
$p = \bar\phi(P)$, making $p$ the generalized
future limit of $\phi[c]$.  Therefore,
$\bar\phi = \widetilde\phi^g$, and the quotient of
$\hat M$ given in Theorem 4.8 for the \h topology of
$\bm$ is precisely the same as the one above for the
$\phi$-topology.

For case (2) we apply Corollary 4.9.  Note that for 
a spacetime, $\hat\partial^g(M) = \hat\partial(M)$,
as a spacetime is regular. The maps and quotients
are exactly the same as above, with the same
result:  The quotient of $\hat\partial(M)$ given in
Corollary 4.9 for the \h topology of $\bdm$ is
precisely the same as for the one above for the
$\phi$-topology.  \qed \enddemo

It follows, for example, that not only is any
generalized past-distinguishing and generalized
future-completing boundary for interior 
Schwarzschild a quotient of $\r^1\times\s^2$ in the
\h topology for that completion (Corollary 4.9);
but that, furthermore, any completion for interior
Schwarzschild obtained by embedding it in another
manifold so that the embedding continuously extends
to the Future Chronological Boundary---the
singularity---in a manner which maps the
singularity properly onto its image, must be a
quotient of $\r^1\times\s^2$ in the induced
topology from the embedding (Theorem 5.3).  Thus is
$\r^1\times\s^2$ (the Future Chronological Boundary)
universal for boundaries on interior Schwarzschild.


\Refs
\widestnumber \key{GKP}

\ref
\key BE
\by J. K. Beem and P. E. Ehrlich
\book Global Lorentzian Geometry
\publ Marcel Dekker, New York \yr 1981
\endref

\ref 
\key BS
\by R. Budic and R. K. Sachs
\paper Causal boundaries for general relativistic
space-times
\jour J. Math. Phys. \vol 15 \yr 1974 \pages 1302--1309
\endref

\ref
\key GKP
\by R. P. Geroch, E. H. Kronheimer, and R. Penrose
\paper Ideal points in space-time
\jour Proc. Roy. Soc. Lond. A \vol 327 \yr 1972
\pages 545--567
\endref

\ref
\key H
\by S. G. Harris
\paper Universality of the Future Chronological
Boundary
\jour J. Math. Phys. \vol 39 \yr 1998 \pages
5427--5445
\endref

\ref
\key HD
\by S. G. Harris and T. Dray
\paper The causal boundary of the trousers space
\jour Class. Quantum Grav. \vol 7 \yr 1990 \pages
149-161
\endref

\ref
\key HE
\by S. W. Hawking and G. F. R. Ellis
\book The Large Scale Structure of Space-Time
\publ Cambridge University, Cambridge \yr 1973
\endref

\ref
\key M
\by S. Mac Lane
\book Categories for the Working Mathematician
\publ Springer-Verlag, New York \yr 1971
\endref

\ref
\key S
\by L. B. Szabados
\paper Causal boundary for strongly causal
spacetimes
\jour Class Quantum Grav. \vol 5 \yr 1988 
\pages 121-134
\endref

\ref
\key W
\by R. B. Wald
\book General Relativity
\publ The University of Chicago Press, Chicago
\yr 1984
\endref

\endRefs
\bye